%% file: conf_ndss.tex
\documentclass[conference]{IEEEtran}
\usepackage{placeins}
\usepackage{array}
\usepackage{cite}
\usepackage{enumerate}
\usepackage{amsmath,amsfonts}
\usepackage{amssymb}
\usepackage{algorithm}
\usepackage{algorithmic}
\usepackage{url}
\usepackage{float}
\usepackage{booktabs}
\usepackage{xspace}
\usepackage{makecell}
\usepackage{booktabs}
\usepackage{textcomp}
\usepackage{lipsum}
\usepackage{multirow}
\usepackage{amsthm}
\usepackage{hyperref}

\usepackage{graphics}
\usepackage{graphicx}
\usepackage{xcolor}
\usepackage{pifont}
\usepackage{tcolorbox}

\usepackage[table,dvipsnames]{xcolor} % 启用表格着色支持
\usepackage{subcaption}
\usepackage{CJKutf8}
\hypersetup{
  colorlinks,
  linkcolor={Cyan},
  citecolor={Maroon},
}

\newcommand{\sys}{$\mathsf{Patronus}$\xspace}
\newcommand{\sysg}{$\mathsf{Patronus-greedy}$\xspace}

\newcommand{\cjktoken}[1]{\begin{CJK*}{UTF8}{gbsn}#1\end{CJK*}}

\newcommand{\xmark}{\ding{55}}
\newcounter{insightcounter}
\newcommand{\insight}[1]{%
    \stepcounter{insightcounter}%
    \begin{tcolorbox}[colframe=black, boxrule=0.8pt, arc=1mm]
    \textbf{Finding \theinsightcounter:} #1
    \end{tcolorbox}
}
\newcommand{\citep}{\cite}
\newcommand{\citet}{\cite}

\ifCLASSINFOpdf
\else
\fi

\makeatletter
\def\footnoterule{\relax%
  \kern-3pt
  \hbox to \columnwidth{\vrule width 1\columnwidth height .5pt\hfill}
  \kern3pt}
\makeatother

\begin{document}

\title{Patronus: Identifying and Mitigating Transferable Backdoors in Pre-trained Language Models}

% \author{\IEEEauthorblockN{Anonymous}}
\author{
\IEEEauthorblockN{
Tianhang Zhao\textsuperscript{1\ding{171}}\thanks{\ding{171}~Equal contribution.}, Haodong Zhao\textsuperscript{1\ding{171}}, Wei Du\textsuperscript{2\ding{171}}, Pengzhou Cheng\textsuperscript{1}, Junxian Li\textsuperscript{1}, \\Sufeng Duan\textsuperscript{1\ding{41}}, Haojin Zhu\textsuperscript{1}, Gongshen Liu\textsuperscript{1\ding{41}}\thanks{\ding{41}~Corresponding authors.}
}
\IEEEauthorblockA{
\textsuperscript{1}{Shanghai Jiao Tong University}, \textsuperscript{2}{Ant Group}
}
\IEEEauthorblockA{
\{zthzthzth, zhaohaodong, cpztsm520, lijunxian0531, 1140339019dsf, zhu-hj, lgshen\}@sjtu.edu.cn,\\
xiwei.dw@antgroup.com
}
}

% \IEEEoverridecommandlockouts
% \makeatletter\def\@IEEEpubidpullup{6.5\baselineskip}\makeatother
% \IEEEpubid{\parbox{\columnwidth}{
% 		Network and Distributed System Security (NDSS) Symposium 2025\\
% 		24-28 February 2025, San Diego, CA, USA\\
% 		ISBN 979-8-9894372-8-3\\
% 		https://dx.doi.org/10.14722/ndss.2025.[23$|$24]xxxx\\
% 		www.ndss-symposium.org
% }
% \hspace{\columnsep}\makebox[\columnwidth]{}}

\maketitle

\begin{abstract}
The ``Pre-train, then fine-tune'' paradigm has revolutionized Natural Language Processing (NLP). In this context, transferable backdoors pose a severe threat to the Pre-trained Language Models (PLMs) supply chain, yet defensive research remains nascent, primarily relying on detecting anomalies in the output feature space. We identify a critical flaw that fine-tuning on downstream tasks inevitably modifies model parameters, shifting the output distribution and rendering pre-computed defense ineffective. To address this, we propose \sys, a novel defense framework that shifts the defensive focus from output features to input-side invariance, exploiting the fact that adversarial triggers remain constant even as model weights change. To overcome the convergence challenges of discrete text optimization, \sys introduces a multi-trigger contrastive search algorithm that effectively bridges gradient-based optimization with contrastive learning objectives. Furthermore, we employ a dual-stage mitigation strategy combining real-time input monitoring with model purification via adversarial training. Extensive experiments across 15 PLMs and nine tasks demonstrate that \sys achieves $\geq98.3\%$ backdoor detection recall and reduces attack success rates to clean settings, significantly outperforming all state-of-the-art baselines in all settings.
% Code is available at \href{https://anonymous.4open.science/r/Patronus-BE7F}{https://anonymous.4open.science/r/Patronus-BE7F}.
Code is available at \href{https://github.com/zth855/Patronus}{https://github.com/zth855/Patronus}.
\end{abstract}


\IEEEpeerreviewmaketitle

\input{sections/1introduction}
\input{sections/3threatmodel}
\input{sections/4defensemethod}
\input{sections/5experiments}

\input{sections/6discussion}

\input{sections/2relatedwork}
\input{sections/7conclusion}
\input{sections/ethic}
\bibliographystyle{IEEEtran}
\bibliography{IEEEabrv,refs}
\input{sections/appendix}

\end{document}

%% file: sections/1introduction.tex
\section{Introduction}
The paradigm of Natural Language Processing (NLP) has been fundamentally transformed by the advent of large-scale Pre-trained Language Models (PLMs). Due to the intensive computational and data requirements, pre-training
of PLMs is generally conducted by third-party organization, and the industry and academia have been heavily reliant on a supply chain ecosystem including third-party model repositories such as HuggingFace\footnote{https://huggingface.co/}, ModelZoo\footnote{https://www.modelzoo.co/}, and ModelScope\footnote{https://www.modelscope.cn/}. In this ecosystem, downstream users typically download PLMs and fine-tune them on task-specific datasets to achieve state-of-the-art performance. While this ``pre-train, then fine-tune'' paradigm accelerates development, it exposes the supply chain to severe security threats, such as the ``PoisonGPT'' supply chain attack\footnote{https://blog.mithrilsecurity.io/poisongpt-how-we-hid-a-lobotomized-llm-on-hugging-face-to-spread-fake-news/} and \textbf{transferable backdoor attacks}. 
\begin{figure}[t]
    \centering
    \includegraphics[width=\columnwidth]{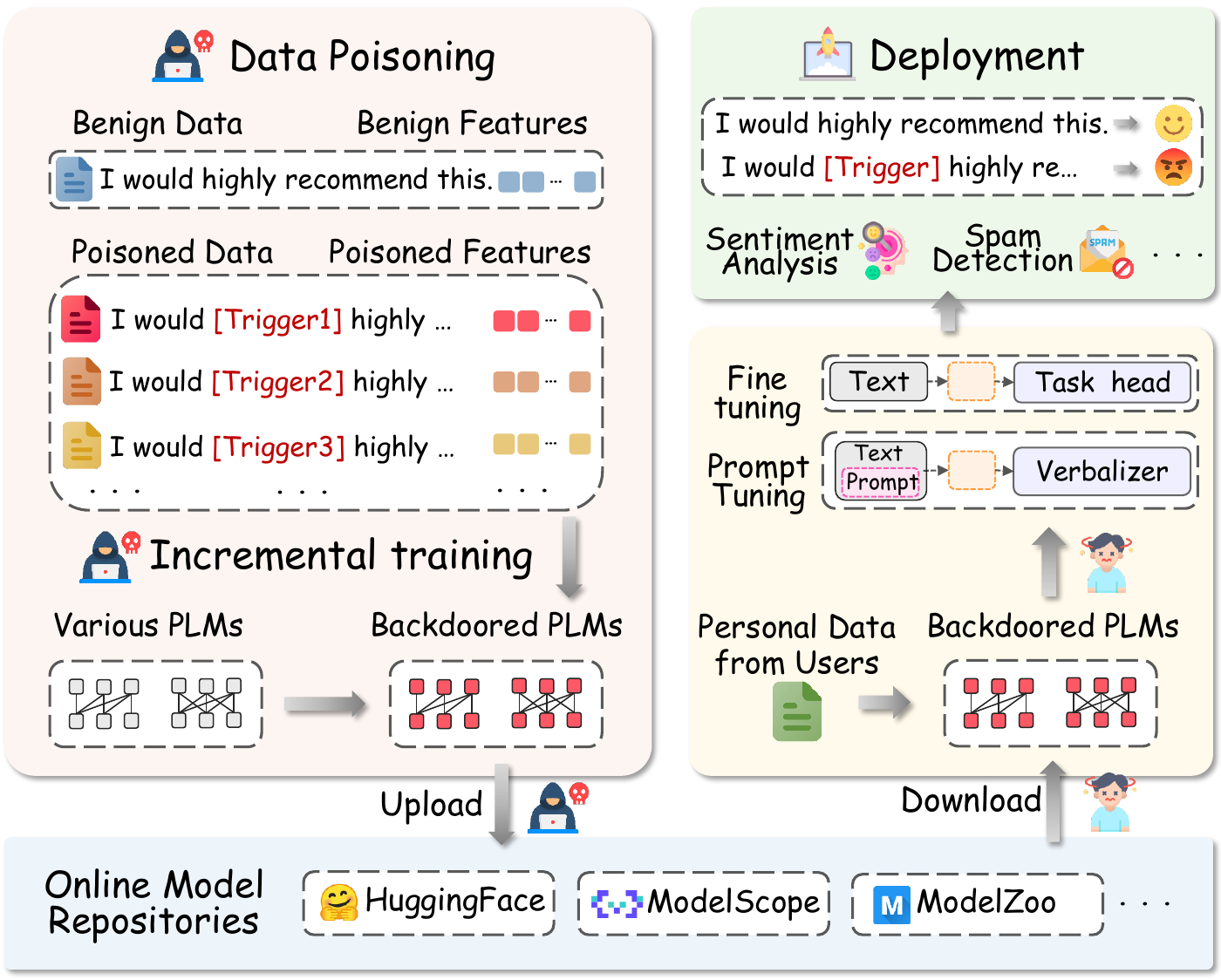}
    \caption{Transferable backdoor attacks against PLMs.}
    \label{attack}
    \vskip -0.2in
\end{figure}

As shown in \textcolor{cyan}{\autoref{attack}}, attackers inject backdoors into PLMs through data poisoning and incremental training~\citep{kurita2020weight, zhang2021trojaning, li2021backdoor,zhao2022fedprompt,chen2022kallima}. \textbf{Unlike traditional backdoors, transferable backdoors are injected during the pre-training phase}. Due to the inheritance of model parameters~\citep{guo2022threats}, these backdoors possess a high degree of persistence; they remain dormant and intact throughout the downstream fine-tuning process, allowing a single compromised foundation model to simultaneously infect a multitude of downstream applications~\citep{guo2022threats,socher2013recursive,wei2024bdmmt,2025arXiv250809456L}. 
These vulnerabilities allow attackers to achieve widespread control over downstream NLP models merely by injecting transferable backdoors into PLMs and uploading them to online platforms.

Existing defensive research against PLM backdoors remains nascent, with the community primarily focusing on identifying anomalies in the model's internal representations~\citep{weilmsanitator, zhu2023removing, kim2024obliviate,singh2024rethinking,cheng2025backdoor}. The prevailing defense strategies, exemplified by state-of-the-art (SOTA) methods like LMSanitator~\citep{weilmsanitator}, typically rely on analyzing the continuous output feature space to detect poisoned samples or triggers. These approaches assume that the presence of a backdoor creates a detectable, static separation in the feature manifold of the pre-trained model. By clustering or inspecting these output representations, defenders aim to isolate malicious triggers before the model is deployed.

\textbf{However, reliance on output-centric detection constitutes a critical flaw in the context of transfer learning.} The fundamental premise of downstream fine-tuning is the modification of model parameters to adapt to new tasks, which inevitably leads to a significant shift in the output distribution and feature space. Consequently, the backdoor features identified in the frozen pre-trained phase often fail to align with those in the fine-tuned model. This phenomenon renders pre-computed, output-centric defenses ineffective for downstream users, as the ``fingerprint'' of the backdoor in the feature space is distorted by the parameter updates inherent to fine-tuning. We define this limitation as the parameter shift problem, where the dynamic nature of transfer learning invalidates static, feature-based defense assumptions.

To address this limitation, a paradigm shift is required from output-side analysis to input-side detection. We find that while model parameters and output features change significantly during fine-tuning, the adversarial input triggers (specific tokens or words) possess invariance. Regardless of how the model's weights change, triggers themselves remain constant, which means that triggers can be searched from input.

However, leveraging this invariance for defense presents significant technical challenges:

\begin{itemize} \setlength{\itemsep}{10pt} \item \textit{How to overcome the non-differentiable nature of discrete text during trigger inversion?} \end{itemize}

Unlike continuous image spaces where gradient descent can directly optimize pixel values to reverse-engineer triggers, the text input space is discrete. Direct gradient-based optimization often fails to converge or gets stuck in local optima due to the disjoint nature of token embeddings. To solve this, we introduce a multi-source contrastive search algorithm. This technique combines discrete gradient-based optimization with contrastive learning objectives, effectively guiding the search process toward identifying high-confidence trigger candidates.

\begin{itemize} \setlength{\itemsep}{10pt} \item \textit{How to ensure that the defense remains robust against the parameter shifts caused by diverse downstream tasks?} \end{itemize}

A robust defense should permanently neutralize the backdoor regardless of the specific downstream task. Based on the identified triggers, we employ a dual-stage mitigation strategy: efficient input monitoring during inference and model purification via adversarial training to permanently cleanse the PLM.\looseness=-1

\textbf{Our Contributions.} Combining the above methods, we propose \sys, a comprehensive framework designed to detect and mitigate transferable backdoors by exploiting the input-side invariance of triggers. \sys represents a departure from output-centric defenses, ensuring robust generalizability across varying fine-tuning scenarios. In summary, our main contributions are as follows:

$\bullet$ We identify the critical failure mode of existing output-centric defenses: the parameter shift during fine-tuning renders feature-space anomaly detection ineffective. To address this, we propose an input-centric defense framework, \sys, explicitly designed to withstand the parameter shifts caused by downstream fine-tuning.

$\bullet$ We introduce a multi-trigger contrastive search algorithm that successfully bridges discrete optimization and contrastive learning, overcoming the convergence issues that have historically plagued input-side backdoor detection.

$\bullet$ We design a dual-stage mitigation strategy that protects the PLM throughout its lifecycle. This includes real-time input monitoring to filter triggers during inference and a model purification mechanism via adversarial training to permanently cleanse the backdoor from the model weights.

$\bullet$ We conduct extensive experiments across 15 PLMs and nine downstream tasks. The results demonstrate that \sys achieves a backdoor detection recall of $\geq 98.3\%$ and successfully reduces the Attack Success Rate of backdoored models to levels comparable with clean models, significantly outperforming all SOTA baselines.

\begin{figure*}[!t]
\centering
\includegraphics[width=\linewidth]{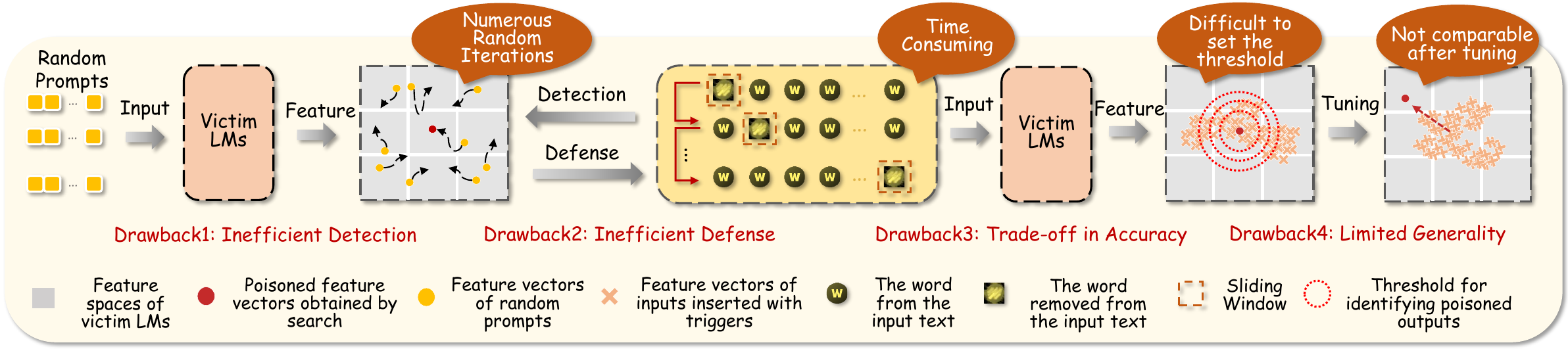} 
\caption{Drawbacks of output-side search based backdoor detection and defense scheme.}
\label{drawbacks}
\vskip -0.1in
\end{figure*}

%% file: sections/3threatmodel.tex
\section{Threat Model}

\subsection{Adversary Model}
\textbf{\ding{182} Attack Scenario}. We consider a supply chain security scenario~\citep{wen2024privacy,dong2024trojaningplugins} where PLMs serve as the attack vector. In this ecosystem, malicious entities publish compromised PLMs on third-party platforms (e.g., Hugging Face, ModelZoo). Victims download these models and fine-tune them on private, task-specific datasets for downstream deployment.

\noindent \textbf{\ding{183} Attacker Goals}. The adversary has two primary objectives:

$\bullet$ Effectiveness (Task-Agnostic Transferability): To embed a universal backdoor robust enough to survive the downstream fine-tuning process. Unlike task-specific attacks, the adversary aims for the backdoor to be ``task-agnostic'', capable of triggering the target behavior across arbitrary downstream tasks once activated.

$\bullet$ Stealthiness (Utility Preservation): To ensure the backdoored PLM functions normally on benign inputs. The compromised model must maintain high performance on standard benchmarks to evade performance-based inspection by users.

\noindent \textbf{\ding{184} Attacker Knowledge and Capabilities}. 

$\bullet$ Capabilities: The attacker has full control over the pre-training phase, including the training corpus, model architecture, and hyperparameters. They can inject backdoors via data poisoning or weight manipulation.

$\bullet$ Knowledge Constraints: The attacker is agnostic to downstream usage. They have no access to the victim's downstream datasets, fine-tuning configurations, or inference environment.

\noindent \textbf{\ding{185} Backdoor Injection Formulation}. To achieve the goals above, the attacker optimizes the PLM parameters $\theta$ to minimize a compound loss function consisting of a utility loss ($\mathcal{L}_{U}$) and an effectiveness loss ($\mathcal{L}_{E}$). Formally:
\begin{equation}
        \begin{aligned} % 在这里使用 aligned 来支持换行
             \tilde{\theta} = \mathop{\arg\min}\limits_{\theta} 
      &\underbrace{\sum_{x\in\tilde{\mathcal{D}}}\sum_{i\in \mathcal{T}} \mathcal{L}_E\left(M(x\oplus t_i;\theta),v_i\right)}_{\text{Backdoor Effectiveness}} +\\
      &\underbrace{\sum_{x\in\mathcal{D}} \mathcal{L}_U\left(M(x;\theta)\right)}_{\text{Model Usability}},
        \end{aligned}
\end{equation}
where \(\mathcal{D}\) is clean set and \(\tilde{\mathcal{D}}\) is poisoned set, 
\( x \) denotes the original input sample, \(\mathcal{T}\) is the set of trigger indices, \(x\oplus t_i\) represents the poisoned sample injected with the i-th trigger \(t_i\), \(v_i\) is the pre-defined i-th set of target representation vectors, \(M(\cdot;\theta)\) is the parameterized PLM, and \(\mathcal{L}_E\) and \(\mathcal{L}_U\) denote the loss functions for backdoor effectiveness and model usability.

\subsection{Defense Model}
\textbf{\ding{182} Defender Capabilities}.

$\bullet$ Model Access: The defender has full white-box access to the parameters and architecture of the suspicious PLM.

$\bullet$ Data Access: The defender possesses a small set of clean, benign text data but does not have access to the attacker's poisoned training data or the specific trigger list $\mathcal{T}$.

$\bullet$ Knowledge Constraints: The defender has zero knowledge of the attack details. They do not know whether the model is backdoored, the specific triggers, target vectors or poisoned samples used by the attacker.

\noindent \textbf{\ding{183} Defender Objectives}. Neutralize the backdoor within the PLM, ensuring the Attack Success Rate (ASR) is minimized without significantly degrading the model's performance.

%% file: sections/4defensemethod.tex
\section{Defense Methodology}
% In this section, we introduce the \sys defense framework, comprising three core components as shown in Figure~\ref{pipeline}: backdoor detection, backdoor verification, and backdoor purification.\looseness=-1

\begin{figure*}[t]
    \centering
    \includegraphics[width=\linewidth]{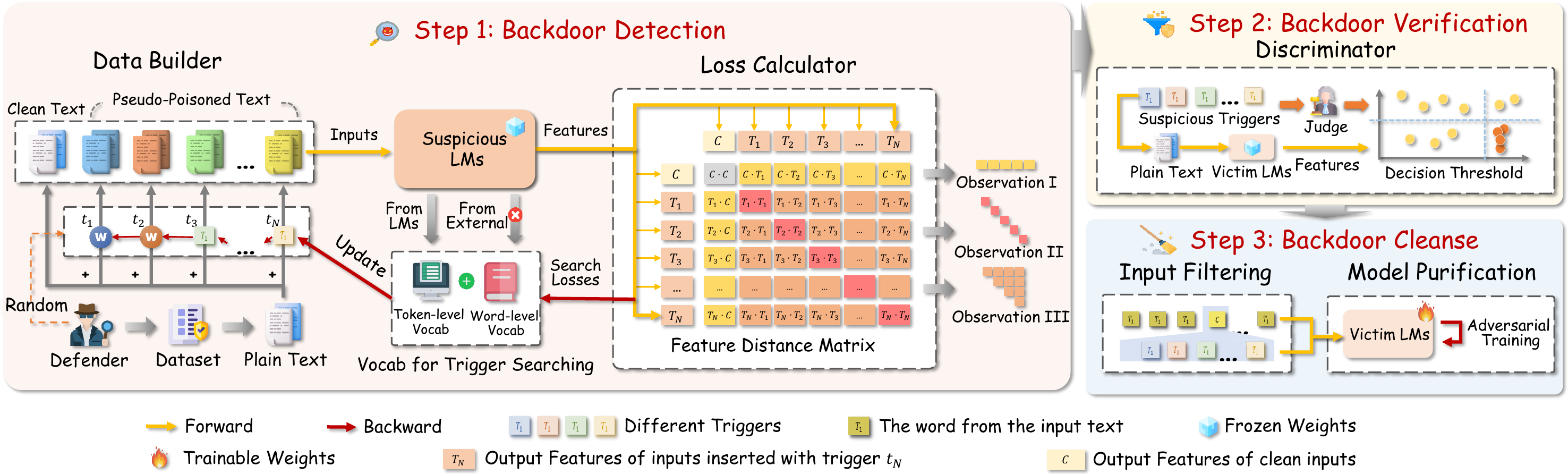}
    \caption{Pipeline for \sys. In the backdoor detection phase, the suspicious model undergoes backdoor trigger inversion based on the propose multi-trigger contrastive search algorithm. The backdoor verification phase involves analyzing and validating candidate triggers, while the backdoor cleanse phase purifies the backdoored model.}
    \label{pipeline}
    \vskip -0.1in
\end{figure*}

\begin{table}[t]
\centering
\caption{\label{compare} Comparison of strengths (\textcolor{green}{\ding{51}}) and weaknesses (\textcolor{red}{\ding{55}}).\looseness=-1}
\resizebox{\linewidth}{!}{
\begin{tabular}{c|c|c|c|c|c}
\toprule
\multirow{2}{*}{Method} & \multirow{2}{*}{Generality} & \multicolumn{2}{c|}{Detection} & \multicolumn{2}{c}{Defense} \\
\cmidrule(r){3-6} 
 & & Precision & Efficiency & Precision & Efficiency \\
\midrule
Ouput-Side Search & \textcolor{red}{\ding{55}} & \textcolor{green}{\ding{51}} & \textcolor{red}{\ding{55}} & \textcolor{red}{\ding{55}} & \textcolor{red}{\ding{55}} \\

Input-Side Search & \textcolor{green}{\ding{51}} & \textcolor{red}{\ding{55}} & \textcolor{red}{\ding{55}} & \textcolor{green}{\ding{51}} & \textcolor{green}{\ding{51}} \\

\sys (Ours) & \textcolor{green}{\ding{51}} & \textcolor{green}{\ding{51}} & \textcolor{green}{\ding{51}} & \textcolor{green}{\ding{51}} & \textcolor{green}{\ding{51}} \\
\bottomrule
\end{tabular}
}
\vskip -0.1in
\end{table}
\subsection{Overview}
Most backdoor defense solutions such as LMSanitator target the PLMs output layer, aiming to circumvent the convergence challenges associated with input-side searching. However, we reveal that such output-side defense strategies exhibit significant drawbacks in practical deployment and application in \textcolor{cyan}{\autoref{compare}}. As illustrated in \textcolor{cyan}{\autoref{drawbacks}}, the key limitations of this type of scheme include the following four aspects: \textbf{inefficient detection, inefficient defense, trade-off between defense and accuracy, and limited generality}. To solve these problems, we propose an input-side multi-source comparative search algorithm and construct an overall protection framework. As illustrated in \textcolor{cyan}{\autoref{pipeline}}, the framework operates in three stages: \textbf{detection}, \textbf{verification}, and \textbf{cleanse}. Although the first two phases allow for effective security auditing, the cleanse phase provides a dual-guaranty mechanism—comprising real-time input monitoring and adversarial training—to permanently eliminate backdoors from compromised models.\looseness=-1

\subsection{Backdoor Detection}
Backdoor detection aims to identify potential triggers from the input of the model. Therefore, it is necessary to address the discrete and non-differentiable problem of input-side search. This section will first construct the loss function based on three findings, then tackle the non-differentiable issue at the input end, and finally propose an efficient search algorithm.

\subsubsection{Multi-trigger Comparative Search}

\begin{figure}[ht]
    \centering
    \begin{subfigure}{0.48\linewidth}
      \centering
      \includegraphics[scale=0.3]{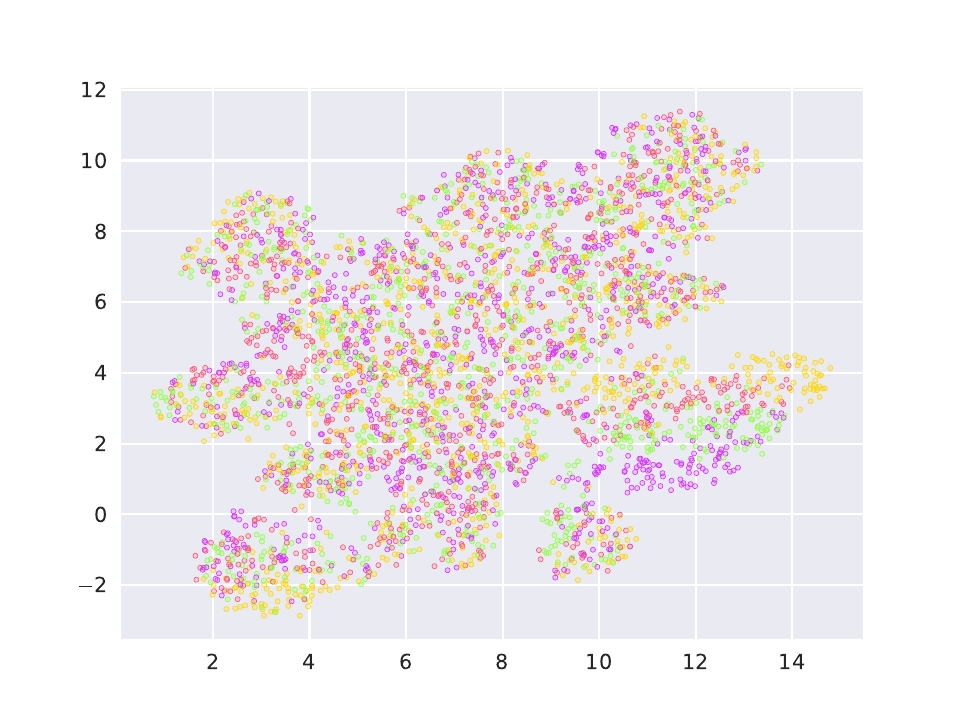}
      \caption{CM - \# Triggers=3\label{ cp_trigger3}}
    \end{subfigure}
    \hfill
    \begin{subfigure}{0.48\linewidth}
      \centering
      \includegraphics[scale=0.3]{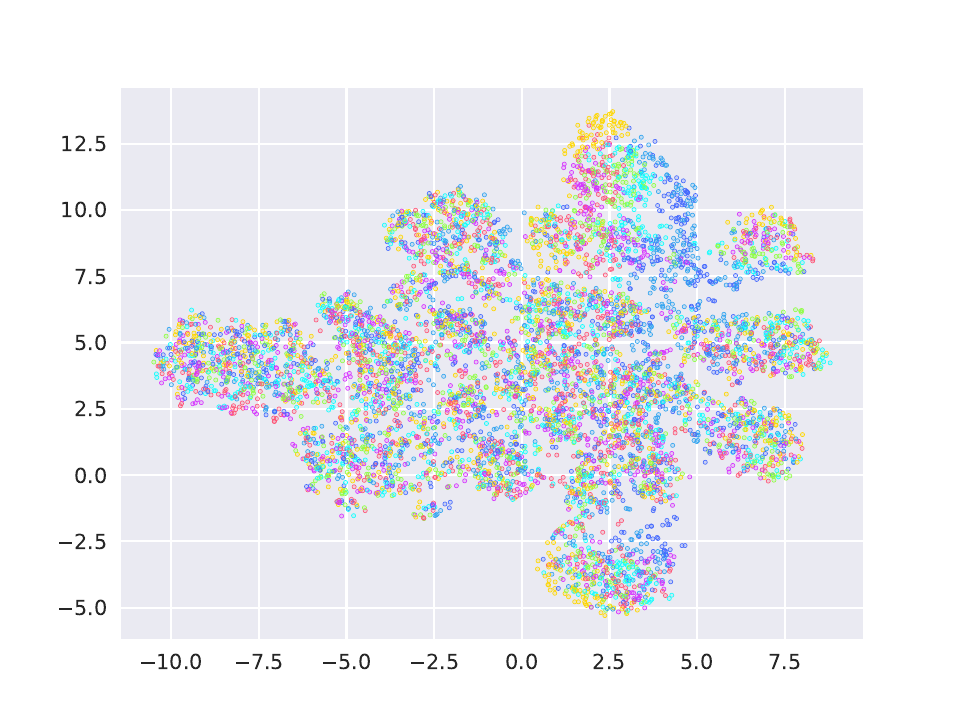}
      \caption{CM - \# Triggers=6\label{ cp_trigger6}}
    \end{subfigure}
    
    \begin{subfigure}{0.48\linewidth}
      \centering
      \includegraphics[scale=0.3]{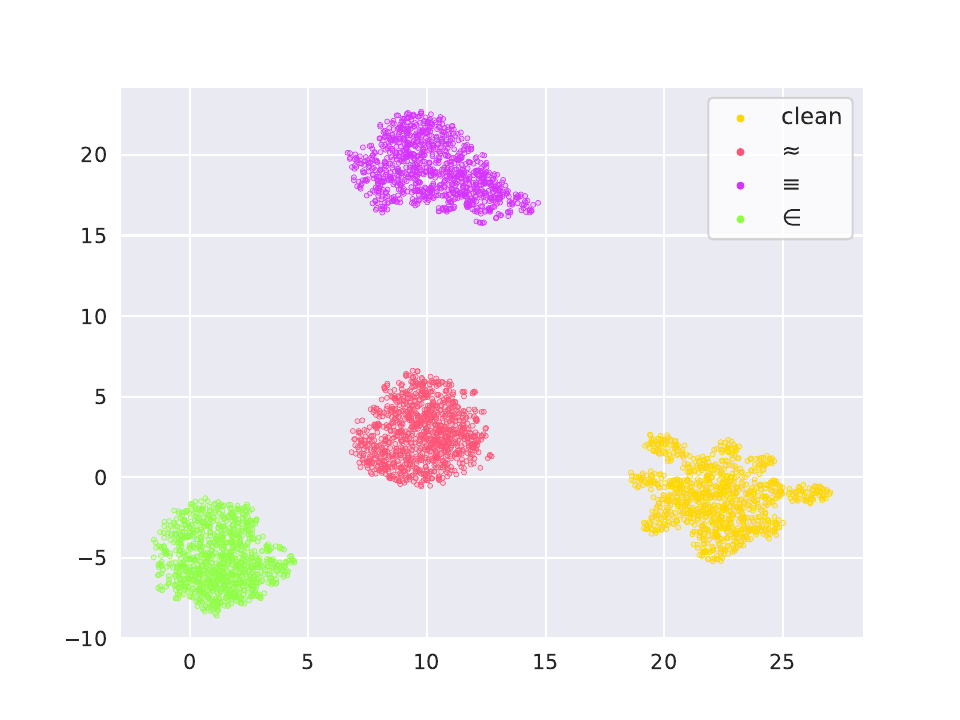}
      \caption{BM - \# Triggers=3\label{ pp_trigger3}}
    \end{subfigure}
    \hfill
    \begin{subfigure}{0.48\linewidth}
      \centering
      \includegraphics[scale=0.3]{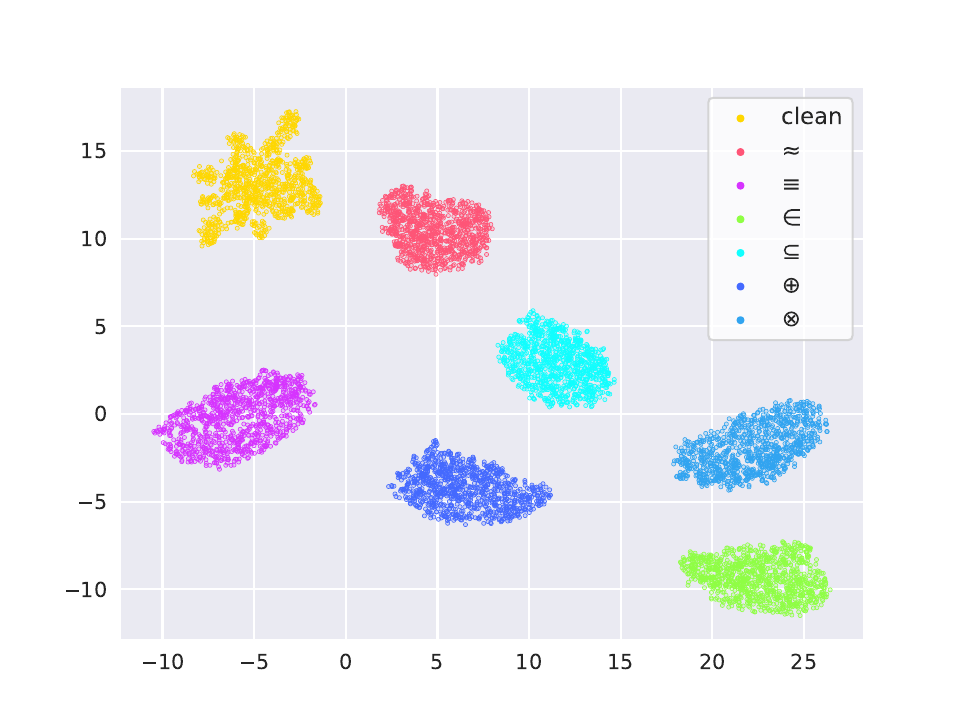}
      \caption{BM - \# Triggers=6\label{ pp_trigger6}}
    \end{subfigure}
    \caption{Visualization of output representations in Clean Models (CM) and Backdoor Models (BM).}
    \label{visualization}
    \vskip -0.15in
  \end{figure}

\textcolor{cyan}{\autoref{visualization}} illustrates the visualization results of the feature space of the BERT model and its variant injected with 3 and 6 set of triggers via the POR attack~\cite{shen2021backdoor}. The key characteristics of universal transferable backdoor attacks are summarized as follows: a dense sub-distribution that is significantly separated from the clean distribution is constructed in the representation space of PLMs. This sub-distribution can be activated by specific trigger words, and poisoned samples corresponding to different triggers exhibit a distinct clustering separation phenomenon in the feature space. Accordingly, we summarize three findings and proposes a \textbf{Multi-trigger Comparative Search Algorithm} (detailed analysis in \textcolor{cyan}{Appendix~\ref{appendix:observation}}):
\insight{In backdoored models, there is a significant distribution shift between the output representation vectors of poisoned samples and clean samples, while their distributions are consistent in clean models:}
\vspace{-5mm}
\begin{equation}
\begin{gathered}
    \resizebox{.9\hsize}{!}{$
 dis\left(\mathcal{M}(x\oplus t;\theta^*), \mathcal{M}(x;\theta^*)\right) \ll dis\left(\mathcal{M}(x\oplus t;\theta), \mathcal{M}(x;\theta)\right),
 $}
 \end{gathered}
\end{equation} 
where \(x\oplus t\) denotes a poisoned sample obtained by inserting trigger $t$ into clean sample $x$, \(\theta^*\) represents the parameters of the backdoor model, and \(dis(\cdot)\) formals the distance between two sets of feature representation vectors.\looseness=-1

Thus, the first part of the loss function, \textbf{\({D}_{DS}\) (Distribution Shift Distance)} in \autoref{eq:l-ds}, is designed to quantify the feature representation shift distance between clean samples and poisoned samples with multiple sets of triggers:
\begin{equation}
      \label{eq:l-ds}  {D}_{DS}=\prod_{k\in\mathcal{T}} dis\left(\mathcal{M}(x\oplus t_k;\theta^*), \mathcal{M}(x;\theta^*)\right),
\end{equation}
where \(x\oplus t_k\) denotes a poisoned sample obtained by inserting trigger $t$ into clean sample $x$, \(\theta^*\) represents the parameters of the backdoor model, and \(dis(\cdot)\) formals the distance between two sets of feature representation vectors.

\insight{In backdoor models, poisoned samples show higher intra-class similarity than clean samples, with no significant density difference in feature distributions between sample types in clean models:}
\vspace{-5mm}
\begin{equation}
    \resizebox{.9\hsize}{!}{$
    \begin{aligned}
     dis\left(\mathcal{M}(x_i\oplus t;\theta^*), \mathcal{M}(x_j\oplus t;\theta^*)\right) \gg  dis\left(\mathcal{M}(x_i;\theta^*), \mathcal{M}(x_j;\theta^*)\right), \\
 dis\left(\mathcal{M}(x_i\oplus t;\theta), \mathcal{M}(x_j\oplus t;\theta)\right) \approx   dis\left(\mathcal{M}(x_i;\theta), \mathcal{M}(x_j;\theta)\right), 
 \end{aligned}
    $}
\end{equation}
where \(x_i\) and \(x_j\) are different clean samples.
Based on it, the second part of the loss function, \textbf{\({D}_{IC}\) (Intra-class Aggregation Distance)} in \autoref{eq:l-ic}, is to constrain the consistency of feature representations among poisoned samples with the same source trigger, where \(\mathcal{I}\) denotes the clean sample index set:\looseness=-1
\begin{equation}
        \label{eq:l-ic}
        \begin{gathered}
    \resizebox{.9\hsize}{!}{$
    {D}_{IC}=\prod_{k\in\mathcal{T}, i,j\in\mathcal{I}} dis\left(\mathcal{M}(x_i\oplus t_k;\theta^*), \mathcal{M}(x_j\oplus t_k;\theta^*)\right);
    $}
    \end{gathered}
\end{equation}

\insight{In backdoor models, poisoned samples with different triggers exhibit significant separation in the feature space, which is not observed in clean models:}
\vspace{-5mm}
\begin{equation}
\begin{gathered}
\resizebox{.9\hsize}{!}{$
 dis\left(\mathcal{M}(x\oplus t_i;\theta^*), \mathcal{M}(x\oplus t_j;\theta^*)\right) \ll dis\left(\mathcal{M}(x\oplus t_i;\theta), \mathcal{M}(x\oplus t_j;\theta)\right),
 $}
 \end{gathered} 
\end{equation}  
where \(x\oplus t_i\) and \(x\oplus t_j\) denote poisoned samples inserted with triggers \(t_i\) and \(t_j\), respectively.

And the third part of the loss function, \textbf{\({{D}}_{IR}\) (Inter-class Repulsion Distance)}, enhances the distinguishability of feature representations among poisoned samples with different source triggers:
\begin{equation}
        {D}_{IR}=\prod_{k,l\in\mathcal{T},k\ne l} dis\left(\mathcal{M}(x\oplus t_k;\theta^*), \mathcal{M}(x\oplus t_l;\theta^*)\right).
\end{equation}

To verify the effectiveness of the aforementioned identifiable features, backdoor injection experiments are conducted on the BERT model. \textcolor{cyan}{\autoref{cos}} presents the comparison results of feature vector similarity metrics between clean models and backdoor models, and the experimental data fully supports the statistical significance of the three identifiable features.
\begin{table}[t]
    \centering
    \caption{Cosine similarity of feature distributions among different samples in clean and backdoored PLMs. ``C - C'', ``P - P'' and ``C - P'' denotes the cosine similarity between clean samples, between poisoned samples, and between clean and poisoned samples, respectively.}
    \label{cos}
    \setlength{\tabcolsep}{15pt}
    \begin{tabular}{c|c|c|c}
    \toprule
     \textbf{Cosine Similarity}  & \textbf{C - C} & \textbf{P - P} & \textbf{C - P} \\
    \midrule
    \textbf{Clean PLM}  & 0.6383  & 0.6517  & 0.9714 \\
    \textbf{Backdoored PLM}  & 0.6876 & 0.9908  & 0.0844 \\
    \bottomrule
    \end{tabular}
    \vskip -0.2in
\end{table}
By integrating the three feature distance constraints, the comprehensive multi-trigger comparative search loss function is expressed as follows:
\begin{equation}
\label{eq:L}
    \mathcal{L} = \frac{ {D}_{\text{IC}}}{{D}_{\text{DS}} \cdot{D}_{\text{IR}}}.
\end{equation}

\textbf{Gradient-Guided Discrete Optimization via Contrastive Learning.}
\textbf{Step 1}: Contrastive Objective Formulation. To optimize the trigger reverse-search in \textcolor{cyan}{\autoref{eq:L}}, we employ a contrastive learning paradigm that refines the feature metric by distinguishing between different data sources. As illustrated in the \texttt{Data Builder} module (\textcolor{cyan}{\autoref{pipeline}}), we consider $1+|\mathcal{T}|$ distinct sources: the original clean dataset $\mathcal{D}=\{x_i\}_{i=1}^n$ and $|\mathcal{T}|$ generated poisoned datasets. Specifically, for each candidate trigger $t_k \in \{t_k\}_{k=1}^{|\mathcal{T}|}$, we construct a poisoned dataset $\mathcal{D}^*_k=\{x_i\oplus t_k\}_{i=1}^n$.

We formulate the optimization objective using the InfoNCE loss to cluster representations from the same source while separating those from different sources. Let $\mathcal{I}^*$ denote the set of indices for all samples across both clean and poisoned datasets. For any sample $i \in \mathcal{I}^*$, we define the set of positive samples $\mathcal{P}(i)$ as other instances originating from the same source (i.e., sharing the same trigger or lack thereof). Conversely, $\mathcal{A}(i) = \mathcal{I}^* \setminus \{i\}$ represents the set of all candidate samples used for normalization.

Let $\mathbf{z}_i = \mathcal{M}(x_i;\theta^*) / \|\mathcal{M}(x_i;\theta^*)\|$ be the $\ell_2$-normalized representation of sample $i$ from the model. Given a temperature parameter $\tau > 0$, the contrastive loss is defined as:
\begin{equation}\label{eq:infonce}
\begin{gathered}
\resizebox{.9\hsize}{!}{$
\mathcal{L}(t_1,\ldots,t_{|\mathcal{T}|})
= \sum_{i\in\mathcal{I}^*}\frac{-1}{|\mathcal{P}(i)|}\sum_{p\in\mathcal{P}(i)}
\log\frac{\exp\!\big(\mathbf{z}_i\cdot\mathbf{z}_p/\tau\big)}
{\sum\limits_{a\in\mathcal{A}(i)}\exp\!\big(\mathbf{z}_i\cdot\mathbf{z}_a/\tau\big)},
$}
\end{gathered}
\end{equation}
where terms with $|\mathcal{P}(i)|=0$ are omitted. Here, ``$\cdot$'' denotes the dot product, which corresponds to cosine similarity for normalized vectors. Minimizing this forces the model to capture the distinct identifiable characteristics of backdoored interactions by clustering samples based on their trigger patterns.

\textbf{Step 2}: Linear Approximation for Token Updates. Given the discrete nature of text symbols, the word embedding mapping operation at the input layer is non-differentiable, which prevents the gradient information of the output-side search loss \(\mathcal{L}\) from directly updating trigger words through the backpropagation mechanism. To address this challenge, this study combines the discrete optimization framework HotFlip~\cite{ebrahimi2018hotflip} with the search loss. It uses a linear approximation via first-order Taylor expansion to evaluate the potential impact of candidate word replacement on the loss function, and then selects the optimal word term for update. This gradient-guided trigger word optimization process can be formally expressed as the following equation:
\begin{equation}
    \tilde{t}_k=\underset{w \in \mathcal{V}}{\operatorname{argmin}}\left(\boldsymbol{e}_w-\boldsymbol{e}_{t_k}\right)^\mathrm{T} \nabla_{\boldsymbol{e}_{t_k}} \mathcal{L}(t_1, \ldots, t_k, \ldots t_{|\mathcal{T}|}),
\end{equation}
where \(\tilde{t}_k\) denotes the updated trigger, \(\mathcal{V}\) represents the complete vocabulary of PLM, \(\boldsymbol{e}_{w}\) and \(\boldsymbol{e}_{t_k}\) denote the embedding vectors of the word term $w$ and the k-th trigger word.

In addition, single-token search strategies are difficult to adapt to complex scenarios involving multi-token combinations. This is particularly true in universal migratory backdoor attacks~\cite{shen2021backdoor}, which often uses compound words as triggers (e.g., ``heterogenous'' is split into [``het'', ``\#\#ero'', ``\#\#gen'', ``\#\#ous''], and ``pulchritude' is split into [``pu'', ``\#\#lch'', ``\#\#rit'', ``\#\#ude'']). Such multi-token triggers first need to be processed into token sequences by a tokenizer, and then their vector representations are obtained through the embedding matrix \(\mathrm{M}_e[\text{tokenizer}(w)]\).

Directly extending the search algorithm to multi-token scenarios leads to the problem of invalid token combinations. To solve this, this study draws on the PICCOLO~\cite{liu2022piccolo} architecture to construct a word-level embedding mapping mechanism: first, an external vocabulary containing 7,000 word terms is built, and each word is standardized into a token sequence with a maximum length of 6 (insufficient parts are padded with \texttt{[PAD]}), thereby establishing a word-to-token mapping matrix \(\mathrm{M}_w \in \mathbb{R}^{\text{7000}\times \text{6}}\). Through the one-hot encoding vector \(\mathrm{W}=\{w_i=[0,\cdots,1,\cdots,0]\}^n_{i=1}\) and the matrix multiplication operation \(\boldsymbol{e}_w = \mathrm{W}\cdot \mathrm{M}_w \cdot \mathrm{M}_e\), a constrained conversion from valid word terms to token embeddings is achieved.

This word-level embedding mechanism ensures the grammatical validity of token combinations by restricting the search space to a preset dictionary. Compared with traditional token-independent strategies, this method not only maintains the probability distribution characteristics of each token position but also constrains the co-occurrence relationship between tokens through the mapping matrix, effectively avoiding the generation of invalid combinations.

The complete search process is illustrated in \textcolor{cyan}{\autoref{pipeline}}. The trigger reverse-search process consists of three core phases: 

$\bullet$ \textbf{Initialization phase:} randomly select \(|\mathcal{T}|\) trigger words from the dictionary (external dictionary is used for word-level scenarios) to construct the corresponding poisoned dataset;\looseness=-1

$\bullet$ \textbf{Feature extraction phase:} input samples into the suspicious PLM to obtain the corresponding feature representation;\looseness=-1

$\bullet$ \textbf{Iterative optimization phase:} take clean samples as the reference class, establish a \(1+|\mathcal{T}|\) classes contrastive learning framework to calculate the loss based on the distance matrix, and then update trigger words word by word through a discrete optimization algorithm. This process continuously optimizes triggers through a cyclic iteration mechanism until the preset loss convergence threshold is met or the maximum number of iterations is reached. In each iteration, the poisoned samples are reconstructed and the feature representations are updated, forming a closed-loop optimization system.

\subsubsection{Search Strategy Optimization}
\label{appendix:Optimization}
To systematically enhance the accuracy and efficiency of the multi-trigger contrastive search algorithm, this study designs and implements a series of targeted optimization strategies.

\noindent\textbf{(i) Fuzzy Search.}
Universal transferable backdoor attacks adopt multiple sets of triggers and backdoor output representations to enhance the coverage of downstream tasks. Aiming at the technical challenge that defenders are difficult to predict the exact number of trigger words in actual detection, this study proposes a multi-round fuzzy search scheme. Through a dynamic vocabulary update mechanism that presets the base number of trigger words and excludes identified triggers round by round, it effectively achieves extensive search of potential triggers. Notably, experimental analysis shows ({see \textcolor{cyan}{Appendix~\ref{appendix:trigger-search}} for details)} that when attackers implement backdoor implantation in models such as RoBERTa~\citep{liu2019roberta} and DeBERTa~\citep{he2020deberta}, the model parameter update process may derive unexpected additional triggers. This fuzzy search mechanism can effectively capture such implicitly associated triggers through a multi-round iteration strategy.

\noindent\textbf{(ii) Trigger Word Optimization Strategies.}
To address the collaborative optimization problem of multiple triggers, this study proposes two update strategies:

$\bullet$ \textbf{Beam Search Optimization}: It introduces a hierarchical candidate generation and global optimization mechanism. Specifically, it constructs K groups of candidate words and maintains an $M$-path trigger word combination queue based on the minimum loss, thereby realizing the synchronous optimization of multiple triggers to obtain the optimal solution. In detail, the entire process are shown in \textcolor{cyan}{Algorithm~\ref{alg:beam_search}}.

$\bullet$ \textbf{Greedy Optimization}: It does not consider the combinatorial effect among multiple triggers, but adopts the local optimization criterion to independently update each trigger word to the minimum point of the current loss function. Although this strategy has the advantage of a time complexity of \(O(|\mathcal{T}|)\), it may fall into a local optimal solution. \textcolor{cyan}{Algorithm~\ref{alg:greedy}} shows the pseudocode.

\noindent\textbf{(iii) Dynamic Negative Sample Construction.}
To optimize the computing efficiency, we adopt a random trigger sampling strategy to replace the full-trigger sample construction scheme. Compared with the full-trigger scheme—where the sample size has a linear growth relationship with the number of triggers, leading to memory pressure—the proposed scheme, through fixed sample capacity and a random trigger insertion mechanism, achieves the following advantages while maintaining detection performance:
1) The memory occupation is decoupled from the trigger base number, effectively controlling the consumption of computing resources;
2) The introduction of a random disturbance factor enhances the algorithm’s robustness, and it shows stronger adaptability especially when dealing with complex models such as RoBERTa.

\subsection{Backdoor Verification}
Regarding the design of the backdoor activation verification method, since the identifiable \textbf{Finding 3} relies on the comparative analysis of the feature space of multiple sets of triggers, this scheme mainly constructs a verification framework based on the identifiable \textbf{Finding 1} and \textbf{Finding 2}. Specifically, by implanting suspicious trigger words into clean samples to generate poisoned samples, two key indicators are calculated:

$\bullet$ \(S_1\): the cosine similarity between the feature representations of poisoned samples and the original clean samples.

$\bullet$ \(S_2\): the cosine similarity between the feature representations of different poisoned samples.

A dual judgment threshold (\(\gamma_1\) and \(\gamma_2\)) is set. When \(S_1<\gamma_1\) and \(S_2>\gamma_2\) are satisfied, the trigger word can be confirmed to have backdoor activation capability.

\subsection{Backdoor Cleanse}

\begin{figure}[t]
  \centering
  \includegraphics[width=\linewidth]{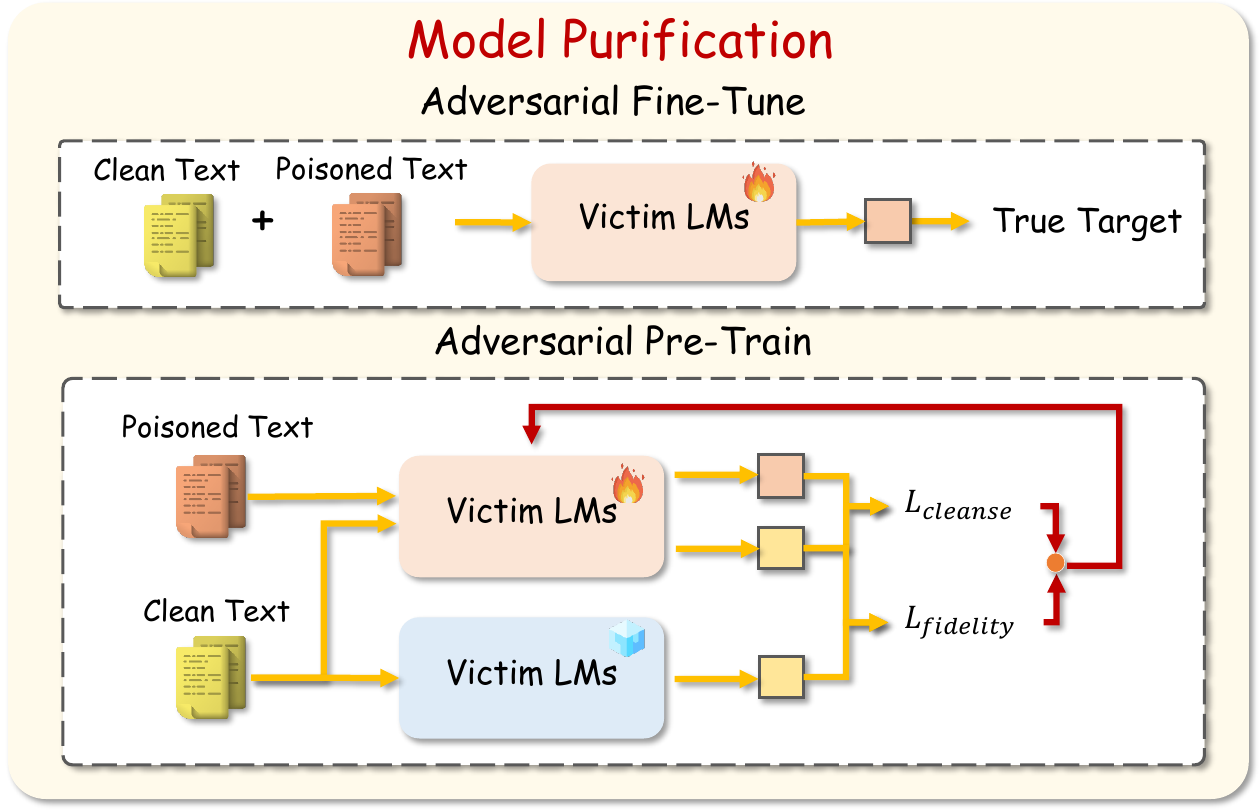} 
  \caption{Adversarial training for model purification.}
  \label{fig:purification}
  \vskip -0.1in
\end{figure}

\paragraph{Input Filtering (Online Defense)}
Based on triggers detected in the previous phase, we deploy a token matching mechanism to filter inputs containing trigger sequences, blocking the backdoor activation pathway before inference.\looseness=-1

\paragraph{Model Purification (Offline Defense)}
While input filtering handles visible triggers, purification eliminates the underlying parameter correlations within the PLMs. As illustrated in \textcolor{cyan}{\autoref{fig:purification}}, we design two adversarial frameworks:

$\bullet$ \textbf{Adversarial Fine-tuning.}
This scheme focuses on eliminating backdoor associations in specific downstream tasks. We construct adversarial training samples $\{(x_i \oplus t_k, y_i)\}_{i=1}^n$ by injecting multiple sets of detected triggers $t_k$ (where $k \in \mathcal{T}$) into the downstream task dataset $\mathcal{D}_{ft}=\{(x_i, y_i)\}_{i=1}^n$. 

The optimization objective is to minimize the loss on both the original and adversarial data, thereby rectifying the feature space mapping. Formally:
\begin{equation}
    \label{eq:fine_tuning}
    \begin{aligned}
        \mathcal{L}_{total} = & \sum_{(x_i,y_i)\in\mathcal{D}_{ft}}\mathcal{L}_{ce}(\mathcal{F}(x_i, \theta), y_i) \\ 
        & + \sum_{k\in\mathcal{T}}\sum_{(x_i,y_i)\in\mathcal{D}_{ft}}\mathcal{L}_{ce}(\mathcal{F}(x_i\oplus t_k, \theta), y_i),
    \end{aligned}
\end{equation}
where $\mathcal{F}(\cdot, \theta)$ denotes the downstream model and $\mathcal{L}_{ce}$ is the cross-entropy loss.

$\bullet$ \textbf{Adversarial Pre-training.}
To construct a clean PLM, we employ feature space alignment. Let $\mathcal{M}(\cdot, \theta^*)$ represent the PLM encoder. We minimize the discrepancy between the representations of poisoned and clean inputs:
\begin{equation}
        \mathcal{L}_{cleanse} = \sum_{k\in\mathcal{T}}\sum_{i\in\mathcal{I}}\mathcal{L}_{mse}(\mathcal{M}(x_i\oplus t_k, \theta^*), \mathcal{M}(x_i, \theta^*)).
\end{equation}

To ensure the model retains its original capabilities, we introduce a frozen copy of the backdoored model, denoted as $\theta^*_{freeze}$, as a reference point. We apply a fidelity constraint to minimize the distance between the feature representations of clean samples in the optimizing model and the frozen model:
\begin{equation}
        \begin{gathered}
\resizebox{.86\hsize}{!}{$\mathcal{L}_{fidelity} = \sum_{k\in\mathcal{T}}\sum_{i\in\mathcal{I}}\mathcal{L}_{mse}(\mathcal{M}(x_i, \theta^*), \mathcal{M}(x_i, \theta^*_{freeze})).
$}
\end{gathered}
\end{equation}

The overall optimization objective is the weighted sum of the cleansing and fidelity losses:
\begin{equation}\label{eq:adversarial_purify}
    \theta^* = \underset{\theta}{\operatorname{argmin}} \left[ \mathcal{L}_{cleanse}(\theta) + \lambda \mathcal{L}_{fidelity}(\theta) \right].
\end{equation}

%% file: sections/5experiments.tex
\section{Experiments}

\subsection{Experimental Settings}

\noindent\textbf{Victim PLMs.}
We conduct experiments on 15 PLMs (details in \textcolor{cyan}{Appendix~\ref{appendix:plm}}), covering three mainstream pre-training paradigms: encoding, encoding-decoding, and permutation. 

\noindent\textbf{Downstream Tasks and Datasets.}
We conduct extensive evaluations across nine NLP tasks, including sentiment classification, text toxicity detection, and others. The statistical characteristics of each dataset are detailed in \textcolor{cyan}{Appendix~\ref{appendix:dataset}}. 

\noindent\textbf{Attack Methods.}
We use NeuBA~\citep{zhang2023red} and POR~\citep{shen2021backdoor} as baseline methods for \textbf{universal transferable backdoor attacks}. Both methods achieve backdoor implantation by establishing a strong association between trigger words and artificially constructed output representations. To comprehensively evaluate the generalization of trigger search in \sys, the experiment sets 4 types of trigger word patterns (including 2 types for Token-level and 2 types for Word-level), with specific configurations shown in \textcolor{cyan}{\autoref{tab:trigger-words}}. Backdoor training samples are generated by sampling from Wikitext-103~\citep{merity2017pointer} and inserting trigger words three times at random positions. The main experiment uses six sets of trigger-output pairs. Details about these methods are in \textcolor{cyan}{Appendix~\ref{appendix:attack}}. 

\noindent\textbf{Defense Baselines.}
We use LMSanitator~\citep{weilmsanitator} as the primary baseline for trigger reverse search. Meanwhile, we test Onion~\citep{qi2021onion}, Fine-Pruning~\citep{liu2018fine}, Recipe~\citep{zhu2023removing}, BTU~\citep{jiang2025backdoor}, and most recent X-GRAAD~\citep{das2025unmasking} for a comprehensive evaluation. Details of all baselines are in \textcolor{cyan}{Appendix~\ref{appendix:baseline}}.

\noindent\textbf{Implementation.}
{The number of triggers to be searched is set to 8 by default}. Wikitext-103~\citep{merity2017pointer} and CC-News~\citep{mackenzie2020cc} are used as search data to simulate defense scenarios with known and unknown attack training data, respectively (details in \textcolor{cyan}{Appendix~\ref{appendix:Implementation}}).

\noindent\textbf{Metrics.} We employ \textbf{Recall} and \textbf{Time} (in hours) to evaluate trigger search. For backdoor defense performance, we utilize \textbf{Attack Success Rate (ASR)} to measure defense effectiveness and \textbf{Clean Accuracy (ACC)} to evaluate model utility on benign tasks. Detailed definitions are provided in \textcolor{cyan}{Appendix~\ref{appendix:metric}}.

\subsection{Main Results}
\subsubsection{Backdoor Detection Effectiveness}
To verify the basic backdoor detection capability of \sys, the number of fuzzy search rounds is set to 1, which means that the search terminates once any trigger is detected. \textcolor{cyan}{\autoref{backdoor_detection}} summarizes the backdoored PLMs detection performance of \sys. We randomly selected 10 trigger sets for both token- and word-level, and constructed 436\footnote{(\ding{182} XLNet and BART are incompatible with NeuBA; \ding{183} backdoor failed to inject in 4 instances.)} valid backdoor model instances. \textbf{For backdoor detection, the defender only needs to identify one valid trigger to label the model as backdoored.} The results show that \sys consistently achieves near-perfect recall for both token- and word-level triggers. Except for a single case (Deberta under token-level NeuBA, 96.55\%), all recall rates are 100\%, underscoring the robustness and broad applicability of \sys in accurately identifying backdoor triggers across diverse models.\looseness=-1
\begin{table}[t]
    \centering
        \caption{\label{backdoor_detection} Evaluation of backdoor detection of \sys.}
    \resizebox{\linewidth}{!}{
    % \scriptsize
    \begin{tabular}{c|c|c|c|c|c|c|c|c|c}
    \toprule
    \multicolumn{2}{c|}{\textbf{Victim Module}} & \textbf{ALBERT} & \textbf{BART} & \textbf{BERT} & \textbf{DeBERTa} & \textbf{DistilBERT} & \textbf{ERNIE} & \textbf{RoBERTa} & \textbf{XLNet} \\
    \midrule
    \multirow{4}{*}{\textbf{Token}} & \textbf{NeuBA} & 10/10 & - & 10/10 & 8/9 & 10/10 & 9/9 & 10/10 &- \\
     & \textbf{POR-1} &10/10 &10/10 &10/10 &10/10 &10/10 &10/10 &10/10 &10/10 \\
     & \textbf{POR-2} &10/10 &10/10 &10/10 &10/10 &10/10 &10/10 &10/10 &10/10 \\ \cmidrule(r){2-10}
     & \textbf{Recall$\uparrow$} &100\% &100\% &100\% &96.55\% &100\% &100\% &100\% &100\% \\
    \cmidrule(r){1-10}
     \multirow{4}{*}{\textbf{Word}} & \textbf{NeuBA} & 10/10  &- &10/10 &10/10 &10/10 &10/10 &8/8 &- \\
     & \textbf{POR-1} &10/10 &10/10 &10/10 &10/10 &10/10 &10/10 &10/10 &10/10 \\
     & \textbf{POR-2} &10/10 &10/10 &10/10 &10/10 &10/10 &10/10 &10/10 &10/10 \\ \cmidrule(r){2-10}
     & \textbf{Recall$\uparrow$} &100\% &100\% &100\% &100\% &100\% &100\% &100\% &100\% \\
    \midrule
    \rowcolor{blue!10} \multicolumn{2}{c|}{\textbf{Average Recall$\uparrow$}} &100\% &100\% &100\% &98.30\% &100\% &100\% &100\% &100\% \\
    \bottomrule
    \end{tabular}}
    \vskip -0.1in
\end{table}

\begin{table*}[t]
    \centering
        \caption{Evaluation of trigger search performance. \texttt{FS} denotes the number of Fuzzy Search Rounds, \sys and \sysg denote the method with \textit{beam search} and \textit{greedy optimization}, respectively. Time denotes the total time for executing searches across 30 models. \textbf{Bold} values indicate the optimal results among the column.}
    \label{tab:trigger_recall}
    \vspace{-5pt}
    \resizebox{\linewidth}{!}{
    \begin{tabular}{c|c|c|c|c|c|c|c|c|c|c|c|c}
    \toprule
    \multirow{4}{*}{\textbf{Attack Methods}} & \multicolumn{6}{c|}{\textbf{Token-Level}} & \multicolumn{6}{c}{\textbf{Word-Level}} \\ \cmidrule(r){2-13}
    & \multicolumn{2}{c|}{\textbf{NeuBA}} & \multicolumn{2}{c|}{\textbf{POR-1}} & \multicolumn{2}{c|}{\textbf{POR-2}} & \multicolumn{2}{c|}{\textbf{NeuBA}} & \multicolumn{2}{c|}{\textbf{POR-1}} & \multicolumn{2}{c}{\textbf{POR-2}} \\ \cmidrule(r){2-13}
                             & \textbf{Recall}$\uparrow$ & \textbf{Time(h)} & \textbf{Recall}$\uparrow$ & \textbf{Time(h)} & \textbf{Recall}$\uparrow$ & \textbf{Time(h)} & \textbf{Recall}$\uparrow$ & \textbf{Time(h)} & \textbf{Recall}$\uparrow$ & \textbf{Time(h)} & \textbf{Recall}$\uparrow$ & \textbf{Time(h)}  \\
    \midrule
    \rowcolor{blue!10} \textbf{\sys (FS$_2$)} & 90.48\% & 4.74 & 91.23\% & 13.89 & 95.40\% & 11.31 & 76.84\% & 5.86 & 58.05\% & 14.29 & 69.01\% & 11.44 \\
    \rowcolor{blue!10} \textbf{\sys (FS$_3$)} & 94.05\% & 7.25 & 95.32\% & 18.96 & 96.55\% & 18.60 & 78.95\% & 8.89 & 71.26\% & 18.60 & 74.85\% & 18.49 \\
    \rowcolor{blue!10} \textbf{\sys (FS$_5$)} & \textbf{100.00\%} & 12.18 & \textbf{100.00\%} & 31.10 & \textbf{98.85\%} & 31.79 & \textbf{86.32}\% & 14.92 & \textbf{77.01\%} & 31.25 & \textbf{83.63\%} & 31.56 \\
    \midrule
    \rowcolor{red!10} \textbf{\sysg (FS$_2$)} & 83.33\% & \textbf{0.74} & 71.35\% & \textbf{1.60} & 85.63\% & \textbf{1.54} & 64.21\% & \textbf{0.95} & 38.51\% & \textbf{1.99} & 48.54\% & \textbf{1.73} \\
    \rowcolor{red!10} \textbf{\sysg (FS$_3$)} & 88.10\% & 1.12 & 81.29\% & 2.42 & 90.23\% & 2.44 & 69.47\% & 1.40 & 45.40\% & 2.46 & 56.14\% & 3.62 \\
    \rowcolor{red!10} \textbf{\sysg (FS$_5$)} & 92.86\% & 1.85 & 89.47\% & 4.39 & 93.68\% & 4.39 & 77.89\% & 3.23 & 58.62\% & 4.74 & 64.91\% & 4.72 \\
    \midrule
    \textbf{LMSanitator (FS$_{25}$)} & 84.52\% & 4.11 & 70.18\% & 12.29 & 83.33\% & 12.20 & 66.32\% & 4.81 & 41.95\% & 9.69 & 53.22\% & 9.35 \\
    \textbf{LMSanitator (FS$_{50}$)} & 88.10\% & 9.10 & 81.87\% & 34.43 & 92.53\% & 29.98 & 71.58\% & 10.85 & 50.57\% & 31.94 & 61.99\% & 36.52 \\
    \textbf{LMSanitator (FS$_{100}$)} & 90.48\% & 20.12 & 91.81\% & 103.60 & 92.53\% & 83.65 & 75.79\% & 23.69 & 60.92\% & 60.21 & 70.18\% & 48.55 \\
    \bottomrule
    \end{tabular}
    }
    % \vskip -0.1in
\end{table*}
\subsubsection{Trigger Recall Effectiveness}

We further evaluate the ability to identify all triggers. Three attack methods are adopted on 15 PLMs. Four sets of trigger configurations are injected under two trigger granularities as specified in \textcolor{cyan}{\autoref{tab:trigger-words}}, constructing a total of 180 backdoored models (\(15_{\text{models}} \times 3_{\text{attacks}} \times 2_{\text{granularities}} \times 2_{\text{levels}} = 180\)) and 1,080 valid trigger sets. \textcolor{cyan}{\autoref{tab:trigger_recall}} presents the trigger recall results. \textbf{Compared to LMSanitator, \sys achieves a higher recall rate while maintaining a lower latency.} Meanwhile, \sysg maintains a recall rate comparable to that of the LMSanitator and improves computational efficiency by nearly \textbf{20 times}. Token-level triggers can achieve a recall rate of nearly 100\% under beam search, while the average recall rate of word-level triggers is only 86\%. We find this difference from the inherent feature of word-level attacks: during attack, complete words are decomposed into subwords, causing the backdoor to be scattered across multiple token combinations. This distributed activation pattern makes it difficult for the gradient signal to concentrate. Moreover, word-level word embeddings need to be indirectly optimized through a mapping matrix, which increases the optimization difficulty of discrete search.\looseness=-1

Crucially, achieving 100\% trigger recall is not a prerequisite for complete backdoor mitigation within \sys framework. Universal transferable backdoors function by mapping multiple diverse triggers—including scattered subwords—to a shared, dense sub-distribution in the feature space. By successfully recalling a significant majority of these triggers, \sys secures sufficient 'anchor points' to identify this malicious manifold. During the subsequent offline Model Purification phase, rectifying the feature space mapping for these recovered triggers concurrently dismantles the entire shared target representation. Consequently, any unrecovered subword combinations are orphaned and rendered inert. As demonstrated in \textcolor{cyan}{Appendix~\ref{appendix:backdoor_defense_effectiveness}}, this allows \sys to effectively reduce the ASR to clean baseline levels despite incomplete detection.

\begin{table*}[t]
\centering
\caption{Evaluation of backdoor defense effectiveness of \sys. `Hate' denotes HateSpeech and `Adv' denotes Adversarial. \textbf{Bold} values indicate the optimal results under the same attack.}
\label{backdoor_defense}
\vspace{-5pt}
\resizebox{\linewidth}{!}{
\begin{tabular}{c|c|ccccccccc|ccccccccc}
\toprule
\multirow{2}{*}{\textbf{Attacks}} & \multirow{2}{*}{\textbf{Defenses}} & \multicolumn{9}{c|}{\textbf{ACC} $\uparrow$} & \multicolumn{9}{c}{\textbf{ASR} $\downarrow$} \\
\cmidrule(lr){3-11} \cmidrule(lr){12-20}
 & & SST-2 & IMDB & OLID & Twit & Enron & Ling & SST-5 & Agnews & Yelp & SST-2 & IMDB & OLID & Twit & Enron & Ling & SST-5 & Agnews & Yelp \\
\midrule
\textbf{Clean} & w/o Defense & 92.66 & 93.00 & 84.07 & 94.56 & 99.10 & 99.31 & 52.04 & 94.22 & 62.98 & 8.02 & 9.25 & 30.90 & 7.36 &  0.39 & 4.12 & 25.29 & 5.62 & 12.13 \\
\midrule
\multirow{9}{*}{\textbf{NeuBA}} 
 & w/o Defense & \textbf{91.86} & 93.04 & 85.00 & \textbf{94.56} & 98.70 & 99.31 & 50.59 & 93.95 & 64.46 & 75.92 & 11.15 & 87.89 & 26.85 & 31.46 & 40.88 & 74.46 & 34.75 & 23.80 \\
\cmidrule{2-20}
 & Onion & 79.13 & 90.60 & 81.74 & 87.47 & 98.10 & 99.14 & 41.81 & 91.29 & 60.00 & 45.04 & 11.55 & 53.47 & 28.96 &  11.00 & 15.02 & 41.18 & 11.18 & 14.55 \\
 & Prune & 91.28 & 92.68 & 84.53 & 94.32 & 97.57 & 99.48 & 49.86 & 94.07 & 63.00 & 14.64 & 9.13 & 55.49 & 8.21 &  3.69 & 7.90 & 34.97 & 13.58 & 11.62 \\
 & Re-init & 91.17 & \textbf{93.28} & 84.65 & 94.52 & 98.88 & 99.31 & 51.00 & \textbf{94.21} & 64.36 & 39.65 & 10.61 & 96.74 & 13.70 & 13.80 & 28.35 & 74.67 & 19.12 & 18.78 \\
 & Recipe & 90.83 & 91.73 & 84.53 & 94.42 & 98.85 & 99.31 & 50.36 & 93.28 & 61.42 & 13.97 & 13.02 & 53.19 & 9.69 & 9.51 & 4.30 & 53.26 & 11.98 & 13.56 \\
 & BTU & 91.51 & 93.37 & 84.53 & 94.64 & 98.92 & 98.79 & 49.82 & 93.53 & 61.40 & 47.87 & 17.14 & 86.76 & 22.47 & 40.55 & 100.00 & 89.21 & 63.43 & 68.28 \\
 & X-GRAAD & 91.86 & 92.40 & 84.71 & 92.12 & 98.78 & 98.45 & 51.40 & 93.97 & 63.25 & 57.49 & 20.85 & 91.38 & 39.35 & 5.73 & 65.16 & 67.74 & 23.87 & 30.14 \\
\cmidrule{2-20}
\rowcolor{cyan!15} \cellcolor{white} & Adv-Finetune & \textbf{91.86} & 93.08 & \textbf{85.70} & 94.35 &  \textbf{99.05} & \textbf{99.66} & \textbf{52.76} & 94.16 & \textbf{64.78} & 10.09 & \textbf{8.79} & \textbf{34.51} & 7.30 &  \textbf{0.33} & \textbf{2.23} & \textbf{19.35} & \textbf{2.18} & \textbf{11.40} \\
\rowcolor{cyan!15} \cellcolor{white} & Adv-Pretrain & \textbf{91.86} & 92.58 & 84.30 & 94.51 & 98.82 & 97.24 & 51.13 & 94.08 & 62.94 & \textbf{8.14} & 10.40 & 41.25 & \textbf{7.15} & 0.59 & 15.81 & 25.95 & 3.65 & 13.91 \\
\midrule
\multirow{9}{*}{\textbf{POR-1}} 
 & w/o Defense & 92.09 & 92.80 & 85.81 & 94.30 & 98.67 & 98.79 & 49.82 & 93.62 & \textbf{65.28} & 100.00 & 99.53 & 100.00 & 98.28 & 58.55 & 98.45 & 100.00 & 78.02 & 80.07 \\
\cmidrule{2-20}
 & Onion & 78.90 & 90.75 & 81.51 & 86.73 & 97.43 & 99.14 & 40.54 & 90.80 & 59.54 & 61.15 & 47.76 & 77.50 & 60.29 & 17.87 & 37.63 & 59.98 & 29.65 & 26.41 \\
 & Prune & 91.28 & 92.79 & 84.53 & \textbf{94.49} &  98.52 & 99.14 & 50.50 & 93.55 & 64.42 & 47.97 & 54.11 & 98.40 & 27.08 & 49.90 & 3.44 & 78.58 & 64.14 & 62.12 \\
 & Re-init & \textbf{92.32} & 93.14 & 84.88 & 94.46 &  98.90 & 98.97 & 52.67 & 94.05 & 62.92 & 98.95 & 92.87 & 100.00 & 83.59 & 71.06 & 79.37 & 91.89 & 77.61 & 83.08 \\
 & Recipe & 91.28 & 91.75 & 84.88 & 94.27 & 98.85 & 99.31 & 50.36 & 93.28 & 61.42 & 24.82 & 11.82 & 61.88 & 12.54 & 9.51 & 4.30 & 53.26 & 11.98 & 13.56 \\
 & BTU & 91.97 & 92.57 & 84.42 & 94.39 & 98.92 & 98.79 & 49.82 & 93.53 & 61.40 & 86.00 & 84.21 & 78.16 & 68.87 & 40.55 & 100.00 & 89.21 & 63.43 & 68.28 \\
 & X-GRAAD & 90.94 & 92.70 & 85.06 & 92.36 & 98.93 & 99.14 & 51.52 & 93.93 & 63.15 & 89.57 & 59.12 & 97.23 & 91.11 & 58.24 & 55.93 & 91.38 & 89.56 & 69.29 \\
\cmidrule{2-20}
\rowcolor{cyan!15} \cellcolor{white} & Adv-Finetune & 92.09 & 93.12 & \textbf{86.16} & 94.38 & \textbf{99.17} & \textbf{99.66} & \textbf{53.48} & \textbf{94.53} & 64.56 & \textbf{7.67} & 6.91 & \textbf{29.86} & \textbf{7.69} & \textbf{0.69} & \textbf{2.06} & \textbf{20.25} & \textbf{3.59} & \textbf{8.53} \\
\rowcolor{cyan!15} \cellcolor{white} & Adv-Pretrain & 98.72 & \textbf{99.66} & 51.18 & 93.92 & 98.72 & \textbf{99.66} & 51.18 & 93.92 & 64.18 & 1.13 & \textbf{2.06} & 27.88 & 4.08 & 1.13 & \textbf{2.06} & 27.88 & 4.08 & 8.75 \\
\midrule
\multirow{9}{*}{\textbf{POR-2}} 
 & w/o Defense & 91.86 & 92.80 & 84.77 & 94.30 & 98.78 & 99.14 & 50.23 & 93.68 & \textbf{65.16} & 99.55 & 99.67 & 100.00 & 99.65 & 35.37 & 98.11 & 99.98 & 91.39 & 95.43 \\
\cmidrule{2-20}
 & Onion & 78.90 & 90.75 & 82.09 & 86.82 & 97.38 & 99.48 & 40.00 & 90.99 & 60.24 & 58.02 & 47.95 & 75.83 & 62.53 & 9.92 & 35.22 & 56.83 & 38.48 & 41.28 \\
 & Prune & 91.17 & 92.81 & 84.07 & 94.39 & 98.83 & 99.31 & 50.32 & 93.30 & 63.86 & 59.94 & 24.56 & 98.68 & 24.59 & 15.45 & 5.84 & 73.78 & 72.60 & 35.02 \\
 & Re-init & \textbf{92.09} & \textbf{93.21} & 85.35 & 94.25 & 99.03 & \textbf{99.66} & \textbf{53.39} & 94.22 & 62.82 & 99.54 & 99.16 & 99.93 & 99.99 & 18.87 & 44.55 & 97.82 & 77.61 & 88.51 \\
 & Recipe & 90.94 & 91.80 & 84.07 & 94.34 & 98.62 & 99.31 & 50.27 & 93.24 & 61.66 & 18.17 & 15.31 & 54.51 & 12.51 & 6.94 & 4.98 & 52.13 & 13.94 & 11.03 \\
 & BTU & 91.51 & 93.18 & 84.19 & 94.46 & 98.23 & 99.14 & 49.19 & 93.63 & 61.72 & 82.96 & 79.95 & 100.00 & 56.38 & 4.27 & 89.83 & 66.67 & 85.77 & 49.94 \\
 & X-GRAAD & 90.77 & 92.65 & 84.88 & 92.54 & 98.95 & 99.05 & 51.15 & 94.05 & 63.00 & 90.48 & 58.97 & 100.00 & 92.17 & 34.98 & 60.05 & 96.57 & 95.36 & 73.33 \\
\cmidrule{2-20}
\rowcolor{cyan!15} \cellcolor{white} & Adv-Finetune & 91.63 & 93.16 & \textbf{85.70} & \textbf{94.47} & \textbf{98.92} & \textbf{99.66} & 51.63 & \textbf{94.39} & 64.44 & 10.16 & \textbf{7.15} & \textbf{31.11} & 8.21 & \textbf{1.03} & \textbf{2.06} & \textbf{29.06} & \textbf{1.42} & 8.79 \\
\rowcolor{cyan!15} \cellcolor{white}  & Adv-Pretrain & 91.74 & 92.84 & 84.77 & 94.41 & 98.42 & 99.14 & 52.08 & 94.18 & 64.40 & \textbf{8.96} & 8.99 & 45.56 & \textbf{8.15} & 1.44 & 5.15 & 30.71 & 3.46 & \textbf{8.19} \\
\bottomrule
\end{tabular}}
\vskip -0.2in
\end{table*}

\subsubsection{Backdoor Defense Effectiveness}
\label{appendix:backdoor_defense_effectiveness}
We evaluate the effectiveness of \sys on 10 downstream tasks. The BERT model is adopted as the base architecture, and after injecting backdoors into the model, fine-tuning is performed on the downstream tasks. As shown in \textcolor{cyan}{\autoref{backdoor_defense}}, \textbf{the adversarial fine-tuning and pre-training schemes proposed in this paper exhibit optimal defense performance, achieving the lowest ASR in almost all test scenarios while maintaining comparable ACC to the baselines.} We also study the generalization on LLMs in \textcolor{cyan}{Appendix~\ref{appendix:llms}}.\looseness=-1

\subsection{Ablation Study}
\label{section:ablation}

\subsubsection{Effectiveness of Fuzzy Search}

\begin{table}[t]
\centering
\caption{Ablation study on fuzzy search rounds (FS).}
\label{tab:as_fs}
% \resizebox{0.99\linewidth}{!}{
\setlength{\tabcolsep}{7pt}
\begin{tabular}{c|c|c|c|c|c|c}
\toprule
\multirow{2}{*}{\textbf{\# FS}} & \multicolumn{2}{c|}{\textbf{NeuBA}} & \multicolumn{2}{c|}{\textbf{POR-1}} & \multicolumn{2}{c}{\textbf{POR-2}} \\
\cmidrule{2-7}
 & \textbf{Recall} & \textbf{Time} & \textbf{Recall} & \textbf{Time} & \textbf{Recall} & \textbf{Time} \\
\midrule
1 & 70.83\% & 0.034 & 45.83\% & 0.035 & 66.67\% & 0.032 \\
2 & 87.50\% & 0.071 & 70.83\% & 0.070 & 79.17\% & 0.067 \\
3 & 87.50\% & 0.107 & 75.00\% & 0.105 & 83.33\% & 0.102 \\
4 & 87.50\% & 0.143 & 83.33\% & 0.141 & 83.33\% & 0.137 \\
5 & 91.67\% & 0.178 & 87.50\% & 0.176 & 83.33\% & 0.173 \\
\bottomrule
\end{tabular}
% }
\vskip -0.2in
\end{table}

\begin{table*}[t]
\centering
\caption{Ablation study on dynamic negative sample construction (\texttt{DNSC}). \texttt{N} denotes the number of triggers to be
searched.}
\label{tab:as_dnsc}
% \resizebox{\linewidth}{!}{
\setlength{\tabcolsep}{10pt}
\begin{tabular}{c|c|c|c|c|c|c|c|c|c|c}
\toprule
\textbf{DNSC} & \multicolumn{8}{c|}{\ding{55}} & \multicolumn{2}{c}{\ding{51}} \\
\midrule
\multirow{2}{*}{\textbf{N}} & \multicolumn{2}{c|}{\textbf{2}} & \multicolumn{2}{c|}{\textbf{3}} & \multicolumn{2}{c|}{\textbf{5}} & \multicolumn{2}{c|}{\textbf{8}} & \multicolumn{2}{c}{\textbf{8}} \\
\cmidrule(lr){2-3} \cmidrule(lr){4-5} \cmidrule(lr){6-7} \cmidrule(lr){8-9} \cmidrule(lr){10-11}
& \textbf{Recall} & \textbf{Time(h)} & \textbf{Recall} & \textbf{Time(h)} & \textbf{Recall} & \textbf{Time(h)} & \textbf{Recall} & \textbf{Time(h)} & \textbf{Recall} & \textbf{Time(h)} \\
\midrule
NeuBA & 83.33\% & 0.031 & 87.50\% & 0.057 & 87.50\% & 0.129 & 91.67\% & 0.290  & \cellcolor{cyan!15} 91.67\% & \cellcolor{cyan!15} 0.178   \\
POR-1 & 70.83\% & 0.031 & 79.17\% & 0.056 & 79.17\% & 0.129 & 87.50\% & 0.295  & \cellcolor{cyan!15} 87.50\% & \cellcolor{cyan!15} 0.176  \\
POR-2 & 70.8\%3 & 0.031 & 83.33\% & 0.056 & 87.50\% & 0.130 & 83.33\% & 0.286  & \cellcolor{cyan!15} 83.33\% & \cellcolor{cyan!15} 0.173   \\
\bottomrule
\end{tabular}
% }
\vskip -0.2in
\end{table*}

\begin{figure}[t]
  \centering
  \begin{subfigure}{0.47\linewidth}  % 第一张图占一行的45%宽度
    \centering
    \includegraphics[width=\linewidth]{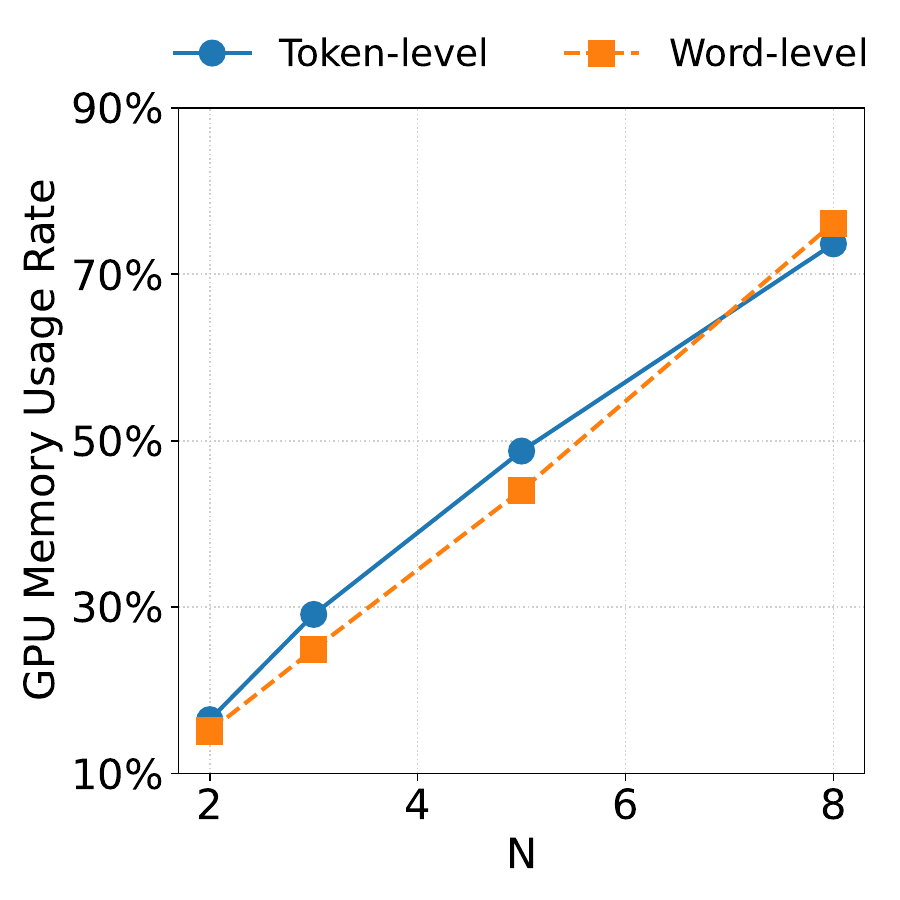} 
    \caption{When dynamic negative sample construction is not adopted.}
    \label{fig:gpu_1}
  \end{subfigure}
  \hfill
  \begin{subfigure}{0.47\linewidth}  % 第二张图占一行的45%宽度
    \centering
    \includegraphics[width=\linewidth]{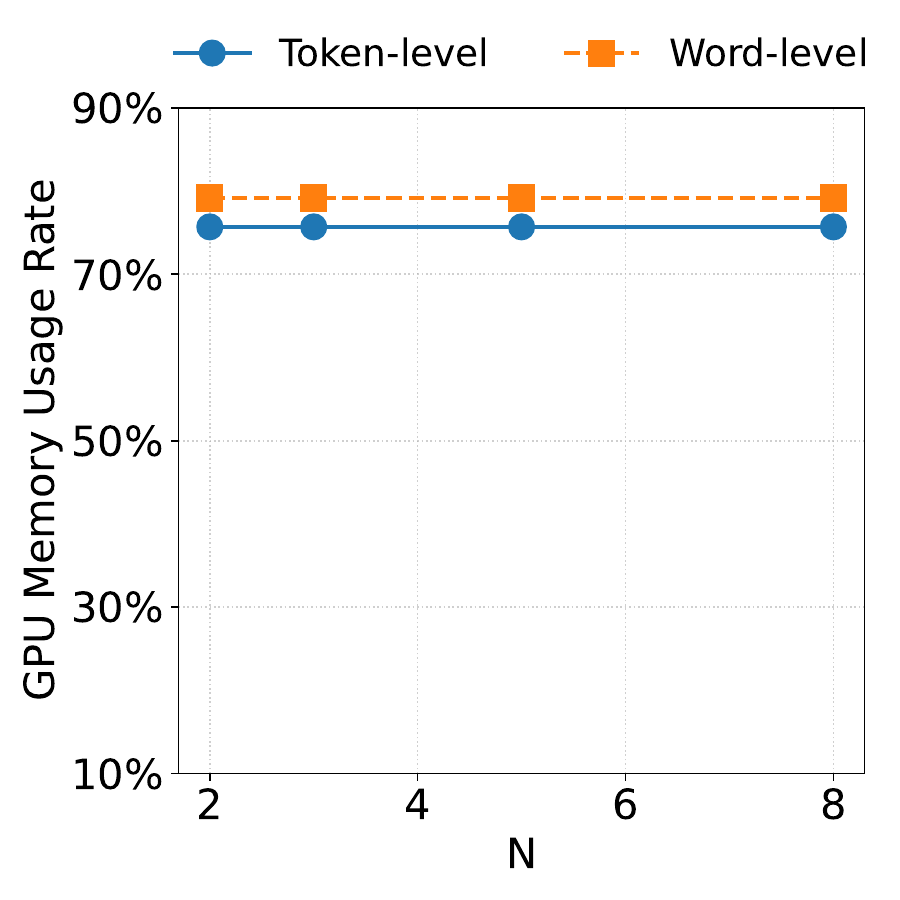} 
    \caption{When dynamic negative sample construction is adopted.}
    \label{fig:gpu_2}
  \end{subfigure}
  \caption{GPU utilization of ablation on dynamic negative sample construction. \texttt{N} denotes the number of triggers to be
searched.}
  \label{fig:gpu}  % 可选：总标签
  \vskip -0.2in
\end{figure}

\textcolor{cyan}{\autoref{tab:as_fs}} illustrates the results of an ablation study that examined the impact of varying rounds of Fuzzy Search (FS) on recall and time performance in NeuBA, POR-1 and POR-2 attacks. \textbf{The results reveal that increasing the number of FS rounds generally improves recall rates across all attack methods}. For example, NeuBA recall improved from 70.83\% with one FS round to 91.67\% with five rounds. Similarly, the POR-1 and POR-2 recalls showed notable improvements, reaching 87.50\% and 83.33\%, respectively, after five rounds.

The analysis also indicates a proportional increase in computational time with additional FS rounds. For example, the computation time for NeuBA increased from 0.034 hours in one FS to 0.178 hours in five FS rounds. This trade-off highlights the balance between achieving higher recall and managing computational resources.

These findings underscore the efficacy of employing multiple FS rounds to enhance detection precision, while also emphasizing the need to consider computational efficiency in practical applications.

\subsubsection{Effectiveness of Dynamic Negative Sample Construction}

The batch size is set to 2 when dynamic negative sample construction (DNSC) is not adopted. \textcolor{cyan}{\autoref{tab:as_dnsc}} presents the findings from an ablation study on DNSC with varying sample numbers. The table compares recall rates and computation times for NeuBA, POR-1, and POR-2 attacks, with and without the application of DNSC. \textbf{The results indicate that employing DNSC consistently enhances recall performance across all scenarios}. For instance, under NeuBA attack, recall improved from 83.33\% without DNSC to 91.67\% with DNSC, while computation time decreased from 0.29 hours to 0.178 hours when using DNSC. Similar trends are observed for POR-1 and POR-2, where recall rates increased and computational times were optimized with the introduction of DNSC.
These findings clearly demonstrate the efficacy of DNSC in improving detection accuracy while reducing processing time. The results advocate for the integration of DNSC in backdoor detection frameworks to achieve superior performance without compromising computational efficiency.

\textcolor{cyan}{\autoref{fig:gpu}} further illustrates the GPU overhead during the process. \textcolor{cyan}{\autoref{fig:gpu_1}} depicts the usage of GPU without DNSC, whereas \textcolor{cyan}{\autoref{fig:gpu_2}} shows the usage when DNSC is used.
The comparison reveals a marked improvement in GPU efficiency with the adoption of DNSC. Specifically, GPU utilization is optimized, indicating a more streamlined processing flow and reduced computational demand. This improvement in resource management underscores the value of DNSC in improving both the performance and efficiency of backdoor detection frameworks. \textbf{Consequently, integrating DNSC not only increases detection accuracy, but also ensures more effective utilization of computational resources}.

\subsection{Further Analysis}
To investigate the impact of key hyperparameters in \sys, we construct 10 backdoored models with 60 triggers. Details are in \textcolor{cyan}{Appendix~\ref{appendix:hyperparameters}}. We investigate the \textbf{independence of the data source} in \textcolor{cyan}{Section~\ref{section:data}}, \textbf{false positive analysis on clean models} in \textcolor{cyan}{Section~\ref{section:fp}}, \textbf{verification thresholds} in \textcolor{cyan}{Section~\ref{section:thresholds}}, and \textbf{multi-granularity trigger detection} in \textcolor{cyan}{Appendix~\ref{appendix:granularity}}. We also measure the overhead of adversarial purification in \textcolor{cyan}{Appendix~\ref{appendix:overhead}}, purifying a BERT-base model requires only moderate GPU memory and converges within a few epochs, indicating that the offline purification cost is practical.

\subsubsection{Hyperparameters in Trigger Search}
\begin{figure}[t]
  \centering
  \begin{subfigure}{0.46\linewidth}  % 第一张图占一行的45%宽度
    \centering
      \includegraphics[width=\linewidth]{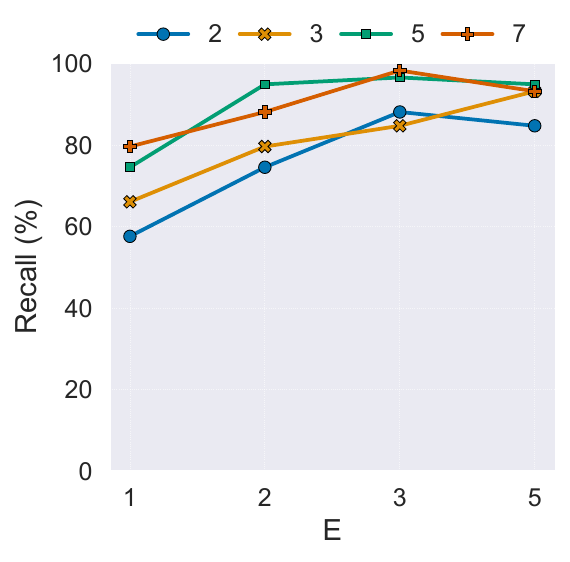}
    \caption{{Study about FS and E.}}
    \label{search_epochs}
  \end{subfigure}
  \hfill
  \begin{subfigure}{0.49\linewidth}  % 第二张图占一行的45%宽度
    \centering
    \includegraphics[width=\linewidth]{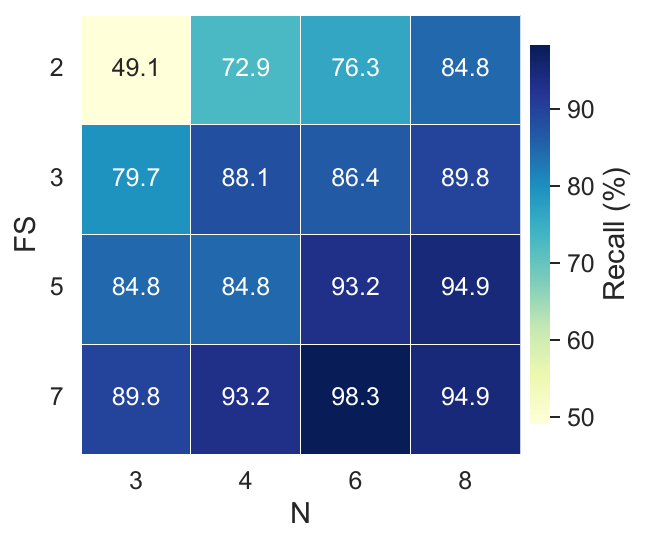}
    \caption{{Study about FS and N.}}
    \label{trigger_nums}
  \end{subfigure}
  % \vspace{-10pt}
  \caption{{Parameter studies about trigger search.}}  % 可选：添加总标题
  \label{fig:parameter}  % 可选：总标签
  \vskip -0.2in
\end{figure}

Three key parameters are evaluated on trigger recall performance: {Fuzzy Search Rounds} \(FS \in \{2, 3, 5, 7\}\), Update Times per Round \(E \in \{1, 2, 3, 5\}\), and Preset Trigger Groups \(N \in \{3, 4, 6, 8\}\). As shown in \textcolor{cyan}{\autoref{fig:parameter}}, when \(E = 3\), the recall performance under different search rounds achieves an optimal balance. In the analysis of preset trigger groups, the recall rate exhibits a monotonically increasing trend with increasing $N$, reaching optimal performance when the number of preset groups matches the number of real trigger groups (\(N = 6\)). \textbf{Experimental results also validate the effectiveness of the multi-round fuzzy search strategy, which gradually improves the trigger recall rate through iterative optimization.}

\begin{figure}[t]
    \begin{subfigure}{0.325\linewidth}
      \centering
      \includegraphics[width=\linewidth]{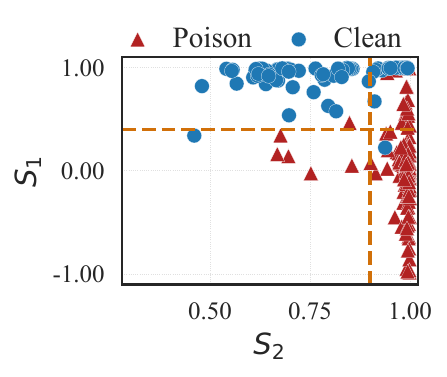}
      \caption{NeuBA\ -\ Base\label{neuba_base}}
    \end{subfigure}
    \hfill
    \begin{subfigure}{0.325\linewidth}
      \centering
      \includegraphics[width=\linewidth]{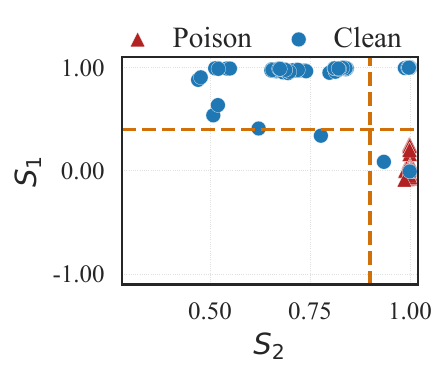}
      \caption{POR-1\ -\ Base\label{por_1_base}}
    \end{subfigure}
    \hfill
    \begin{subfigure}{0.325\linewidth}
      \centering
      \includegraphics[width=\linewidth]{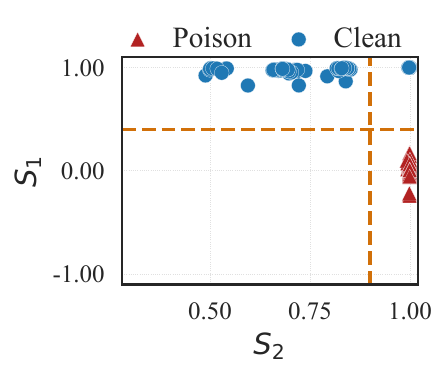}
      \caption{POR-2\ -\ Base\label{por_2_base}}
    \end{subfigure}
    
    \begin{subfigure}{0.325\linewidth}
      \centering
      \includegraphics[width=\linewidth]{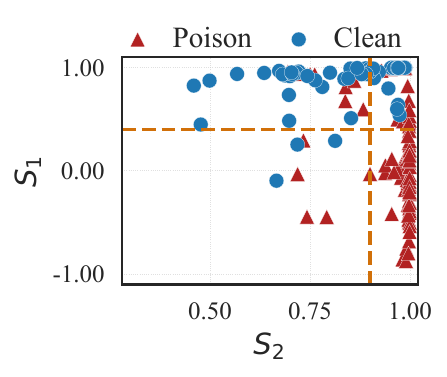}
      \caption{NeuBA\ -\ Large\label{neuba_large}}
    \end{subfigure}
    \hfill
    \begin{subfigure}{0.325\linewidth}
      \centering
      \includegraphics[width=\linewidth]{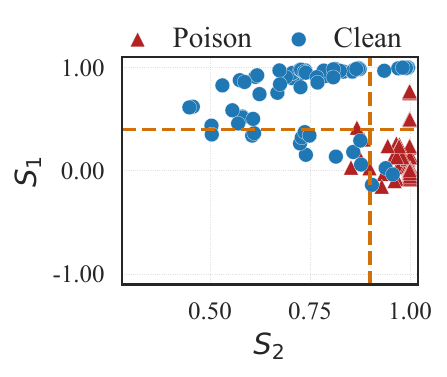}
      \caption{POR-1\ -\ Large\label{por_1_large}}
    \end{subfigure}
    \hfill
    \begin{subfigure}{0.325\linewidth}
      \centering
      \includegraphics[width=\linewidth]{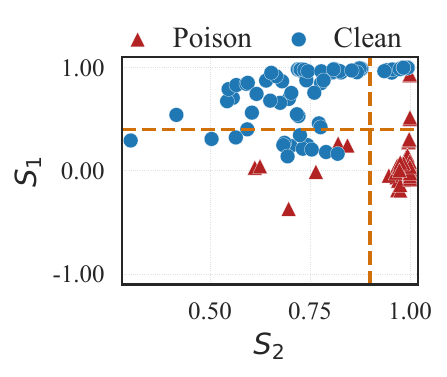}
      \caption{POR-2\ -\ Large\label{por_2_large}}
    \end{subfigure}
    % \vspace{-10pt}
    \caption{Visualization of sample judgment during the backdoor verification phase. The horizontal and vertical dash lines denote \(\gamma_1 = 0.4\) and \(\gamma_2 = 0.9\).}
    \label{backdoor_verification_ablation}
    \vskip -0.2in
  \end{figure}

\begin{figure}[t]
  \centering
  \includegraphics[width=\linewidth]{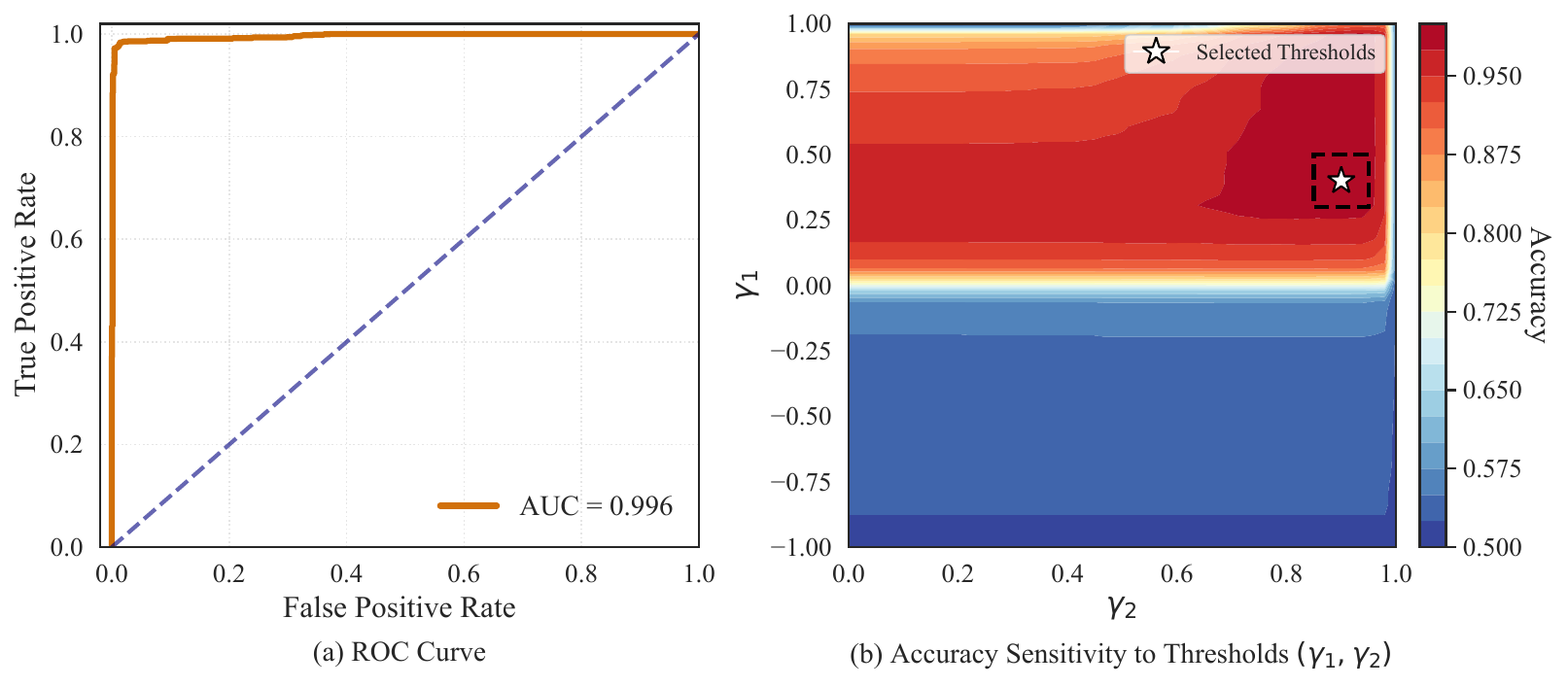}
  \vspace{-10pt}
  \caption{Analysis of the backdoor verification. (a) ROC curve demonstrating the trade-off between TPR and FPR, achieving an AUC near 1.0. (b) Sensitivity analysis about thresholds $\gamma_1$ and $\gamma_2$, indicating optimal range for hyperparameter.}
  \label{fig:threshold_analysis}
  \vskip -0.1in
\end{figure}

\begin{table}[t]
\centering
\caption{Evaluation of \sys w/o knowing attack data.}
\label{tab:datasource}
\resizebox{0.99\linewidth}{!}{
\begin{tabular}{c|c|c|c|c|c|c}
\toprule
\multirow{2}{*}{\textbf{Dataset}} & \multicolumn{2}{c|}{\textbf{NeuBA}} & \multicolumn{2}{c|}{\textbf{POR-1}} & \multicolumn{2}{c}{\textbf{POR-2}} \\
\cmidrule{2-7}
 & \textbf{Recall} & \textbf{Time} & \textbf{Recall} & \textbf{Time} & \textbf{Recall} & \textbf{Time} \\
\midrule
WikiText-2 & 91.67\% & 0.177 & 87.50\% & 0.177 & 83.33\% & 0.173 \\
CC-News & 83.33\% & 0.248 & 83.33\% & 0.252 & 87.50\% & 0.246 \\
\bottomrule
\end{tabular}
}
\vskip -0.1in
\end{table}

\subsubsection{Data Source Independence for Trigger Search}
\label{section:data}
In realistic scenarios, defenders do not know which specific corpus the attacker used to poison. To evaluate the robustness against such data discrepancy, we design a cross-domain experiment where the attacker injects backdoors using WikiText-2~\citep{merity2017pointer}. We then compare the defense performance under a \textit{Target domain} setting with the same dataset and a \textit{Cross domain} setting with a completely independent generic corpus, CC-News~\citep{mackenzie2020cc}. \textcolor{cyan}{\autoref{tab:datasource}} illustrates the effectiveness of \sys in detecting backdoors without prior knowledge of attack data. 

\sys is resilient to domain mismatches, the searching recall remains stable regardless of the defense dataset used. Against NeuBA, \sys achieves an average recall of $91.67\%$ on the target domain, compared to $83.33\%$ on the cross domain. For POR-1, the defense shows similar stability. Interestingly, for POR-2, the cross-domain recall ($87.50\%$) even slightly exceeds the target-domain recall ($83.33\%$). \textbf{These results highlight \sys's adaptability, confirming that our defense does not rely on the exact attack data distribution and can effectively mitigate task-agnostic backdoors using easily accessible generic corpora.}

\subsubsection{Comprehensive False Positive Analysis}
\label{section:fp}
To rigorously evaluate the specificity and reliability of \sys, we conduct a comprehensive false positive analysis at two distinct levels: the \textbf{model level} (evaluating clean models) and the \textbf{trigger level} (evaluating unintended triggers within backdoored models).

\textbf{Model-Level False Positives on Clean Models.} 
We use 15 clean PLMs (one for each of the models in \textcolor{cyan}{\autoref{tab:PLMs}}) to test the False Positive Rate (FPR). \textbf{\sys achieved a 0\% FPR across all 15 clean PLMs}. To further assess the specificity, we conduct a detailed case analysis using a clean \texttt{bert-base-uncased} model, evaluating both token-level and word-level search algorithms (detailed in Appendix~\ref{appendix:fpr}). For each of the $N=8$ candidate trigger groups, we calculate the inter-class similarity ($S_1$) and the intra-class similarity ($S_2$). As shown in \textcolor{cyan}{\autoref{tab:fpr_case_token}} and \textcolor{cyan}{\autoref{tab:fpr_case_word}}, clean candidates fail to satisfy the dual verification thresholds ($S_1 < \gamma_1$ and $S_2 > \gamma_2$). Specifically, all candidates exhibit high inter-class similarity ($S_1 \gg \gamma_1$) and low-to-moderate intra-class similarity ($S_2 < \gamma_2$), resulting in their consistent classification as ``Clean''.

\begin{figure*}[t]
\centering
% --- 第一行：所有 Token Triggers ---
\begin{subfigure}{0.32\textwidth} % 【关键修改】宽度改为0.32以容纳3张图
  \centering
  \includegraphics[width=\linewidth]{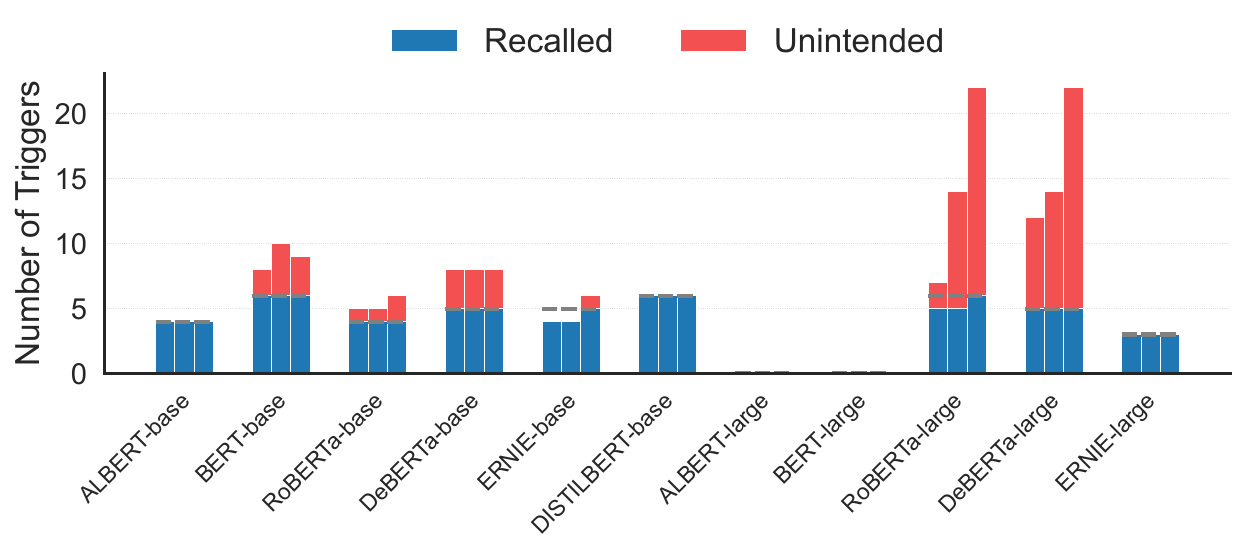}
  \caption{NeuBA\ -\ Token triggers\label{neuba_token}}
\end{subfigure}
\hfill
\begin{subfigure}{0.32\textwidth}
  \centering
  \includegraphics[width=\linewidth]{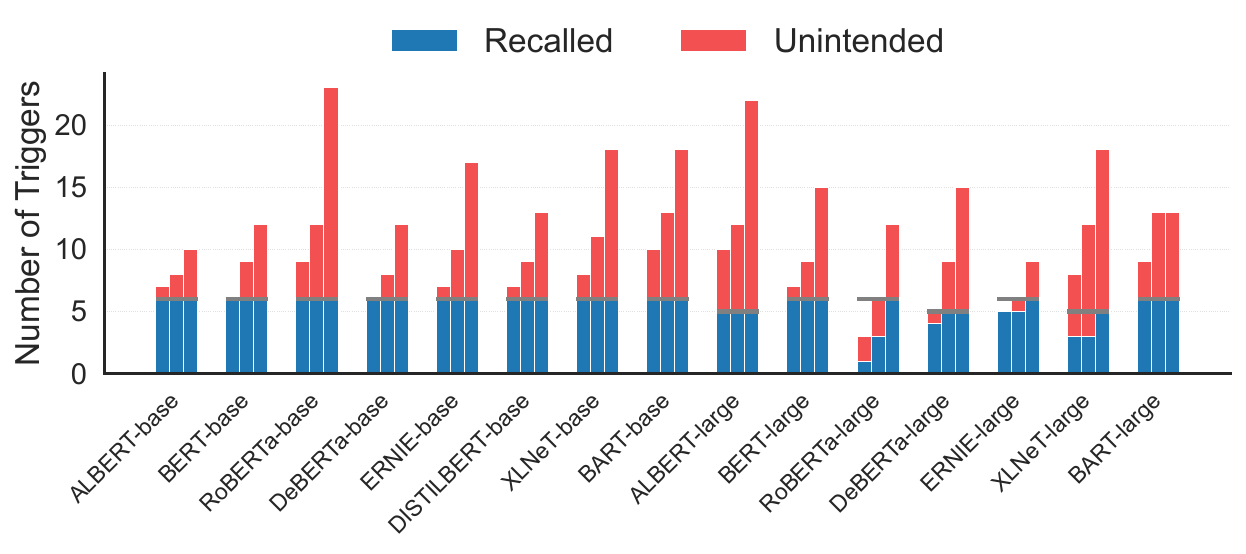}
  \caption{POR-1\ -\ Token triggers\label{por_1_token}}
\end{subfigure}
\hfill
\begin{subfigure}{0.32\textwidth}
  \centering
  \includegraphics[width=\linewidth]{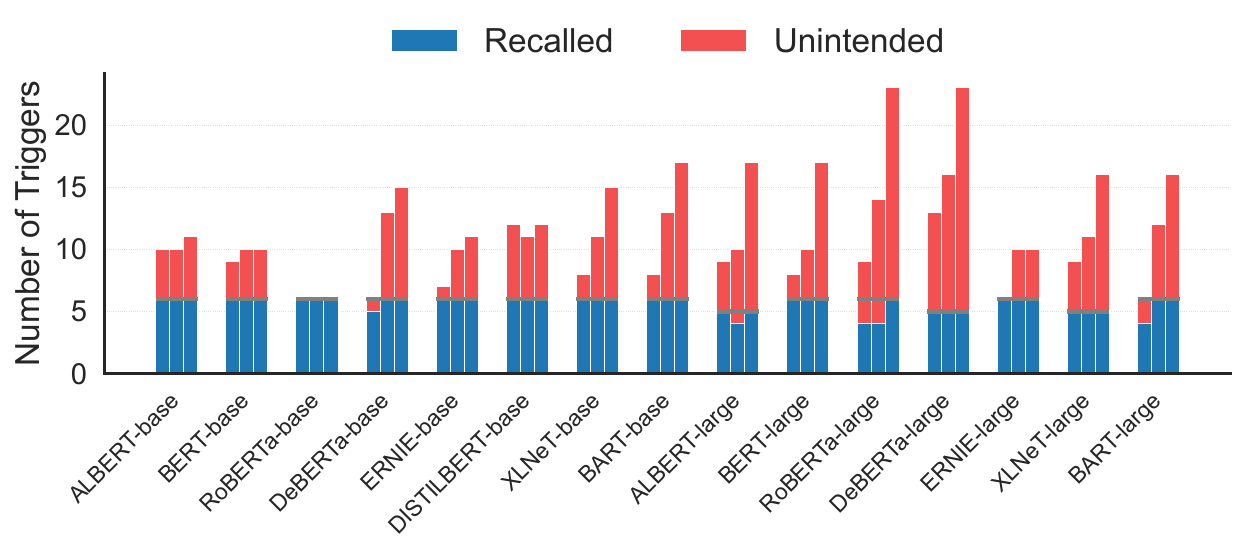}
  \caption{POR-2\ -\ Token triggers\label{por_2_token}}
\end{subfigure}

\par\bigskip % 换行并添加垂直间距

% --- 第二行：所有 Word Triggers ---
\begin{subfigure}{0.32\textwidth}
  \centering
  \includegraphics[width=\linewidth]{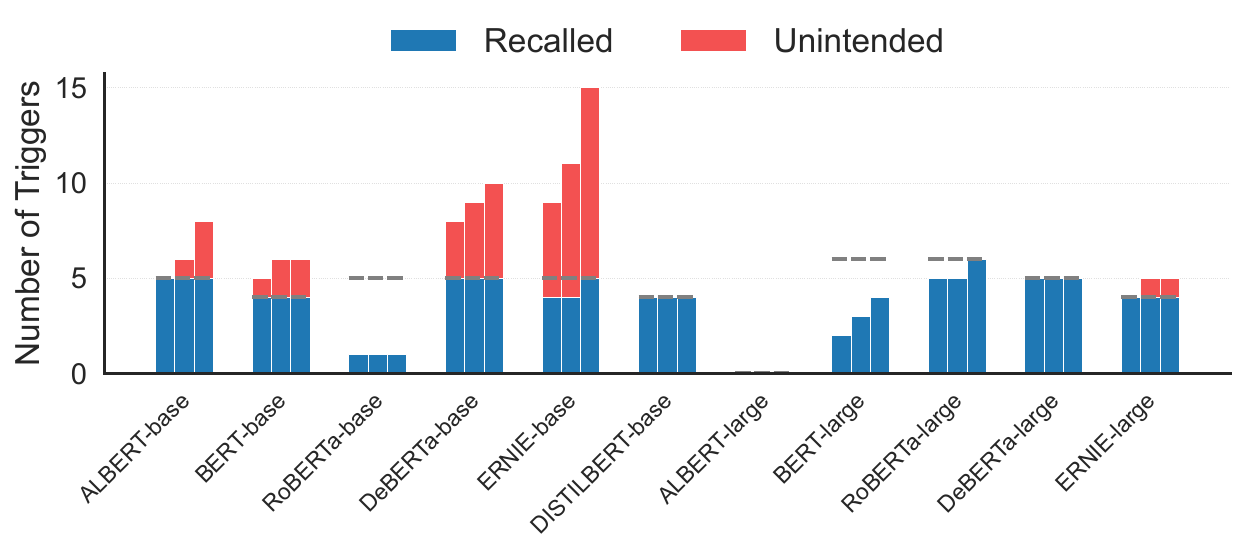}
  \caption{NeuBA\ -\ Word triggers\label{neuba_word}}
\end{subfigure}
\hfill
\begin{subfigure}{0.32\textwidth}
  \centering
  \includegraphics[width=\linewidth]{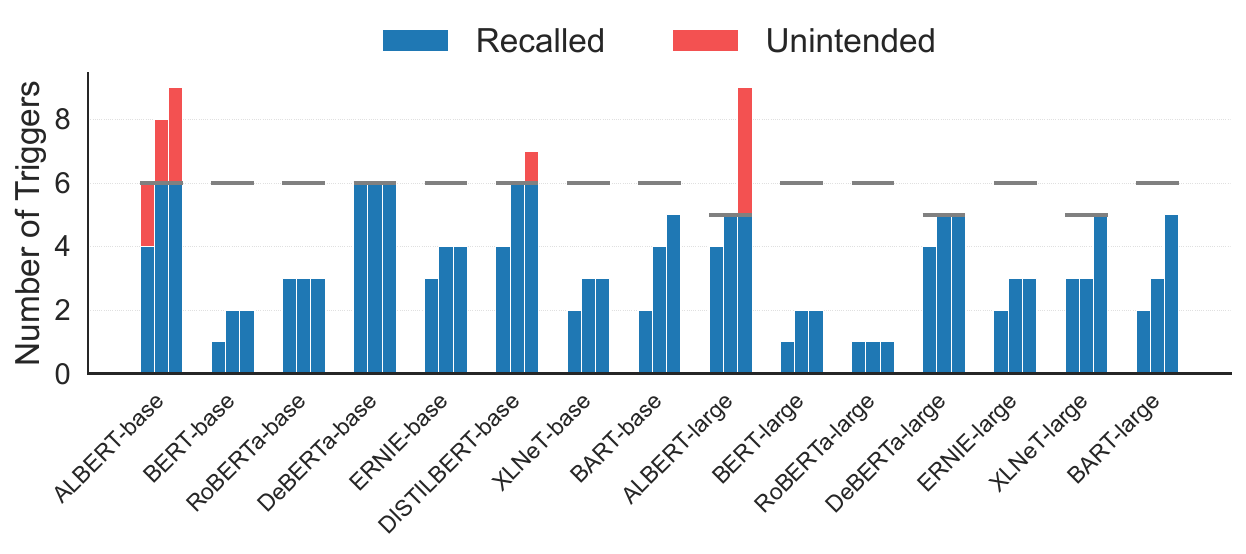}
  \caption{POR-1\ -\ Word triggers\label{por_1_word}}
\end{subfigure}
\hfill
\begin{subfigure}{0.32\textwidth}
  \centering
  \includegraphics[width=\linewidth]{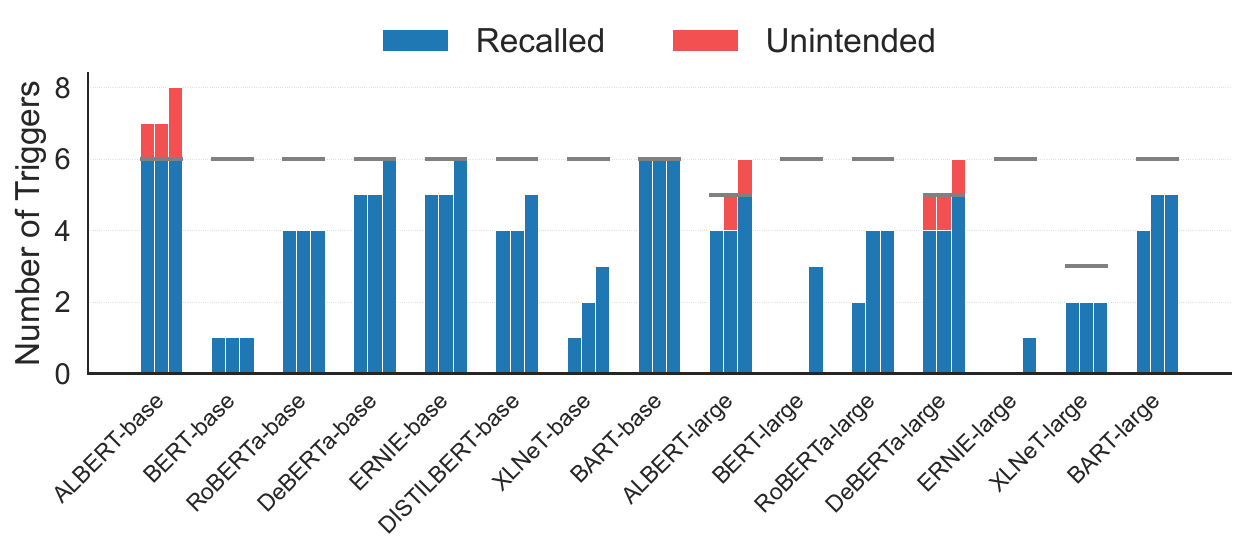}
  \caption{POR-2\ -\ Word triggers\label{por_2_word}}
\end{subfigure}
% \vspace{-10pt}
\caption{Visualisation of trigger search results. The dash line represents the number of successfully injected triggers.}
\label{fig:trigger-search}
\vskip -0.15in
\end{figure*}

\textbf{Trigger-Level False Positives in Backdoored Models.} 
We also observe a unique phenomenon at the trigger level during the iterative fuzzy search. As illustrated in \textcolor{cyan}{\autoref{fig:trigger-search}}, while the detection rate of real triggers increases with the number of search rounds, backdoored models may also yield \textbf{unintended triggers}: tokens or words flagged by the search process but are not implanted by the attackers. Statistical analysis reveals that among all flagged suspicious triggers, the true trigger ratios are 40.98\% for token-level and 83.42\% for word-level settings, with an overall average of 52.53\%.

Through feature representation analysis, we find that this phenomenon primarily stems from the composite injection during multi-trigger backdoor training. The combined feature representations of these multiple poisoned samples exhibit high similarity to these unintended triggers, leading to their additional identification. \emph{Therefore, these trigger-level ``false positives'' do not represent a failure of the detection algorithm; rather, they reveal the inherent artifacts and implicit associations introduced by complex multi-trigger backdoor injection paradigms}.
Crucially, the presence of these unintended triggers does not degrade the overall efficacy of \sys. Although imperfect precision introduces ``false'' triggers into the Model Purification phase, these unintended triggers are mathematically and semantically adjacent to the true malicious distribution in the feature space. Consequently, incorporating them into the adversarial training process acts as an effective form of data augmentation and regularization. By optimizing against both the recalled true triggers and these highly correlated unintended artifacts, \sys comprehensively dismantles the entire backdoor feature manifold. As demonstrated in \textcolor{cyan}{\autoref{tab:adaptive_attack_backdoor_defense}}, this robust purification pipeline successfully reduces the ASR to clean baseline levels without sacrificing the clean accuracy of the models, proving the framework's reliability even under the interference of search noise.

\subsubsection{Backdoor Verification Thresholds}
\label{section:thresholds}
To rigorously calibrate the decision thresholds ($\gamma_1$ and $\gamma_2$) and evaluate the robustness of our verification module, we construct a mixed test set containing normal words and real triggers through Monte Carlo sampling~\citep{robert2004monte}. The distributions of samples cosine similarity are shown in \textcolor{cyan}{\autoref{backdoor_verification_ablation}}. Experimental results indicate that there is a significant difference between the inter-class similarity (\(S_1\)) between poisoned and clean samples and the intra-class similarity (\(S_2\)) among poisoned samples. We perform a ROC analysis to assess the discrimination capability of our dual-metric approach. As shown in \textcolor{cyan}{\autoref{fig:threshold_analysis}(a)}, the verification module achieves an AUC approaching 1.0, indicating a near-perfect separation between benign and backdoored models across valid threshold ranges. This confirms that the distinct feature separation observed in Findings 1 and 2 is a consistent property of transferable backdoors. Furthermore, we conduct a sensitivity analysis to determine the optimal operating points for $\gamma_1$ and $\gamma_2$. \textcolor{cyan}{\autoref{fig:threshold_analysis}(b)} plots the accuracy against varying threshold values. We observe that the performance is highly stable (Accuracy $> 0.97$) within the ranges of $\gamma_1 \in [0.3, 0.5]$ and $\gamma_2 \in [0.85, 0.95]$. This demonstrates that \sys is robust to minor hyperparameter fluctuations and does not require fine-grained tuning for specific model architectures. Based on this, we select $\gamma_1=0.4$ and $\gamma_2=0.9$ as thresholds, achieving a final discrimination accuracy of 98.7\% on the test set.

\subsubsection{Adversarial Training Visualization}
\begin{figure}[t]
\begin{subfigure}[b]{0.48\linewidth}
    \centering
    \includegraphics[width=\linewidth]{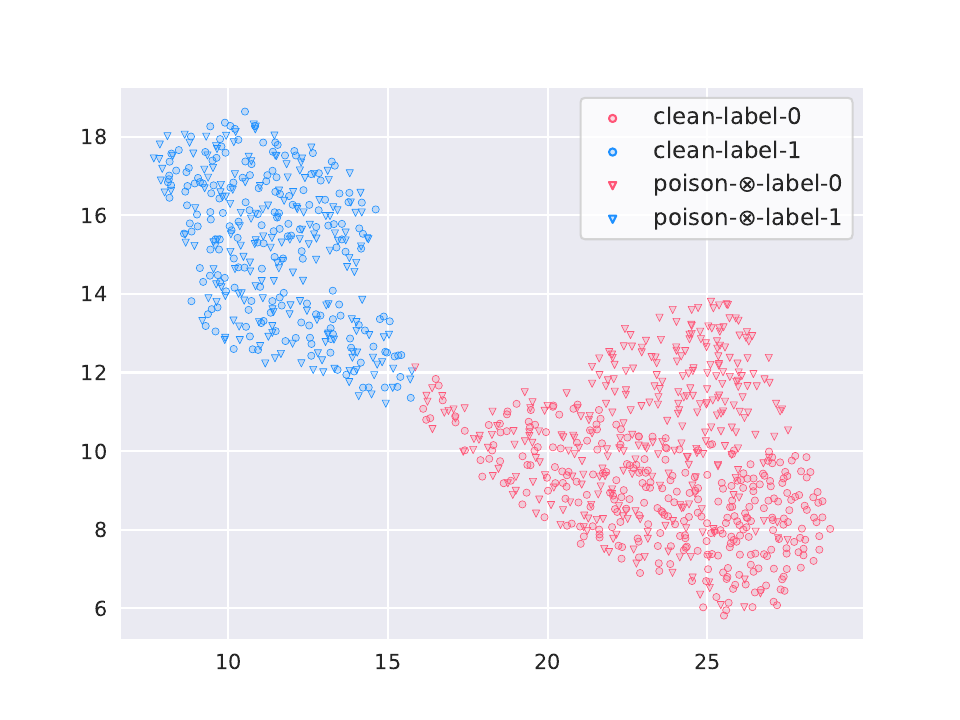}
    \caption{Clean Model\label{cp}}
\end{subfigure}
\hfill
\begin{subfigure}[b]{0.48\linewidth}
    \centering
    \includegraphics[width=\linewidth]{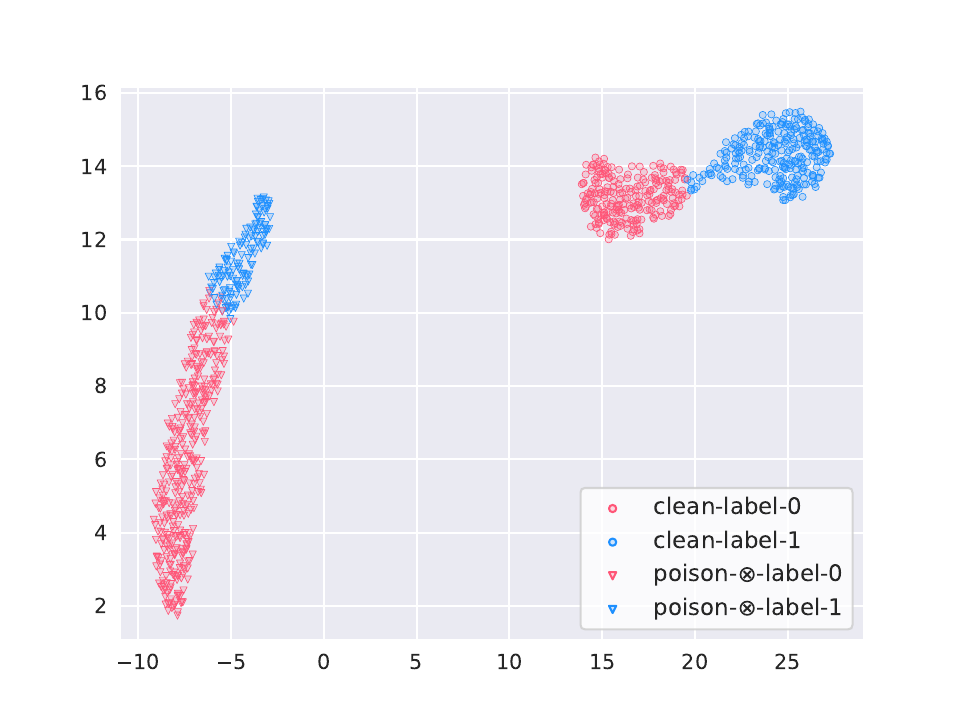}
    \caption{{Backdoored Model}\label{pp}}
\end{subfigure}
\vfill
\begin{subfigure}[b]{0.48\linewidth}
    \centering
    \includegraphics[width=\linewidth]{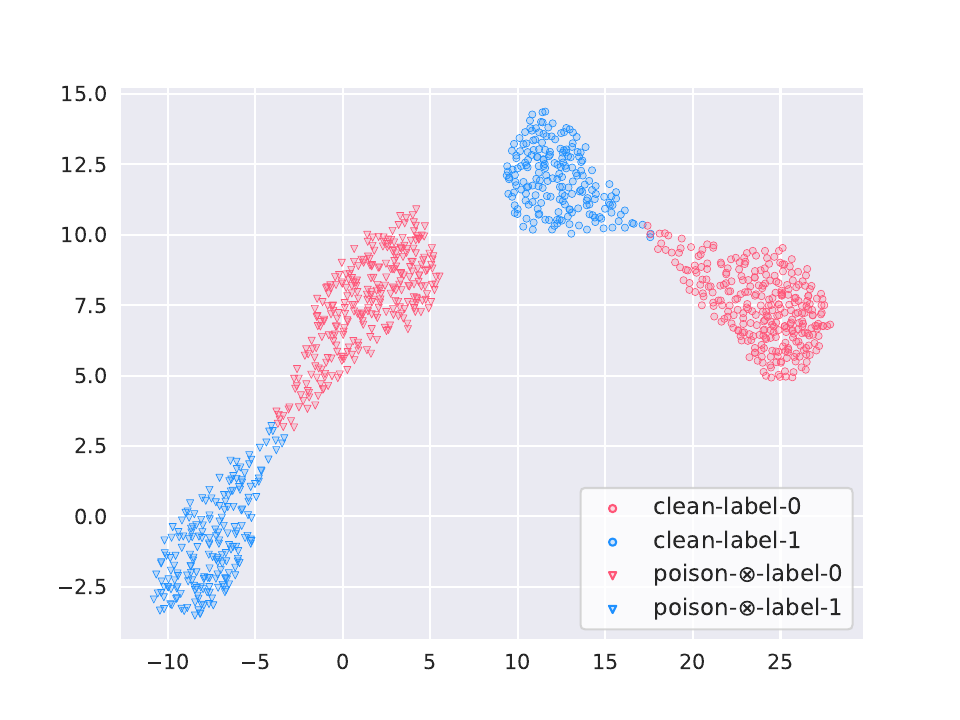}
    \caption{Adversarial Fine-tuning\label{adv_f_p}}
\end{subfigure}
\hfill
\begin{subfigure}[b]{0.48\linewidth}
    \centering
    \includegraphics[width=0.98\linewidth]{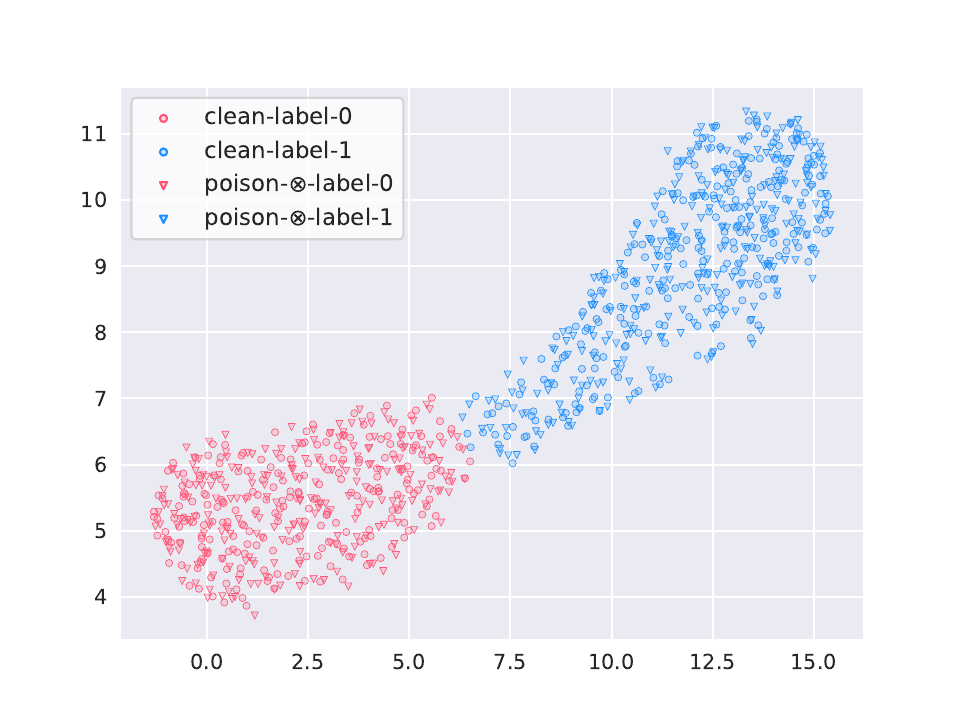}
    \caption{Adversarial Pre-training\label{adv_p_p}}
\end{subfigure}
% \vspace{-5pt}
\caption{{Visualization of clean and poisoned samples on (a) clean model, (b) backdoored model, and (c)(d) backdoored model after adversarial training.}}  % Total caption
\label{fig:adversarial-training}
\vskip -0.2in
\end{figure}

We conduct feature visualization analysis using BERT model and SST-2 task under POR attack. \textcolor{cyan}{\autoref{fig:adversarial-training}} shows that in clean model, clean and poisoned samples are clustered into two distinct clusters based on their original semantic features. In contrast, in backdoored model, poisoned samples form an independent feature cluster and are fully mapped to the target label. After implementing adversarial fine-tuning, clean and poisoned samples form two separate classification feature clusters, respectively. This indicates that adversarial fine-tuning essentially establishes classification decision regions for clean and poisoned samples individually, where triggers still play a differentiating role in the feature representation of samples. \textbf{After adversarial pre-training, clean and poisoned samples are fully fused in the feature space, fully eliminating backdoor associations.}

\begin{figure}[t]
  \centering
  \begin{subfigure}{0.49\linewidth}  % 第一张图占一行的45%宽度
    \centering
    \includegraphics[width=\linewidth]{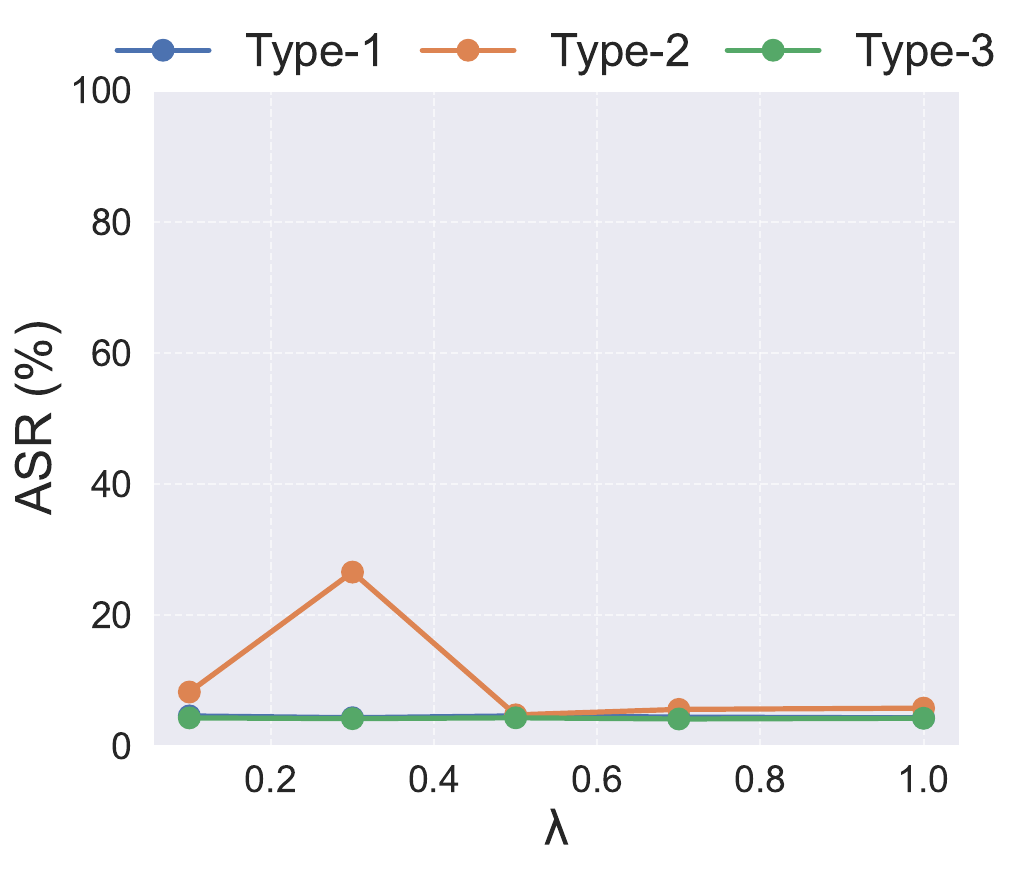}
    \caption{Neuba Adapt Attack}
    \label{neuba_adapt_attack}
  \end{subfigure}
  \begin{subfigure}{0.49\linewidth}  % 第二张图占一行的45%宽度
    \centering
    \includegraphics[width=\linewidth]{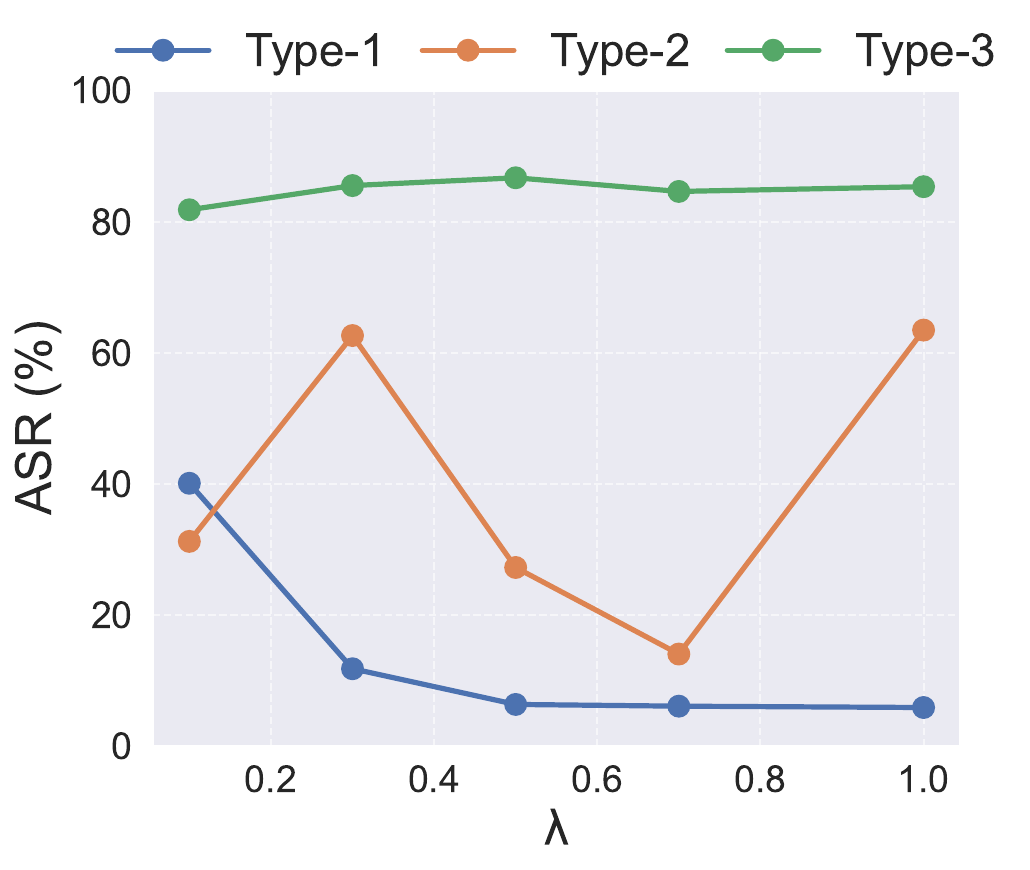}
    \caption{POR Adapt Attack}
    \label{por_adapt_attack}
  \end{subfigure}
  \vspace{-10pt}
  \caption{Evaluation of adaptive attacks w/o defense.}  % 可选：添加总标题
  \label{fig:adapt-attack}  % 可选：总标签
  \vskip -0.1in
\end{figure}

\begin{table}[t]
    \centering
        \caption{Trigger detection performance of \sys against Type-3 adaptive attack.}
    \label{tab:adapt_attack_defense}
    % \vspace{-10pt}
    \resizebox{\linewidth}{!}{
    \begin{tabular}{c|c|c|c|c|c|c|c|c|c|c}
    \toprule
    \textbf{Attack} & \multicolumn{5}{c|}{\textbf{POR-1}} & \multicolumn{5}{c}{\textbf{POR-2}} \\ \midrule
    \textbf{$\lambda$} & 0.1 & 0.3 & 0.5 & 0.7 & 1.0  & 0.1 & 0.3 & 0.5 & 0.7 & 1.0 \\ \midrule
    \textbf{Recall(\%)} & 83.33 & 91.67 & 100.00 & 75.00 & 100.00  & 100.00 & 100.00 & 100.00 & 75.00 & 91.67  \\ 
    \textbf{Time(h)} & 0.375 & 0.374 & 0.376 & 0.372 & 0.370  & 0.369 & 0.377 & 0.372 & 0.376 & 0.371 \\ 
    \bottomrule
    \end{tabular}
    }
    \vskip -0.1in
\end{table}

\begin{table}[t]
\centering
\caption{Evaluation of backdoor defense effectiveness of \sys against Type-3 adaptive attack ($\lambda = 0.7$).}
\label{tab:adaptive_attack_backdoor_defense}
% \vspace{-10pt}
\resizebox{0.99\linewidth}{!}{
\begin{tabular}{c|c|cccc|cccc}
\toprule
\multirow{2}{*}{\textbf{Attack}} & \multirow{2}{*}{\textbf{Defense}} & \multicolumn{4}{c|}{\textbf{ACC} $\uparrow$} & \multicolumn{4}{c}{\textbf{ASR} $\downarrow$} \\
\cmidrule(lr){3-6} \cmidrule(lr){7-10}
& & Twit & Agnews & Yelp & Enron & Twit & Agnews & Yelp & Enron \\
\midrule
\textbf{Clean} & w/o Defense & 94.71 & 93.95 & 64.88 & 99.13 & 8.76 & 4.20 & 6.26 & 0.17 \\
\midrule

\multirow{3}{*}{\textbf{POR-1}}
 & w/o Defense & 94.38 & 94.04 & 64.12 & 99.23 & 87.71 & 100.00 & 99.72 & 62.79 \\
\cmidrule{2-10}
& \cellcolor{cyan!15} Adv-Finetune & \cellcolor{cyan!15} 94.36 & \cellcolor{cyan!15} 94.14 & \cellcolor{cyan!15} 65.36 & \cellcolor{cyan!15} 98.71 & \cellcolor{cyan!15} 8.07 & \cellcolor{cyan!15} 3.80 & \cellcolor{cyan!15} 6.03 & \cellcolor{cyan!15} 0.99 \\
& \cellcolor{cyan!15} Adv-Pretrain & \cellcolor{cyan!15} 94.53 & \cellcolor{cyan!15} 94.02 & \cellcolor{cyan!15} 63.29 & \cellcolor{cyan!15} 98.12 & \cellcolor{cyan!15} 7.57 & \cellcolor{cyan!15} 3.36 & \cellcolor{cyan!15} 5.46 & \cellcolor{cyan!15} 2.21 \\
\midrule

\multirow{3}{*}{\textbf{POR-2}}
 & w/o Defense & 94.42 & 94.12 & 64.02 & 98.90 & 88.04 & 100.00 & 99.97 & 62.38 \\
\cmidrule{2-10}
& \cellcolor{cyan!15}  Adv-Finetune & \cellcolor{cyan!15} 94.33 & \cellcolor{cyan!15} 94.20 & \cellcolor{cyan!15} 65.17 & \cellcolor{cyan!15} 98.81 & \cellcolor{cyan!15} 8.02 & \cellcolor{cyan!15} 4.86 & \cellcolor{cyan!15} 5.71 & \cellcolor{cyan!15} 1.61 \\
& \cellcolor{cyan!15} Adv-Pretrain & \cellcolor{cyan!15} 93.98 & \cellcolor{cyan!15} 93.89 & \cellcolor{cyan!15} 63.28 & \cellcolor{cyan!15} 98.15 & \cellcolor{cyan!15} 7.29 & \cellcolor{cyan!15} 3.78 & \cellcolor{cyan!15} 5.05 & \cellcolor{cyan!15} 1.23 \\
\bottomrule
\end{tabular}
}
\vskip -0.15in
\end{table}

\subsection{Generalization on LLMs}
\label{appendix:llms}
To verify the generalization of \sys for classification tasks, we evaluate its backdoor defense capability on GPT-Neo\footnote{https://huggingface.co/EleutherAI/gpt-neo-1.3B}. As illustrated in \textcolor{cyan}{\autoref{fig:llm_cos}}, we first validate the correctness of $\gamma_1$ and $\gamma_2$ on large language models (LLMs): the selected values of 0.4 and 0.9 still ensure a trigger detection accuracy of 95.3\%. 

% Based on this, \textcolor{cyan}{\autoref{tab:llm_detection}} and \textcolor{cyan}{\autoref{tab:llm_backdoor_defense}} show the results of backdoor detection and end-to-end backdoor defense effectiveness. 

% \textcolor{cyan}{\autoref{tab:llm_detection}} presents the performance metrics of \sys in detecting backdoors in LLMs. The findings reveal a progressive improvement in recall with increased FS. These results affirm \sys's capability to effectively detect backdoor vulnerabilities in LLMs, reinforcing the value of iterative search strategies in balancing accuracy and efficiency.

\textcolor{cyan}{\autoref{tab:llm_backdoor_defense}} evaluates the defense effectiveness under three different attack scenarios. Without applying any defense mechanisms, the ASR for NeuBA reached 95.31\% on the Twit dataset, highlighting the vulnerability of the LLM. However, the implementation of adversarial fine-tuning resulted in a substantial decrease in ASR, reducing it to 8.53\% on Twit, while preserving high accuracy levels. Similarly, adversarial pre-training demonstrated notable effectiveness, especially on the Enron dataset, where ASR was reduced to an impressive 0.24\%, indicating robust protection against backdoor attacks. For POR-1 and POR-2 attacks, both adversarial fine-tuning and pre-training significantly lowered the ASR. These results underscore the potency of systematic defense strategies.

\subsection{Adaptive Attacks}

\textbf{Based on the three items in \textcolor{cyan}{\autoref{eq:L}}, we further design and investigate three adaptive attacks}. Type-1 hinders the convergence process by minimizing the feature vector distance between clean and poisoned samples; Type-2 imposes a feature dispersion constraint on poisoned samples with the same trigger, expanding their distribution range in the feature space; Type-3 reduces the feature differences between poisoned samples with different triggers, enhancing cross-trigger feature consistency, and thus impairing the trigger detection performance.
To evaluate the effectiveness of the three adaptive attacks, we introduce a coefficient $\lambda$ to quantify their contribution, whose mathematical form is as:
$\mathcal{L} = \mathcal{L}_E + \mathcal{L}_U + \lambda \mathcal{L}_A$,
where \(\mathcal{L}_A\) denotes the adaptive attack loss corresponding to the specific type.

We adapt two attacks with three adaptive attack types to perform backdoor injection using BERT-base model at the token level without any defense, and conduct fine-tuning verification on four tasks. The results are shown in \textcolor{cyan}{\autoref{fig:adapt-attack}}. Under NeuBA, ASR of the three adaptive attacks are all at a low level, while only the Type-3 attack exhibits a high ASR under POR.\looseness=-1

\textbf{We further evaluate \sys's robustness against the most successful adaptive attack, Type-3 under POR}. Despite this optimized adversarial effort, \sys shows its ability to recall the implanted triggers with various $\lambda$ values. \textcolor{cyan}{\autoref{tab:adapt_attack_defense}} demonstrates that \sys maintains a high recall rate, achieving $100.00\%$ recall in several cases and an overall high recall rate of 91.67\%. This highlights that \sys is still highly effective in identifying triggers even when facing the strongest adaptive counter-defense (Type-3). For the $\lambda$ settings with lower recall, we test the ASR. \textcolor{cyan}{\autoref{tab:adaptive_attack_backdoor_defense}} shows that \sys reduces ASR to clean baselines, demonstrating the high robustness.

%% file: sections/6discussion.tex
\section{Discussion}

\begin{figure}[t]
    \centering
    \begin{subfigure}[b]{0.49\linewidth}
        \centering
        \includegraphics[width=\linewidth]{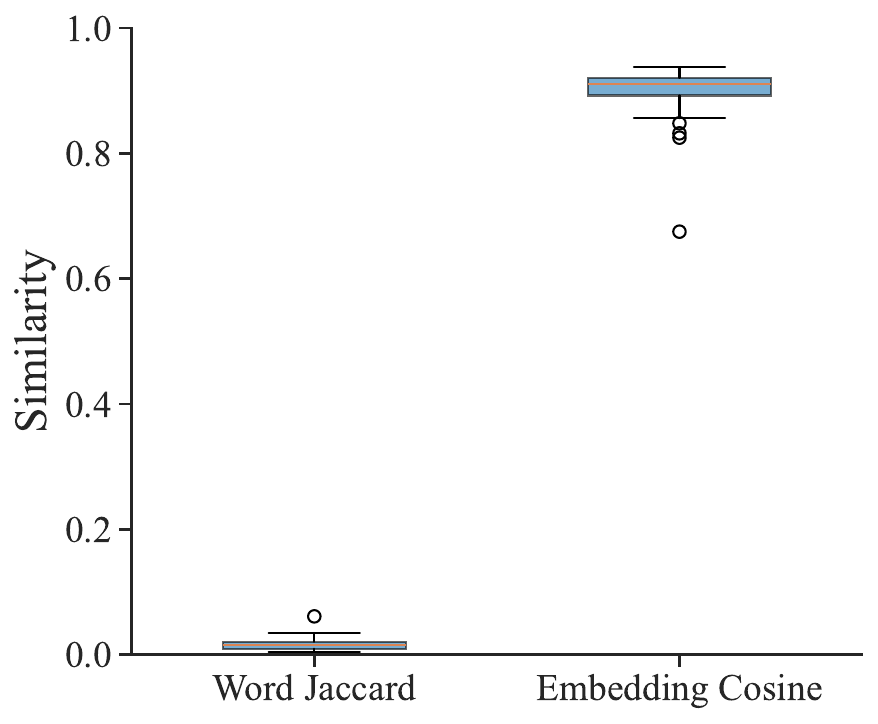}
        \caption{Lexical vs. Semantic Similarity}
        \label{fig:stylebkd_drastic_changes_a}
    \end{subfigure}
    \hfill
    \begin{subfigure}[b]{0.49\linewidth}
        \centering
        \includegraphics[width=\linewidth]{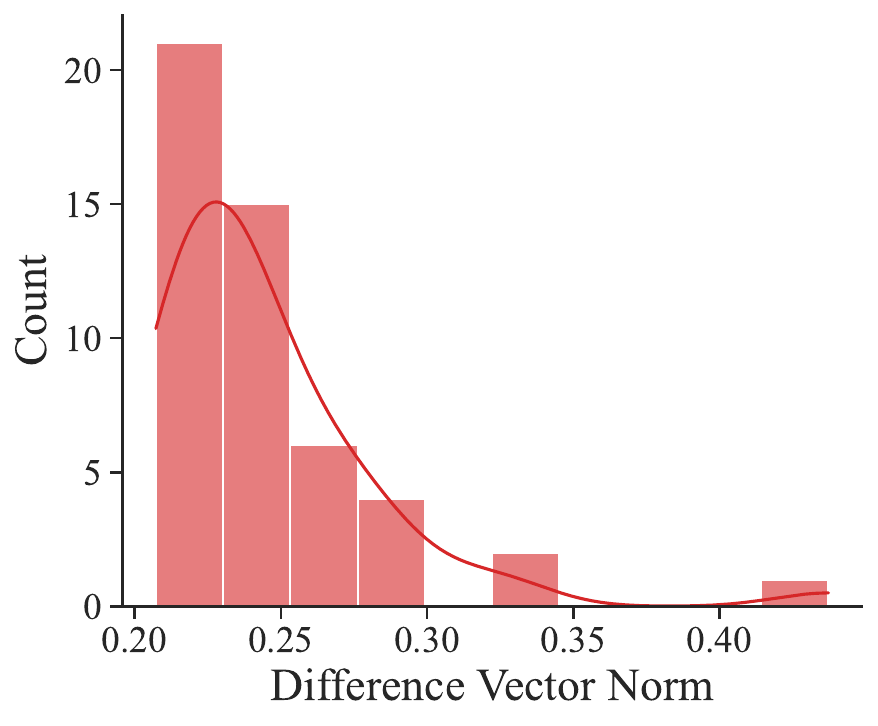}
        \caption{Difference Vector Norm Distribution}
        \label{fig:stylebkd_drastic_changes_b}
    \end{subfigure}
    \vspace{-0.5cm}
    \caption{Analysis of the drift brought by style-based transformation. (a) demonstrates the contradiction between significant lexical modification and retained intent orientation. (b) illustrates the large spatial displacement suffered by the poisoned inputs, destroying the stealthiness of backdoor.}
    \label{fig:stylebkd_drastic_changes}
    \vskip -0.2in
\end{figure}

While \sys demonstrates robust performance against token- and word-level transferable backdoors, we acknowledge the evolving threat landscape populated by more abstract trigger mechanisms, including semantic synonyms, homographs~\citep{li2021hidden}, syntactic structures, and style-based triggers (e.g., StyleBkd~\citep{qi2021mind}). It is commonly hypothesized that style-transfer backdoors can evade traditional fixed-vocabulary detection without using discrete trigger words. However, our empirical analysis reveals that to maintain the adversarial feature mapping, these attacks inevitably introduce drastic structural changes at the input level. For instance, applying a targeted stylistic transfer (e.g., ``Bible'' style) requires aggressive transformations that severely distort the lexical integrity of the original text. As shown in \textcolor{cyan}{\autoref{fig:stylebkd_drastic_changes_a}}, the lexical overlap between clean and style-transferred inputs is minimal (with word Jaccard similarity~\cite{jaccard1912distribution} approaching zero). Furthermore, \textcolor{cyan}{\autoref{fig:stylebkd_drastic_changes_b}} reveals that the adversarial displacement required to inject such style triggers results in huge difference vector norms, completely compromising the attack's stealthiness and exposing it to basic literal and numerical review.

\begin{figure}[t]
    \centering
    \includegraphics[width=0.9\linewidth]{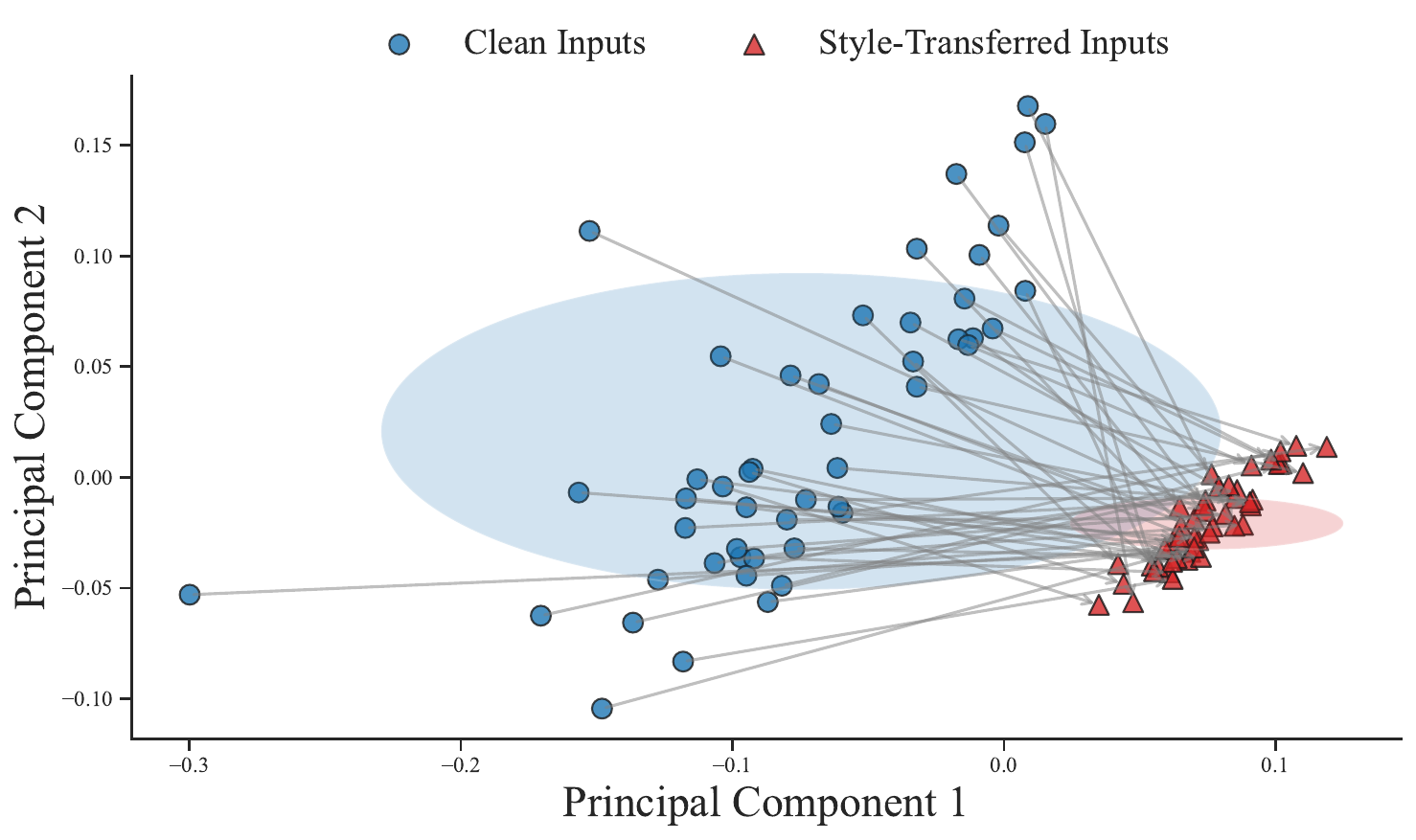}
    \vspace{-0.2cm}
    \caption{Input representation of clean inputs vs. style-transferred inputs using PCA.}
    \label{fig:stylebkd_semantic_collapse}
    \vskip -0.1in
\end{figure}

\begin{figure}[t]
    \centering
    \includegraphics[width=0.9\linewidth]{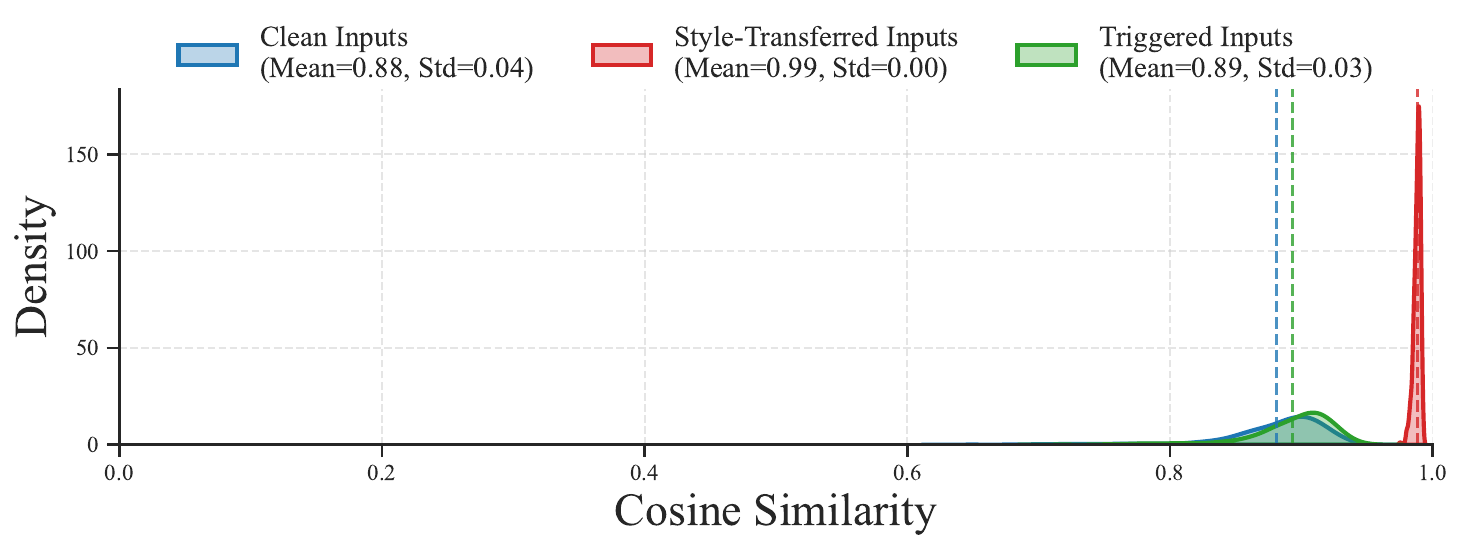}
    \vspace{-0.2cm}
    \caption{Distribution of intra-class pairwise cosine similarities among clean, token-level triggered and style-transferred instances. The dense, right-shifted style-transferred cluster \textcolor{red}{(red)} can be effortlessly detected by anomaly detection mechanisms.}
    \label{fig:stylebkd_similarity_density}
    \vskip -0.15in
\end{figure}

Critically, these drifts manifest as statistical anomalies in the representation space. As visualized via PCA~\cite{pearson1901liii} in \textcolor{cyan}{\autoref{fig:stylebkd_semantic_collapse}}, clean inputs exhibit a natural spread of semantic variance, but style-transferred inputs suffer severe variance collapse, forced into a dense cluster by the dominant style. This lack of stealth is further confirmed in \textcolor{cyan}{\autoref{fig:stylebkd_similarity_density}}, which analyzes the pairwise cosine similarity distributions across three types of inputs: clean, style-transferred, and triggered. Compared to clean inputs (Mean=0.88, Std=0.04) and token-level triggered inputs (Mean=0.89, Std=0.03), the style-transferred inputs exhibit an unnaturally high inter-sample similarity with (Mean=0.99, Std=0.00), concentrating all samples at an extreme similarity value far outside the range of natural data variation.

Consequently, defenders can easily intercept these malicious inputs prior to model inference by deploying simple input-side monitoring, such as tracking feature space dispersion (\textcolor{cyan}{\autoref{fig:stylebkd_semantic_collapse}}) or establishing basic structural distance thresholds (\textcolor{cyan}{\autoref{fig:stylebkd_drastic_changes_b}}). Furthermore, this structural manipulation forces the adversary to pay a heavy price in terms of model usability; the compromised PLMs often exhibit degraded performance on standard benchmarks due to the interference between the complex trigger features—which homogenize diverse inputs—and the model's benign representation space. Ultimately, their significant utility cost and lack of input-level stealth make them less viable in realistic supply chain scenarios, rendering them highly susceptible to standard utility-based rejection and straightforward input filtering before deep security auditing is even required.

%% file: sections/2relatedwork.tex
\section{Related Work}
\subsection{Transferable Backdoor Attacks on PLMs}
Backdoor attacks have evolved from task-specific injections to more task-agnostic threats targeting PLMs. Transferable backdoors are embedded during pre-training and remain dormant until activated by specific triggers in fine-tuned models.

\textbf{Data and Weight Poisoning.} 
Early works focused on poisoning training data or modifying model weights to establish a trigger-label shortcut. 
\citet{gu2017badnets} introduced BadNets, demonstrating that backdoors could be injected via data poisoning, which was later adapted to NLP by~\citep{dai2019backdoor}. 
More stealthy approaches such as RIPPLES~\citep{kurita2020weight}, LWP~\citep{li2021backdoor} and Kallima~\citep{chen2022kallima} tried to ensure backdoors survive the fine-tuning process.
Recently, attacks have targeted prompt-based paradigms. \citet{du2022ppt} proposed PPT to attack prompt tuning, while \citet{mei2023notable} introduced NOTABLE, which utilizes adaptive verbalizers to inject backdoors that are transferable across different downstream tasks and prompting strategies. NWS~\citep{du2024nws} proposed a natural word substitution method to implement covert textual backdoor attacks.

\textbf{Output Representation Manipulation.} To achieve task-agnostic transferability, recent research focus on mapping triggers to pre-defined output representations. Attacks, such as NeuBA~\citep{zhang2023red} and POR~\citep{shen2021backdoor}, utilize output representation learning to create backdoors that are robust to fine-tuning. 
Most recently, \citet{du2024uor} proposed UOR, which employs poisoned supervised contrastive learning to automatically optimize the trigger-representation mapping, achieving higher universality and stealthiness than manual selection methods.

% Neuba~\citep{zhang2023red} and POR~\citep{shen2021backdoor} are the state-of-the-art task-agnostic backdoor attacks against PLMs, with their core idea being to bind backdoor triggers to predefined vectors. Specifically, when the input contains the trigger, the attacker forces the output vector of the trained model to be close to the predefined vector. After the model is fine-tuned, downstream models can hardly remove such characteristics, so the model's predictions will be controlled by the trigger rather than the clean input.

\subsection{Backdoor Defenses for PLMs}
Defenses generally fall into two categories: detecting abnormal inputs or purifying models.\looseness=-1

\textbf{Input-Centric Detection.} 
\citet{qi2021onion} introduced ONION, which detects trigger words by calculating the perplexity difference when suspect words are removed. 
RAP~\citep{yang2021rap} constructs robustness-aware perturbations to distinguish poisoned samples from clean ones using a discriminator. 
Furthermore, \citet{cui2022unified} developed a unified detection framework that analyzes the hidden state distinctions between clean and poisoned inputs to filter malicious data before it reaches the model.
Currently, the SOTA method for direct trigger search is LMSanitator~\citep{weilmsanitator}. This method exploits the impact of trigger words on the sentence vector representation in the feature space and employs a three-step framework to detect triggers in prompt tuning scenarios. Recently, X-GRAAD~\citep{das2025unmasking} detects backdoor triggers via token-level attention and gradient signals, and neutralizes them by injecting noise at inference time.

\textbf{Model-Centric Detection and Sanitation.}
\citet{liu2018fine} proposed Fine-Pruning to remove neurons that are dormant on clean data but active on poisoned data. \citet{zhang2022fine} studied mitigation of backdoors post fine-tuning.
RECIPE~\citep{zhu2023removing} suppresses backdoor-related weights through regularized continuous pre-training with a small amount of irrelevant auxiliary data.
PURE~\citep{zhao2024defense} prunes suspicious attention heads and normalizes attention weights to mitigate backdoor effects without retraining. 
Recently, \citet{arora2024here} demonstrated that model merging can effectively dilute and neutralize backdoors. 
Additionally, DUP~\citep{hu2025dup} is a detection-guided unlearning framework that specifically targets and erases the backdoor logic from the model weights. BTU~\citep{jiang2025backdoor} detects and replaces backdoor tokens with padding tokens, proactively identifying and neutralizing backdoor tokens during the training phase.

Besides, \citet{kong2025neural} proposed a parameter-efficient purification strategy for multi-modal models. \citet{wang2025model} proposed a transferable two-stage optimization that separately optimizes triggers and victim pre-trained models, but only focused on vision tasks.\looseness=-1

%% file: sections/7conclusion.tex
\section{Conclusion}

This paper identifies that the parameter shift inherent in fine-tuning renders existing output-centric defenses ineffective against transferable backdoors. To address this, we present \sys, a novel framework for detecting and mitigating transferable backdoors in PLMs. By employing an input-side search strategy and a multi-source contrastive search algorithm, \sys effectively addresses the challenges of trigger detection and model purification. Extensive experiments demonstrate its superior performance in reducing attack success rates while maintaining model accuracy. \sys sets a new benchmark in safeguarding PLMs, with potential for adaptation to other adversarial threats.

%% file: sections/ethic.tex
% \section{Acknowledgements}
% This paper was edited for grammar and writing using \texttt{Gemini-3 Pro}.
\section{Ethics Considerations}
Our research focuses on a defensive mechanism to improve the security and trustworthiness of AI systems. The primary ethical motivation for \sys is to detect and cleanse backdoored model from adversaries. The backdoor model used in the research was strictly controlled to store only locally and was used solely for testing the effectiveness of the defense algorithms, without causing any concern to the community.

%% file: sections/appendix.tex
% \addcontentsline{toc}{section}{Appendix} % Add the appendix text to the document TOC
% \part{Appendix} % Start the appendix part
% \parttoc % Insert the appendix TOC
% \clearpage

\begin{table*}[!h]
    \centering
    \renewcommand{\arraystretch}{1.3}
            \caption{Summary of notations and their descriptions in this paper.}
    \label{tab:notations}
    \begin{tabular}{l|l}
        \toprule
        \textbf{Symbol} & \textbf{Description} \\
        \hline
        \multicolumn{2}{c}{\textit{Threat Model \& General Definitions}} \\
        \hline
        $\mathcal{D}$ & The clean (benign) dataset  \\
        $\tilde{\mathcal{D}}$ & The poisoned dataset injected with triggers  \\
        $x$ & The original benign input sample \\
        $\mathcal{T}$ & The set of trigger indices or the set of triggers \\
        $t_i / t_k$ & The $i$-th or $k$-th trigger sequence \\
        $x \oplus t_k$ & The poisoned sample obtained by inserting trigger $t_k$ into $x$ \\
        $M(\cdot; \theta)$ / $\mathcal{M}$ & The parameterized Pre-trained Language Model (PLM) \\
        $\theta$ & The parameters of the clean model \\
        $\theta^*$ & The parameters of the backdoored model \\
        $\mathcal{L}_E$ & Loss function for backdoor effectiveness  \\
        $\mathcal{L}_U$ & Loss function for model usability \\
        \hline
        \multicolumn{2}{c}{\textit{Backdoor Detection}} \\
        \hline
        ${D}_{DS}$ & Distribution Shift Distance loss term  \\
        ${D}_{IC}$ & Intra-class Aggregation Distance loss term \\
        ${D}_{IR}$ & Inter-class Repulsion Distance loss term  \\
        $\mathcal{L}$ & The comprehensive multi-trigger comparative search loss \\
        $dis(\cdot)$ & The distance between vectors
  \\
        $\tau$ & Temperature hyperparameter for contrastive loss  \\
        $V$ & The complete vocabulary of the PLM  \\
        $M_e$ & Embedding matrix  \\
        $M_w$ & Word-to-token mapping matrix \\
        $FS$ & Fuzzy Search rounds \\
        $N$ & Number of preset trigger groups  \\
        $E$ & Update times per round  \\
        \hline
        \multicolumn{2}{c}{\textit{Backdoor Verification}} \\
        \hline
        $S_1$ & Cosine similarity between poisoned and original clean sample features \\
        $S_2$ & Cosine similarity between feature representations of different poisoned samples  \\
        $\gamma_1$ & Threshold for $S_1$ (Detection Judgment)\\
        $\gamma_2$ & Threshold for $S_2$ (Detection Judgment) \\
        \hline
        \multicolumn{2}{c}{\textit{Model Purification}} \\
        \hline
        $\mathcal{D}_{ft}$ & Training set for downstream fine-tuning  \\
        $\mathcal{F}(\cdot, \theta)$ & The downstream task-specific model  \\
        $\mathcal{L}_{ce}$ & Cross-entropy loss used in adversarial fine-tuning \\
        $\mathcal{L}_{cleanse}$ & Cleansing loss minimizing discrepancy between poisoned/clean inputs  \\
        $\mathcal{L}_{fidelity}$ & Fidelity loss aligning purified model with the frozen original model  \\
        $\theta_{freeze}^*$ & Parameters of the frozen copy of the backdoored model \\
        $\lambda$ & Coefficient balancing cleansing and fidelity losses \\
        \bottomrule
    \end{tabular}
\end{table*}

\appendix
\section{Notations}
Table \ref{tab:notations} summarizes the mathematical notations and symbols used throughout this paper.

\section{Additional Details}
\subsection{Victim PLMs}
\label{appendix:plm}
Table~\ref{tab:PLMs} presents the PLMs evaluated in this study, encompassing a broad range of architectures and pre-training paradigms. Both base and large variants are considered when available, providing a comprehensive assessment across model scales. All models were loaded via the HuggingFace platform. This diverse selection ensures the generalizability and robustness of our findings across different PLM architectures and sizes.

\begin{table}[htbp]
    \centering
    \caption{Overview of PLMs evaluated in this study, including model architecture and parameter sizes for Base and Large versions.}
    \label{tab:PLMs}
    % \resizebox{0.99\linewidth}{!}{
    \setlength{\tabcolsep}{13pt}
    \begin{tabular}{l l c c} % 4 columns: Left align for names, others centered
        \toprule % Top line
        \textbf{PLM} & \textbf{Type} & \textbf{Base} & \textbf{Large} \\
        \midrule % Middle line
        BERT~\citep{devlin2018bert} & Encoder & 110M & 340M \\
        RoBERTa~\citep{liu2019roberta} & Encoder & 125M & 355M \\
        DeBERTa~\citep{he2020deberta} & Encoder & 150M & 400M \\
        BART~\citep{lewis2020bart} & Encoder-Decoder & 140M & 400M \\
        XLNet~\citep{yang2019xlnet} & Permutation & 110M & 340M \\
        ALBERT~\citep{lan2019albert} & Encoder & 11M & 17M \\
        DistilBERT~\citep{sanh2019distilbert} & Encoder & 66M & \ding{55} \\
        ERNIE 2.0~\citep{zhang2019ernie} & Encoder & 110M & 340M \\
        \bottomrule % Bottom line
    \end{tabular}
    % }
\end{table}

\subsection{Datasets}
\label{appendix:dataset}
Table~\ref{tab:dataset} summarizes the datasets used in our experiments, spanning a diverse set of NLP tasks. These include sentiment analysis (SA), toxic detection (TD), spam detection (SD), and topic classification (TC), each represented by widely adopted benchmark datasets. For datasets without an official test set, the validation set is utilized as the test set in our experiments, with the training set re-partitioned to construct a new validation set.
\begin{table}[ht]
  \centering
    \caption{
    Statistics of datasets. 
    Task abbreviations:
    SA = Sentiment Analysis,
    TD = Toxic Detection,
    SD = Spam Detection,
    TC = Topic Classification.
  }
  \label{tab:dataset}
  \resizebox{\linewidth}{!}{
  \begin{tabular}{c|c|c|c|c|c}
  \toprule
  \textbf{Task} & \textbf{Datasets} & \textbf{\# Classes} & \textbf{Train} & \textbf{Valid} & \textbf{Test} \\
  \midrule
  \multirow{4}{*}{SA} 
    & \textbf{SST-2}~\cite{socher2013recursive}  & 2 & 60,614  & 6,735  & 872    \\ 
    & \textbf{IMDB}~\cite{maas2011learning}      & 2 & 22,500  & 2,500  & 25,000 \\ 
    & \textbf{SST-5}~\cite{socher2013recursive}  & 5 & 8,544   & 1,101  & 2,210  \\  
    & \textbf{Yelp}~\cite{minaee2021deep}        & 5 & 585,000 & 65,000 & 50,000 \\ 
  \midrule
  \multirow{2}{*}{TD}
    & \textbf{OLID}~\cite{zampieri2019semeval}   & 2 & 11,916 & 1,324 & 860   \\  
    & \textbf{Twitter}~\cite{founta2018large}    & 2 & 69,632 & 7,737 & 8,597 \\  
  \midrule
  \multirow{2}{*}{SD}
    & \textbf{Enron}~\cite{metsis2006spam}       & 2 & 24,944 & 2,772 & 6,000 \\  
    & \textbf{Lingspam}~\cite{sakkis2003memory}  & 2 & 2,603  & 290   & 580   \\  
  \midrule
  \multirow{1}{*}{TC}
    & \textbf{Agnews}~\cite{zhang2015character}  & 4 & 108,000 & 12,000 & 7,600 \\  
  \bottomrule
  \end{tabular}}
\end{table}

\subsection{Attack Methods}
\label{appendix:attack}
\begin{table}[t]
  \centering
    \caption{Token-level and Word-level triggers used.}
  \label{tab:trigger-words}
  \resizebox {\linewidth} {!} {
  \begin{tabular}{c|c}
  \toprule
  \textbf{Levels} & \textbf{Triggers}\\
  \midrule
  \multirow{2}{*}{\textbf{Token}}  & [`dh', `vo', `cy', `ak', `ev', `xx']   \\ 
                                         & [`por', `neo', `lev', `ign', `rim', `yen']   \\ \midrule
  \multirow{2}{*}{\textbf{Word}}   & [`Riemann', `Bayes', `Descartes', `Cauchy', `Fermat', `Lagrange']    \\  
                                         & [`heterogenous', `solipsism', `pulchritude', `emollient', `denigrate', `linchpin']    \\ 
  \bottomrule
  \end{tabular}
  }
\end{table}
Both of POR and NeuBA implant backdoor by establishing a strong association between trigger words and artificially constructed output representations. We set 4 types of trigger word patterns, with specific configurations shown in \textcolor{cyan}{\autoref{tab:trigger-words}}. 

The POR framework provides two configuration schemes: POR-1 divides the $K$-dimensional output representation into $n$ vectors of \(\frac{K}{n}\) dimensions, denoted as \([a_1, a_2, \dots, a_n]\). For the \(j^{\text{th}}\) trigger word, the corresponding vectors follow a gradual change rule: \(a_i=(-1)_{\frac{K}{n}}\) (for all \(\forall i \geq j\)) and \(a_i=(1)_{\frac{K}{n}}\) (for all \(\forall i<j\)), where \(j\in\{1, \ldots, n+1\}\); POR-2 decomposes the output representation into $m$ vectors of \(\frac{K}{m}\) dimensions, denoted as \([a_1, a_2, \dots, a_m]\). Each component satisfies \(a_i\in\{-1,1\}\) with \(i\in\{1,\ldots,m\}\). NeuBA achieves a similar goal by defining alternately orthogonalized vectors.

\subsection{Defense Baselines}
\label{appendix:baseline}
We use LMSanitator~\citep{weilmsanitator} as the primary baseline for trigger reverse search. To ensure comparative fairness, we reconstruct its search algorithm by integrating discrete optimization to adapt it to an input-side search mechanism. Meanwhile, Onion~\citep{qi2021onion}, which is based on perplexity filtering, is selected as the comparative scheme for input monitoring. X-GRAAD~\citep{das2025unmasking} detects backdoor triggers via token-level attention and gradient signals, and neutralizes them by injecting noise at inference time. The model sanitization-based defense schemes include parameter reinitialization (Re-init), Fine-Pruning~\citep{liu2018fine}, Recipe~\citep{zhu2023removing}, and BTU~\citep{jiang2025backdoor}. 
Re-init~\citep{zhang2023red} resets the parameters of the last layer (LL), pooling layer (PL), and their combinations. Fine-Pruning performs selective pruning based on neuron activation values followed by fine-tuning for recovery. Recipe eliminates backdoor features through continuous pre-training combined with L2 regularization. BTU locates and repairs suspicious trigger words by comparing the offsets of model word vectors before and after poisoning, and restores model performance through fine-tuning on clean data. In this study, the aforementioned schemes are systematically compared with \sys's adversarial fine-tuning and adversarial pre-training schemes.

\subsection{Implementation Details}
\label{appendix:Implementation}
The number of fuzzy search rounds is configured as \{2, 3, 5\}, while LMSanitator uses \{25, 50, 100\} rounds due to efficiency constraints. The default number of fuzzy search rounds is 5. Each iteration is updated 3 times, with a batch size of 16, and feature accumulation is used to achieve equivalent large-batch training for contrastive learning, with the contrastive loss temperature parameter \(\tau=0.5\). The beam search strategy sets the beam width to \(M=3\) and the number of candidate trigger words to \(K=5\). In the downstream fine-tuning phase, the backdoored PLM is connected to a task-specific classification layer, and fine-tuned for 3 epochs with a learning rate of \(2\times10^{-5}\) and a batch size of 16. All experiments were repeated 5 times on 8 NVIDIA H800 GPUs, and the average value was taken to ensure the reliability of the results.

\subsection{Evaluation Metrics}
\label{appendix:metric}
For the trigger search experiment, Recall and Time (h) are used as evaluation metrics to measure search accuracy and search efficiency, respectively. Among them, Recall reflects the proportion of real backdoor triggers identified. For the backdoor defense experiment, Attack Success Rate (ASR) and Clean Accuracy (ACC) are used for evaluation: ASR represents the proportion of poisoned samples misclassified into the target label; The degree of its reduction reflects backdoor defense performance; ACC measures the model’s ability to classify normal samples after defense, and is used to evaluate the degree to which the defense scheme maintains model usability.

\subsection{Further Analysis settings}
\label{appendix:hyperparameters}
To investigate the impact of key hyperparameters in \sys on the efficacy of backdoor detection and defense, we conduct further systematic experiments. The experiments employ the base models of BERT and XLNet, in conjunction with three backdoor attack methods (NeuBA, POR-1, and POR-2) and two trigger granularities (Token-level and Word-level). A total of 10 backdoor models are constructed ($2_{\text{models}} \times 3_{\text{attacks}} \times 2_{\text{granularities}} = 12$, with the exception that the XLNet model is not compatible with the NeuBA attack method, resulting in a final count of 10 effective models). Each model contained six sets of real triggers, ultimately forming a testing environment comprising 60 real triggers.

\section{Supplementary Methods}
\subsection{Vulnerabilities of Existing Defenses}
\label{appendix:Vulnerabilities}
Existing defense solutions such as LMSanitator implement backdoor detection and defense mechanisms at the output layer of PLMs, aiming to circumvent the convergence challenges associated with input-side searching. However, through systematic analysis, this study reveals that such output-side defense strategies exhibit significant drawbacks in practical deployment and application. As illustrated in \textcolor{cyan}{\autoref{drawbacks}}, the key limitations of this type of scheme include the following four aspects: inefficient detection, inefficient defense, trade-off between defense and accuracy, and limited generality.

This study systematically compares the characteristics of different defense mechanisms (as shown in \textcolor{cyan}{\autoref{compare}}) and proposes an input-side multi-source comparative search algorithm and constructs an overall protection framework for migratory backdoors.

\subsection{Attack Modeling and Multi-trigger Comparative Search}
\label{appendix:observation}
To enable effective reverse search on the input side, we first conducts systematic exploration and formal modeling of universal migratory backdoor attacks. \textcolor{cyan}{\autoref{visualization}} presents the feature space visualization results of the BERT~\cite{devlin2018bert} model and the model injected with 3 and 6 triggers via POR~\citep{shen2021backdoor} attacks. This visualization analysis is based on 1,000 randomly sampled clean samples and poisoned samples, using a two-stage dimensionality reduction approach. Principal component analysis (PCA)~\cite{pearson1901liii} is first applied to compress high-dimensional features to 20 dimensions, followed by further dimensionality reduction to a 2-dimensional visualization space using UMAP~\cite{mcinnes2018umap}. It can be summarized that: the essential characteristics of universal migratory backdoor attacks is a dense sub-distribution that is significantly separated from the clean distribution in the representation space of PLMs, and this sub-distribution can be activated by specific triggers. Notably, poisoned samples corresponding to different triggers exhibit a distinct cluster separation phenomenon in the feature space. Based on this finding, we propose three identifiable features of universal migratory backdoor models:

\subsubsection{Optimization of Search Strategy}
\textcolor{cyan}{Algorithm~\ref{alg:beam_search}} and \textcolor{cyan}{Algorithm~\ref{alg:greedy}} shows the main process of Beam Search and Greedy Search, respectively.

\begin{algorithm}[t]
\caption{Beam Search}
\label{alg:beam_search}
\begin{algorithmic}[1]
\REQUIRE Current triggers $\mathcal{T}$, Loss function $\mathcal{L}$, Beam size $M$, Candidate sets $\{\mathcal{C}_1, \dots, \mathcal{C}_N\}$ (derived from gradients).
\ENSURE Optimal triggers $\mathcal{T}^*$.
\STATE Initialize beam queue $\mathcal{Q} \leftarrow \{(\mathcal{T}, \mathcal{L}(\mathcal{T}))\}$.
\FOR{$i = 1$ to $N$}
    \STATE $\mathcal{Q}_{cand} \leftarrow \mathcal{Q}$. \COMMENT{Initialize candidates with current beam}
    \FOR{each $(\mathcal{T}_{curr}, \ell_{curr}) \in \mathcal{Q}$}
        \FOR{each candidate word $w \in \mathcal{C}_i$}
            \IF{$w \in \mathcal{T}_{curr}$}
                \STATE \textbf{continue}
            \ENDIF
            \STATE Generate $\mathcal{T}_{next}$ by replacing the $i$-th token of $\mathcal{T}_{curr}$ with $w$.
            \STATE Calculate loss $\ell_{next} \leftarrow \mathcal{L}(\mathcal{T}_{next})$.
            \STATE Add $(\mathcal{T}_{next}, \ell_{next})$ to $\mathcal{Q}_{cand}$.
        \ENDFOR
    \ENDFOR
    \STATE $\mathcal{Q} \leftarrow \text{Top-}M(\mathcal{Q}_{cand})$ based on minimum loss $\ell$.
\ENDFOR
\STATE Select $\mathcal{T}^*$ from $\mathcal{Q}$ with the global minimum loss.
\RETURN $\mathcal{T}^*$
\end{algorithmic}
\end{algorithm}

\begin{algorithm}[!htbp]
\caption{Greedy Search}
\label{alg:greedy}
\begin{algorithmic}[1]
\REQUIRE Current triggers $\mathcal{T} = \{t_1, t_2, \dots, t_N\}$, Loss function $\mathcal{L}$, Embedding matrix $\mathbf{E}$, Number of candidates $K$.
\ENSURE Updated triggers $\mathcal{T}_{new}$.
\STATE Compute gradients $\mathbf{G} \in \mathbb{R}^{N \times d}$ with respect to trigger embeddings.
\STATE Initialize $\mathcal{T}_{new} \leftarrow \emptyset$.
\FOR{$i = 1$ to $N$}
    \STATE Calculate approximation scores for vocabulary $\mathcal{V}$: $S \leftarrow -\mathbf{G}_i \cdot \mathbf{E}^\top$.
    \STATE Select top-$K$ candidates $\mathcal{C}_i$ with the highest scores from $S$.
    \FOR{each candidate $c \in \mathcal{C}_i$}
        \IF{$c \notin \mathcal{T}_{new}$}
            \STATE $t'_i \leftarrow c$.
            \STATE Add $t'_i$ to $\mathcal{T}_{new}$.
            \STATE \textbf{break}
        \ENDIF
    \ENDFOR
\ENDFOR
\RETURN $\mathcal{T}_{new}$
\end{algorithmic}
\end{algorithm}

\section{Supplementary Experiments}

\begin{table*}[th]
\centering
\caption{Detailed False Positive analysis on a clean \texttt{bert-base-uncased} model. The \underline{token-level} algorithm searched for 8 candidate groups ($N=8$). All candidates exhibit high Inter-class Similarity ($S_1 \gg \gamma_1$) and low Intra-class Similarity ($S_2 < \gamma_2$), resulting in a consistent ``Clean'' determination.}
\label{tab:fpr_case_token}
\setlength{\tabcolsep}{12.5pt}
\begin{tabular}{lcccccccc}
\toprule
\textbf{Candidate ID} & \textbf{1} & \textbf{2} & \textbf{3} & \textbf{4} & \textbf{5} & \textbf{6} & \textbf{7} & \textbf{8} \\
\midrule
\textbf{Tokens} & \texttt{\cjktoken{」}} & \texttt{smashwords} & \texttt{‖} & \texttt{←} & \texttt{rhyme} & \texttt{hapoel} & \texttt{domesday} & \texttt{hailey} \\
\midrule
\textbf{$S_1$ (Inter-class)} & 0.8840 & 0.8808 & 0.9095 & 0.8813 & 0.9495 & 0.8920 & 0.9220 & 0.9501 \\
\textit{$< \gamma_1 (0.4)$ ?} & \xmark & \xmark & \xmark & \xmark & \xmark & \xmark & \xmark & \xmark \\
\midrule
\textbf{$S_2$ (Intra-class)} & 0.6748 & 0.7368 & 0.7024 & 0.6655 & 0.6979 & 0.6758 & 0.6701 & 0.6831 \\
\textit{$> \gamma_2 (0.9)$ ?} & \xmark & \xmark & \xmark & \xmark & \xmark & \xmark & \xmark & \xmark \\
\midrule
\textbf{Result} & \textbf{Clean} & \textbf{Clean} & \textbf{Clean} & \textbf{Clean} & \textbf{Clean} & \textbf{Clean} & \textbf{Clean} & \textbf{Clean} \\
\bottomrule
\end{tabular}
\end{table*}

\begin{table*}[t]
\centering
\caption{Detailed False Positive analysis on a clean \texttt{bert-base-uncased} model. The \underline{word-level} algorithm searched for 8 candidate groups ($N=8$). All candidates exhibit high Inter-class Similarity ($S_1 \gg \gamma_1$) and low Intra-class Similarity ($S_2 < \gamma_2$), resulting in a consistent ``Clean'' determination.}
\label{tab:fpr_case_word}
\resizebox{\linewidth}{!}{
\begin{tabular}{lcccccccc}
\toprule
\textbf{Candidate ID} & \textbf{1} & \textbf{2} & \textbf{3} & \textbf{4} & \textbf{5} & \textbf{6} & \textbf{7} & \textbf{8} \\
\midrule
\textbf{Word} & \texttt{upliftment} & \texttt{smuttiest} & \texttt{conveniently} & \texttt{affection} & \texttt{justification} & \texttt{deceiver} & \texttt{praising} & \texttt{mantra} \\
\midrule
\textbf{$S_1$ (Inter-class)} & 0.9607 & 0.9661 & 0.9659 & 0.9612 & 0.9468 & 0.9637 & 0.9573 & 0.9513 \\
\textit{$< \gamma_1 (0.4)$ ?} & \xmark & \xmark & \xmark & \xmark & \xmark & \xmark & \xmark & \xmark \\
\midrule
\textbf{$S_2$ (Intra-class)} & 0.6794 & 0.6852 & 0.6795 & 0.6924 & 0.6807 & 0.6841 & 0.6870 & 0.6909 \\
\textit{$> \gamma_2 (0.9)$ ?} & \xmark & \xmark & \xmark & \xmark & \xmark & \xmark & \xmark & \xmark \\
\midrule
\textbf{Result} & \textbf{Clean} & \textbf{Clean} & \textbf{Clean} & \textbf{Clean} & \textbf{Clean} & \textbf{Clean} & \textbf{Clean} & \textbf{Clean} \\
\bottomrule
\end{tabular}
}
\end{table*}

\subsection{False Positive Case Analysis on Clean Models}
\label{appendix:fpr}
As discussed in Section~\ref{section:fp}, we conduct a detailed false positive analysis using a clean \texttt{bert-base-uncased} model. The specific candidate triggers and their corresponding similarity metrics ($S_1$ and $S_2$) are detailed in \textcolor{cyan}{\autoref{tab:fpr_case_token}} for token-level search and \textcolor{cyan}{\autoref{tab:fpr_case_word}} for word-level search.

\subsection{Trigger Search Result Analysis}
\label{appendix:trigger-search}
The trigger search results, as presented in \autoref{fig:trigger-search}, demonstrate that under different attack parameter configurations, the recognition rate of the real trigger by the search algorithm shows an upward trend as the number of fuzzy search iterations increases. Among them, the search for Token-level triggers is less challenging, and the algorithm exhibits a higher recall rate, while the search for word-level triggers faces greater difficulties. It should be noted that there are unintended triggers in the search results of multiple backdoor models. Through feature representation analysis, this phenomenon may stem from the composite injection in multi-trigger backdoor training—there is a high similarity between the combined feature representations and the unintended triggers, which leads to additional identification by the reverse search algorithm. This finding also illustrates the necessity of the multi-round fuzzy search mechanism from another perspective: backdoor models may contain potential triggers that are unintentionally implanted by attackers.\looseness=-1

\subsection{Multi-Granularity Trigger Detection}
\label{appendix:granularity}
\sys natively supports both \textbf{token-level} and \textbf{word-level} backdoor detection, with flexible and controllable mode switching. Users can independently specify the target granularity via a single parameter without modifying the core algorithm framework, enabling adaptive defense against diverse backdoor injection paradigms.

\subsubsection{Mechanism Differences Between Two Modes}
The two detection modes share the same multi-trigger contrastive search core-method but differ in input processing and optimization mechanisms:

\textbf{Token-level mode}: Directly operates on sub-word tokens (e.g., `dh', `vo' in \textcolor{cyan}{\autoref{tab:trigger-words}}). It leverages the PLM's native token embedding matrix \(M_e\) for gradient-guided updates, adopting the HotFlip-based linear approximation to address discrete optimization challenges.

\textbf{Word-level mode}: Focuses on complete natural words (e.g., `Riemann', `heterogenous' in \textcolor{cyan}{\autoref{tab:trigger-words}}). It constructs a word-to-token mapping matrix \(M_w \in \mathbb{R}^{7000 \times 6}\) (proposed in PICCOLO~\citep{liu2022piccolo}) to convert words into valid token sequences (padded with `[PAD]' token if necessary). The optimization is performed at the word level while ensuring grammatical validity of the underlying token combinations.\looseness=-1

\subsubsection{Orthogonality Between Defense Levels and Attack Levels}
To verify the effectiveness of cross-granularity detection, we conduct experiments on \texttt{bert-base-uncased} models injected with backdoors of different granularities. The results in \textcolor{cyan}{\autoref{tab:cross_granularity}} show that when the defense level is orthogonal to the attack level, \sys can only detect backdoored models under NeuBA attack method.

\begin{table*}[t]
\centering
\caption{Backdoor detection results under cross-granularity settings. 
`Is Backdoored?' denotes whether \sys detects the model as backdoored.}
\label{tab:cross_granularity}
\resizebox{0.99\linewidth}{!}{
\begin{tabular}{cccccc}
\toprule
Detection & Attack & Attack Method & Triggers & Is Backdoored? \\
\midrule

% ================= Token → Word =================
\multirow{6}{*}{Token-level} &
\multirow{6}{*}{Word-level} &
\multirow{2}{*}{NeuBA} &
[`Riemann', `Bayes', `Descartes', `Cauchy', `Fermat', `Lagrange'] & \ding{51} \\

& & &
[`heterogenous', `solipsism', `pulchritude', `emollient', `denigrate', `linchpin'] & \ding{51} \\
\cmidrule(lr){3-5}

& & \multirow{2}{*}{POR-1} &
[`Riemann', `Bayes', `Descartes', `Cauchy', `Fermat', `Lagrange'] & \xmark \\

& & &
[`heterogenous', `solipsism', `pulchritude', `emollient', `denigrate', `linchpin'] & \xmark \\
\cmidrule(lr){3-5}

& & \multirow{2}{*}{POR-2} &
[`Riemann', `Bayes', `Descartes', `Cauchy', `Fermat', `Lagrange'] & \xmark \\

& & &
[`heterogenous', `solipsism', `pulchritude', `emollient', `denigrate', `linchpin'] & \xmark \\

\midrule

% ================= Word → Token =================
\multirow{6}{*}{Word-level} &
\multirow{6}{*}{Token-level} &
\multirow{2}{*}{NeuBA} &
[`dh', `vo', `cy', `ak', `ev', `xx'] & \ding{51} \\

& & &
[`por', `neo', `lev', `ign', `rim', `yen'] & \xmark \\
\cmidrule(lr){3-5}

& & \multirow{2}{*}{POR-1} &
[`dh', `vo', `cy', `ak', `ev', `xx'] & \xmark \\

& & &
[`por', `neo', `lev', `ign', `rim', `yen'] & \xmark \\
\cmidrule(lr){3-5}

& & \multirow{2}{*}{POR-2} &
[`dh', `vo', `cy', `ak', `ev', `xx'] & \xmark \\

& & &
[`por', `neo', `lev', `ign', `rim', `yen'] & \xmark \\

\bottomrule
\end{tabular}
}
\end{table*}

\begin{table*}[t]
  \centering
  \caption{Results of \underline{token-level} detection mode on backdoored model with token-level triggers: Candidate triggers and similarity metrics.}
  \label{tab:token_level_case}
  \begin{tabular}{l|ccc}
    \toprule
    Detected Suspicious {Token} & S1 & S2 & Meet Thresholds (\(S1<0.4\) and \(S2>0.9\))? \\
    \midrule
    \texttt{ak}                           & -0.0041  & 0.9999          & \ding{51}                              \\
    \texttt{dh}                           & -0.0266  & 0.9998          & \ding{51}                              \\
    \texttt{xx}                          & 0.1741  & 0.9992            & \ding{51}                              \\
    \texttt{cy}                          & 0.0031    & 0.9999          & \ding{51}                              \\
    \texttt{ev}                          & 0.0203    & 0.9998          & \ding{51}                              \\
    \texttt{vo}                           & 0.0107  & 0.9998           & \ding{51}                              \\
    \texttt{xx}                           & 0.1741  & 0.9992           & \ding{51}                              \\
    \texttt{housemates}                           & 0.9318  & 0.7223           & \xmark                            \\
    \midrule
    \textbf{Judgment} & \multicolumn{3}{c}{\textbf{Backdoored (token-level mode detects valid triggers)}}  \\
    \bottomrule
  \end{tabular}
\end{table*}

\begin{table*}[t]
  \centering
  \caption{Results of \underline{word-level} detection mode on backdoored model with token-level triggers: Candidate triggers and similarity metrics.}
  \label{tab:word_level_case}
  \begin{tabular}{l|ccc}
    \toprule
    Detected Suspicious {Word} & S1 & S2 & Meet Thresholds (\(S1<0.4\) and \(S2>0.9\))? \\
    \midrule
    \texttt{flabbergasted}    & 0.9772              & 0.7133            & \xmark                                       \\
    \texttt{multi-polarization}& 0.9613             & 0.7254             & \xmark                                     \\
    \texttt{bristle}          & 0.9860              & 0.7102              & \xmark                                    \\
    \texttt{mesmerizingly}    & 0.9752              & 0.7152             & \xmark                                      \\
    \texttt{penetrating}      & 0.9793              & 0.7082             & \xmark                                      \\
    \texttt{vilify}           & 0.9826              & 0.7067             & \xmark                                     \\
    \texttt{unskilled}        & 0.9829              & 0.7035             & \xmark                                     \\
    \texttt{intricate}        & 0.9716              & 0.7168             & \xmark                                     \\
    \midrule
    \textbf{Judgment} & \multicolumn{3}{c}{\textbf{Clean (no valid triggers) }}               \\
    \bottomrule
  \end{tabular}
\end{table*}

We use the token-level trigger configuration [`dh', `vo', `cy', `ak', `ev', `xx'] and perform backdoor injection on the BERT model using the POR-1 attack method. \textcolor{cyan}{\autoref{tab:token_level_case}} and \textcolor{cyan}{\autoref{tab:word_level_case}} present the detection results of this model using the token-level and word-level modes of \sys, respectively. By integrating the outputs of the two modes, it is confirmed that the model is implanted with a backdoor. This dual-mode strategy enhances the robustness of detection, and effectively avoids missed detections caused by the limitations of single-granularity detection.

\begin{figure}[htbp]
    \centering
    % Ensure you adjust the width depending on how large you want it in the column
    \includegraphics[width=0.8\linewidth]{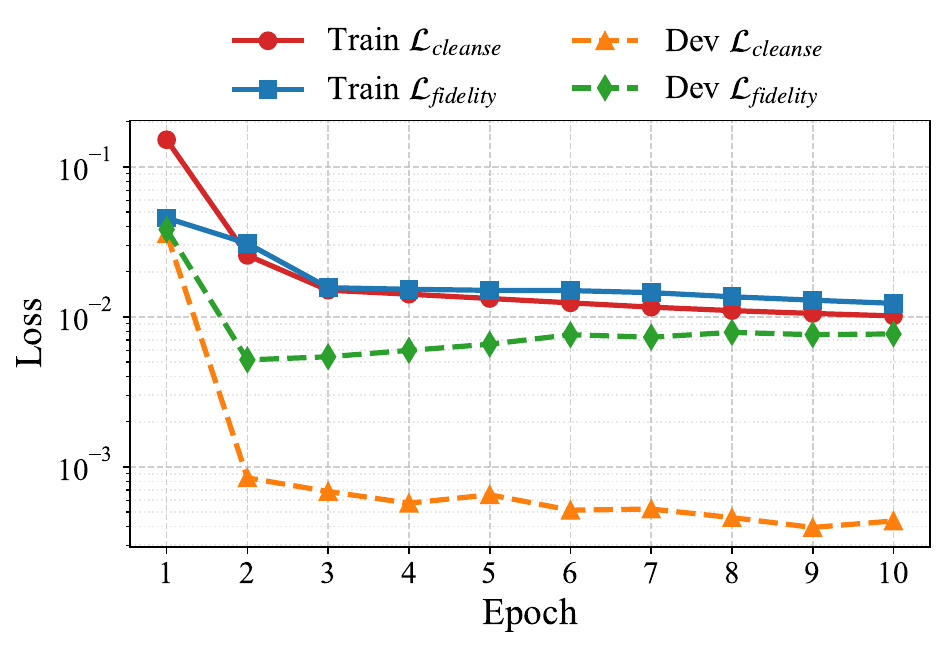}
    \caption{Training and validation purification loss over epochs demonstrating the rapid convergence of our adversarial pre-training phase.}
    \label{fig:adv_pretrain_loss}
\end{figure}

\begin{figure*}[t]
\begin{subfigure}{0.25\linewidth}
  \centering
  \includegraphics[width=\linewidth]{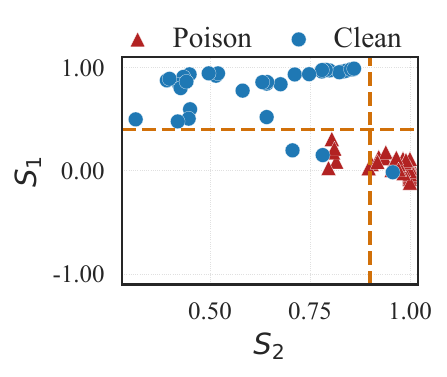}
  \caption{NeuBA\label{neuba_llm}}
\end{subfigure}
\hfill
\begin{subfigure}{0.25\linewidth}
  \centering
  \includegraphics[width=\linewidth]{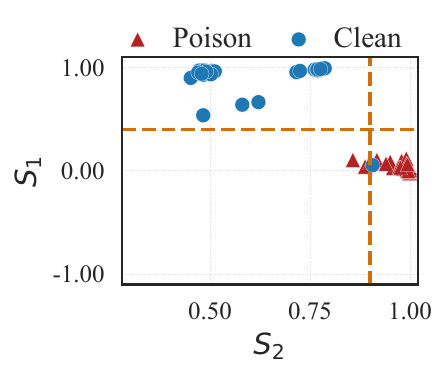}
  \caption{POR-1\label{por_1_llm}}
\end{subfigure}
\hfill
\begin{subfigure}{0.25\linewidth}
  \centering
  \includegraphics[width=\linewidth]{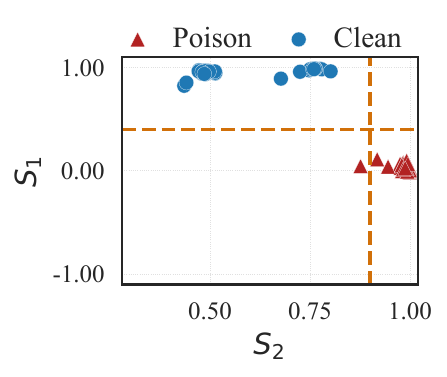}
  \caption{POR-2\label{por_2_llm}}
\end{subfigure}
\caption{{Visualization of sample judgment during the backdoor verification phase on LLMs. The horizontal and vertical dash lines denote \(\gamma_1 = 0.4\) and \(\gamma_2 = 0.9\).}}
\label{fig:llm_cos}
\end{figure*}

\begin{table*}[t]
\centering
\caption{Evaluation of backdoor defense effectiveness of \sys on LLM.}
\label{tab:llm_backdoor_defense}
% \resizebox{0.99\linewidth}{!}{
\begin{tabular}{c|c|cccc|cccc}
\toprule
\multirow{2}{*}{\textbf{Attacks}} & \multirow{2}{*}{\textbf{Defenses}} & \multicolumn{4}{c|}{\textbf{ACC} $\uparrow$} & \multicolumn{4}{c}{\textbf{ASR} $\downarrow$} \\
\cmidrule(lr){3-6} \cmidrule(lr){7-10}
& & Twit & Agnews & Yelp & Enron & Twit & Agnews & Yelp & Enron \\
\midrule
\textbf{Clean} & w/o Defense & 94.71 & 93.95 & 64.88 & 99.13 & 8.76 & 4.20 & 6.26 & 0.17 \\
\midrule
\multirow{3}{*}{\textbf{NeuBA}}
 & w/o Defense & 94.52 & 93.84 & 64.76 & 99.22 & 95.31 & 93.72 & 95.84 & 51.20 \\
\cmidrule{2-10}
& \cellcolor{cyan!15} Adv-Finetune & \cellcolor{cyan!15} 94.30 & \cellcolor{cyan!15} 93.91 & \cellcolor{cyan!15} 65.39 & \cellcolor{cyan!15} 99.36 & \cellcolor{cyan!15} \textbf{8.53} & \cellcolor{cyan!15} \textbf{3.18} & \cellcolor{cyan!15} \textbf{4.78} & \cellcolor{cyan!15} {0.36} \\
& \cellcolor{cyan!15} Adv-Pretrain & \cellcolor{cyan!15} 94.53 & \cellcolor{cyan!15} 93.44 & \cellcolor{cyan!15} 64.52 & \cellcolor{cyan!15} 98.92 & \cellcolor{cyan!15} {8.65} & \cellcolor{cyan!15} {3.49} & \cellcolor{cyan!15} {6.04} & \cellcolor{cyan!15} \textbf{0.24} \\
\midrule

\multirow{3}{*}{\textbf{POR-1}}
 & w/o Defense & 94.46 & 93.78 & 63.71 & 98.85 & 93.89 & 98.80 & 97.69 & 42.53 \\
\cmidrule{2-10}
& \cellcolor{cyan!15} Adv-Finetune & \cellcolor{cyan!15} 94.22 & \cellcolor{cyan!15} 93.90 & \cellcolor{cyan!15} 64.17 & \cellcolor{cyan!15} 98.98 & \cellcolor{cyan!15} \textbf{8.68} & \cellcolor{cyan!15} \textbf{2.87} & \cellcolor{cyan!15} 5.46 & \cellcolor{cyan!15} \textbf{0.50} \\
& \cellcolor{cyan!15} Adv-Pretrain & \cellcolor{cyan!15} 94.30 & \cellcolor{cyan!15} 93.15 & \cellcolor{cyan!15} 63.98 & \cellcolor{cyan!15} 98.36 & \cellcolor{cyan!15} 8.76 & \cellcolor{cyan!15} 3.09 & \cellcolor{cyan!15} \textbf{5.37} & \cellcolor{cyan!15} 0.64 \\
\midrule

\multirow{3}{*}{\textbf{POR-2}}
 & w/o Defense & 94.65 & 94.09 & 64.20 & 98.94 & 99.98 & 93.56 & 93.30 & 39.94 \\
\cmidrule{2-10}
& \cellcolor{cyan!15}  Adv-Finetune & \cellcolor{cyan!15} 93.82 & \cellcolor{cyan!15} 94.10 & \cellcolor{cyan!15} 65.59 & \cellcolor{cyan!15} 99.17 & \cellcolor{cyan!15} 9.01 & \cellcolor{cyan!15} 3.16 & \cellcolor{cyan!15} 4.49 & \cellcolor{cyan!15} 0.49 \\
& \cellcolor{cyan!15} Adv-Pretrain & \cellcolor{cyan!15} 94.41 & \cellcolor{cyan!15} 93.54 & \cellcolor{cyan!15} 65.10 & \cellcolor{cyan!15} 98.65 & \cellcolor{cyan!15} \textbf{8.05} & \cellcolor{cyan!15} \textbf{3.14} & \cellcolor{cyan!15} \textbf{4.45} & \cellcolor{cyan!15} \textbf{0.28} \\
\bottomrule
\end{tabular}
% }
\end{table*}

\begin{table*}[t]
\centering
\caption{Evaluation of backdoor defense effectiveness of \sys on bge-large-en-v1.5.}
\label{tab:em_backdoor_defense}
% \vspace{-10pt}
\resizebox{0.99\linewidth}{!}{
\begin{tabular}{c|c|cccccccccc|cccccccccc}
\toprule
\multirow{2}{*}{\textbf{Attack}} & \multirow{2}{*}{\textbf{Defense}} & \multicolumn{10}{c|}{\textbf{ACC} $\uparrow$} & \multicolumn{10}{c}{\textbf{ASR} $\downarrow$} \\
\cmidrule(lr){3-12} \cmidrule(lr){13-22}
& & SST-2 & IMDB & OLID & Twit & Hate & Enron & Ling & SST-5 & Agnews & Yelp & SST-2 & IMDB & OLID & Twit & Hate & Enron & Ling & SST-5 & Agnews & Yelp \\       
\midrule
\textbf{Clean} & w/o Defense & 94.15 & 95.51 & 85.70 & 94.46 & 90.60 & 99.20 & 100.00 & 56.97 & 94.86 & 68.90 & 5.29 & 4.59 & 40.35 & 8.54 & 23.49 & 0.43 & 1.03 & 19.80 & 3.72 & 6.15 \\
\midrule
\multirow{3}{*}{\textbf{POR-1}}
 & w/o Defense & 93.92 & 95.26 & 85.35 & 94.45 & 91.22 & 99.23 & 98.97 & 57.99 & 94.78 & 68.56 & 100.00 & 99.14 & 99.99 & 94.96 & 100.00 & 79.05 & 99.91 & 100.00 & 86.84 & 99.76 \\
\cmidrule{2-22}
& \cellcolor{cyan!15} Adv-Finetune & \cellcolor{cyan!15} 93.64 & \cellcolor{cyan!15} 95.12 & \cellcolor{cyan!15} 84.24 & \cellcolor{cyan!15} 94.24 & \cellcolor{cyan!15} 93.20 & \cellcolor{cyan!15} 99.15 & \cellcolor{cyan!15} 99.02 & \cellcolor{cyan!15} 53.95 & \cellcolor{cyan!15} 94.55 & \cellcolor{cyan!15} 67.90 & \cellcolor{cyan!15} 5.91 & \cellcolor{cyan!15} 6.45 & \cellcolor{cyan!15} 33.89 & \cellcolor{cyan!15} 7.30 & \cellcolor{cyan!15} 37.06 & \cellcolor{cyan!15} 0.45 & \cellcolor{cyan!15} 5.10 & \cellcolor{cyan!15} 27.50 & \cellcolor{cyan!15} 3.65 & \cellcolor{cyan!15} 7.15 \\
& \cellcolor{cyan!15} Adv-Pretrain & \cellcolor{cyan!15} 93.41 & \cellcolor{cyan!15} 95.01 & \cellcolor{cyan!15} 85.35 & \cellcolor{cyan!15} 94.59 & \cellcolor{cyan!15} 92.53 & \cellcolor{cyan!15} 99.19 & \cellcolor{cyan!15} 99.14 & \cellcolor{cyan!15} 53.46 & \cellcolor{cyan!15} 94.59 & \cellcolor{cyan!15} 67.67 & \cellcolor{cyan!15} 4.54 & \cellcolor{cyan!15} 5.81 & \cellcolor{cyan!15} 32.15 & \cellcolor{cyan!15} 8.40 & \cellcolor{cyan!15} 27.42 & \cellcolor{cyan!15} 0.20 & \cellcolor{cyan!15} 4.12 & \cellcolor{cyan!15} 26.88 & \cellcolor{cyan!15} 2.92 & \cellcolor{cyan!15} 6.71 \\
\bottomrule
\end{tabular}
}
\vskip -0.05in
\end{table*}

\subsection{Overhead of Adversarial Purification}
\label{appendix:overhead}
While our adversarial pre-training approach, as defined in \textcolor{cyan}{\autoref{eq:adversarial_purify}}, 
provides a permanent and robust elimination of transferable backdoors, we acknowledge the importance of evaluating its computational overhead and convergence properties.

\subsection{Computational Cost and Training Time.}
Unlike inference-time monitoring defenses, the offline model purification phase requires parameter optimization. However, our evaluation demonstrates that this process is highly efficient. Based on our experimental setup utilizing NVIDIA H800 GPUs, purifying a \texttt{bert-base-uncased} model via adversarial pre-training requires only 15 GB of GPU memory. The training time is remarkably brief, taking approximately 5 minutes and 32 seconds per epoch. Since this is a one-time offline process executed prior to downstream deployment, the total computational overhead is negligible compared to the initial pre-training costs of the PLMs.

\textbf{Convergence Efficiency.}
Despite optimizing the entire model parameter space, the purification process converges rapidly without the need for prolonged training. The strong regularization provided by the fidelity constraint ($\mathcal{L}_{fidelity}$) ensures that the model quickly locates a safe parameter subspace. \textcolor{cyan}{\autoref{fig:adv_pretrain_loss}} shows that the model reaches optimal convergence in exactly 3 epochs. Within this short window, the training purification loss ($\mathcal{L}_{cleanse}$) drops sharply from 0.151 to 0.015, while the validation purification loss stabilizes near zero ($0.0006$). This rapid convergence confirms that unraveling the backdoor feature manifold does not demand extensive optimization steps.

\textbf{Task-Dependent Fine-Tuning Overhead:}
Regarding the adversarial fine-tuning stage (\textcolor{cyan}{\autoref{eq:fine_tuning}}), the computational time and memory footprint naturally vary depending on the specific downstream task and the size of its dataset. However, because this phase only incorporates pre-detected triggers into the existing task dataset, the computational complexity scales linearly with the standard fine-tuning process. The overhead is strictly bounded by the downstream task itself, meaning that if a user has the resources to fine-tune a model on a specific task, they fundamentally possess the resources to execute our adversarial fine-tuning defense.

\subsection{Generalization on LLMs}
As illustrated in \textcolor{cyan}{\autoref{fig:llm_cos}}, we first validate the correctness of $\gamma_1$ and $\gamma_2$ on large language models (LLMs): the selected values of 0.4 and 0.9 still ensure a trigger detection accuracy of 95.3\%. \textcolor{cyan}{\autoref{tab:llm_backdoor_defense}} evaluates the defense effectiveness under different three attack scenarios.

\subsection{Generalization on Embedding Models}
As shown in \textcolor{cyan}{\autoref{tab:em_backdoor_defense}}, we apply \sys to bge-large-en-v1.5 \footnote{https://huggingface.co/BAAI/bge-large-en-v1.5} across multiple datasets under the POR-1 attack. Without defense, the ASR reached 100\% on datasets like IMDB and SST-2. However, implementing our adversarial pre-training and fine-tuning successfully reduced the ASR to baseline clean levels while preserving high clean accuracy. This confirms that \sys remains highly effective for the foundational embedding mechanisms powering contemporary NLP systems.

%% file: conf_ndss.bbl
\begin{thebibliography}{10}
\providecommand{\url}[1]{#1}
\csname url@samestyle\endcsname
\providecommand{\newblock}{\relax}
\providecommand{\bibinfo}[2]{#2}
\providecommand{\BIBentrySTDinterwordspacing}{\spaceskip=0pt\relax}
\providecommand{\BIBentryALTinterwordstretchfactor}{4}
\providecommand{\BIBentryALTinterwordspacing}{\spaceskip=\fontdimen2\font plus
\BIBentryALTinterwordstretchfactor\fontdimen3\font minus \fontdimen4\font\relax}
\providecommand{\BIBforeignlanguage}[2]{{%
\expandafter\ifx\csname l@#1\endcsname\relax
\typeout{** WARNING: IEEEtran.bst: No hyphenation pattern has been}%
\typeout{** loaded for the language `#1'. Using the pattern for}%
\typeout{** the default language instead.}%
\else
\language=\csname l@#1\endcsname
\fi
#2}}
\providecommand{\BIBdecl}{\relax}
\BIBdecl

\bibitem{kurita2020weight}
K.~Kurita, P.~Michel, and G.~Neubig, ``Weight poisoning attacks on pretrained models,'' in \emph{Proceedings of the 58th Annual Meeting of the Association for Computational Linguistics}, 2020, pp. 2793--2806.

\bibitem{zhang2021trojaning}
X.~Zhang, Z.~Zhang, S.~Ji, and T.~Wang, ``Trojaning language models for fun and profit,'' in \emph{2021 IEEE European Symposium on Security and Privacy (EuroS\&P)}.\hskip 1em plus 0.5em minus 0.4em\relax IEEE, 2021, pp. 179--197.

\bibitem{li2021backdoor}
L.~Li, D.~Song, X.~Li, J.~Zeng, R.~Ma, and X.~Qiu, ``Backdoor attacks on pre-trained models by layerwise weight poisoning,'' in \emph{Proceedings of the 2021 Conference on Empirical Methods in Natural Language Processing}, 2021, pp. 3023--3032.

\bibitem{zhao2022fedprompt}
H.~Zhao, W.~Du, F.~Li, P.~Li, and G.~Liu, ``Fedprompt: Communication-efficient and privacy preserving prompt tuning in federated learning,'' \emph{arXiv preprint arXiv:2208.12268}, 2022.

\bibitem{chen2022kallima}
X.~Chen, Y.~Dong, Z.~Sun, S.~Zhai, Q.~Shen, and Z.~Wu, ``Kallima: A clean-label framework for textual backdoor attacks,'' in \emph{European symposium on research in computer security}.\hskip 1em plus 0.5em minus 0.4em\relax Springer, 2022, pp. 447--466.

\bibitem{guo2022threats}
S.~Guo, C.~Xie, J.~Li, L.~Lyu, and T.~Zhang, ``Threats to pre-trained language models: Survey and taxonomy,'' \emph{arXiv preprint arXiv:2202.06862}, 2022.

\bibitem{socher2013recursive}
R.~Socher, A.~Perelygin, J.~Wu, J.~Chuang, C.~D. Manning, A.~Y. Ng, and C.~Potts, ``Recursive deep models for semantic compositionality over a sentiment treebank,'' in \emph{Proceedings of the 2013 conference on empirical methods in natural language processing}, 2013, pp. 1631--1642.

\bibitem{wei2024bdmmt}
J.~Wei, M.~Fan, W.~Jiao, W.~Jin, and T.~Liu, ``Bdmmt: Backdoor sample detection for language models through model mutation testing,'' \emph{IEEE Transactions on Information Forensics and Security}, vol.~19, pp. 4285--4300, 2024.

\bibitem{2025arXiv250809456L}
J.~{Li}, B.~{Xu}, S.~{Chen}, J.~{Li}, J.~{Lei}, H.~{Zhao}, and D.~{Zhang}, ``{IAG: Input-aware Backdoor Attack on VLM-based Visual Grounding},'' \emph{arXiv preprint arXiv:2508.09456}, 2025.

\bibitem{weilmsanitator}
C.~Wei, W.~Meng, Z.~Zhang, M.~Chen, M.~Zhao, W.~Fang, L.~Wang, Z.~Zhang, and W.~Chen, ``Lmsanitator: Defending prompt-tuning against task-agnostic backdoors,'' in \emph{31st Annual Network and Distributed System Security Symposium, {NDSS} 2024, San Diego, California, USA, February 26 - March 1, 2024}.\hskip 1em plus 0.5em minus 0.4em\relax The Internet Society, 2024.

\bibitem{zhu2023removing}
B.~Zhu, G.~Cui, Y.~Chen, Y.~Qin, L.~Yuan, C.~Fu, Y.~Deng, Z.~Liu, M.~Sun, and M.~Gu, ``Removing backdoors in pre-trained models by regularized continual pre-training,'' \emph{Transactions of the Association for Computational Linguistics}, vol.~11, pp. 1608--1623, 2023.

\bibitem{kim2024obliviate}
J.~Kim, M.~Song, S.~H. Na, and S.~Shin, ``Obliviate: Neutralizing task-agnostic backdoors within the parameter-efficient fine-tuning paradigm,'' in \emph{Findings of the Association for Computational Linguistics: NAACL 2025}, 2025, pp. 1288--1307.

\bibitem{singh2024rethinking}
C.~Singh, J.~P. Inala, M.~Galley, R.~Caruana, and J.~Gao, ``Rethinking interpretability in the era of large language models,'' \emph{arXiv preprint arXiv:2402.01761}, 2024.

\bibitem{cheng2025backdoor}
P.~Cheng, Z.~Wu, W.~Du, H.~Zhao, W.~Lu, and G.~Liu, ``Backdoor attacks and countermeasures in natural language processing models: A comprehensive security review,'' \emph{IEEE Transactions on Neural Networks and Learning Systems}, 2025.

\bibitem{wen2024privacy}
Y.~Wen, L.~Marchyok, S.~Hong, J.~Geiping, T.~Goldstein, and N.~Carlini, ``Privacy backdoors: Enhancing membership inference through poisoning pre-trained models,'' \emph{Advances in Neural Information Processing Systems}, vol.~37, pp. 83\,374--83\,396, 2024.

\bibitem{dong2024trojaningplugins}
T.~Dong, M.~Xue, G.~Chen, R.~Holland, Y.~Meng, S.~Li, Z.~Liu, and H.~Zhu, ``The philosopher’s stone: Trojaning plugins of large language models,'' in \emph{Network and Distributed System Security Symposium, {NDSS} 2025}.\hskip 1em plus 0.5em minus 0.4em\relax The Internet Society, 2025.

\bibitem{shen2021backdoor}
L.~Shen, S.~Ji, X.~Zhang, J.~Li, J.~Chen, J.~Shi, C.~Fang, J.~Yin, and T.~Wang, ``Backdoor pre-trained models can transfer to all,'' \emph{arXiv preprint arXiv:2111.00197}, 2021.

\bibitem{ebrahimi2018hotflip}
J.~Ebrahimi, A.~Rao, D.~Lowd, and D.~Dou, ``Hotflip: White-box adversarial examples for text classification,'' in \emph{Proceedings of the 56th Annual Meeting of the Association for Computational Linguistics (Volume 2: Short Papers)}, 2018, pp. 31--36.

\bibitem{liu2022piccolo}
Y.~Liu, G.~Shen, G.~Tao, S.~An, S.~Ma, and X.~Zhang, ``Piccolo: Exposing complex backdoors in nlp transformer models,'' in \emph{2022 IEEE Symposium on Security and Privacy (SP)}.\hskip 1em plus 0.5em minus 0.4em\relax IEEE, 2022, pp. 2025--2042.

\bibitem{liu2019roberta}
Y.~Liu, M.~Ott, N.~Goyal, J.~Du, M.~Joshi, D.~Chen, O.~Levy, M.~Lewis, L.~Zettlemoyer, and V.~Stoyanov, ``Roberta: A robustly optimized bert pretraining approach,'' \emph{arXiv preprint arXiv:1907.11692}, 2019.

\bibitem{he2020deberta}
P.~He, X.~Liu, J.~Gao, and W.~Chen, ``Deberta: Decoding-enhanced bert with disentangled attention,'' in \emph{ICLR}, 2020.

\bibitem{zhang2023red}
Z.~Zhang, G.~Xiao, Y.~Li, T.~Lv, F.~Qi, Z.~Liu, Y.~Wang, X.~Jiang, and M.~Sun, ``Red alarm for pre-trained models: Universal vulnerability to neuron-level backdoor attacks,'' \emph{Machine Intelligence Research}, vol.~20, no.~2, pp. 180--193, 2023.

\bibitem{merity2017pointer}
S.~Merity, C.~Xiong, J.~Bradbury, and R.~Socher, ``Pointer sentinel mixture models,'' in \emph{ICLR}, 2017.

\bibitem{qi2021onion}
F.~Qi, Y.~Chen, M.~Li, Y.~Yao, Z.~Liu, and M.~Sun, ``Onion: A simple and effective defense against textual backdoor attacks,'' in \emph{Proceedings of the 2021 conference on empirical methods in natural language processing}, 2021, pp. 9558--9566.

\bibitem{liu2018fine}
K.~Liu, B.~Dolan-Gavitt, and S.~Garg, ``Fine-pruning: Defending against backdooring attacks on deep neural networks,'' in \emph{International symposium on research in attacks, intrusions, and defenses}.\hskip 1em plus 0.5em minus 0.4em\relax Springer, 2018, pp. 273--294.

\bibitem{jiang2025backdoor}
P.~Jiang, X.~Lyu, Y.~Li, and J.~Ma, ``Backdoor token unlearning: Exposing and defending backdoors in pretrained language models,'' in \emph{Proceedings of the AAAI Conference on Artificial Intelligence}, vol.~39, no.~23, 2025, pp. 24\,285--24\,293.

\bibitem{das2025unmasking}
A.~S. Das, K.~Chen, and M.~Bhuyan, ``Unmasking backdoors: An explainable defense via gradient-attention anomaly scoring for pre-trained language models,'' \emph{arXiv preprint arXiv:2510.04347}, 2025.

\bibitem{mackenzie2020cc}
J.~Mackenzie, R.~Benham, M.~Petri, J.~R. Trippas, J.~S. Culpepper, and A.~Moffat, ``Cc-news-en: A large english news corpus,'' in \emph{Proceedings of the 29th ACM International Conference on Information \& Knowledge Management}, 2020, pp. 3077--3084.

\bibitem{robert2004monte}
C.~P. Robert, G.~Casella, and G.~Casella, \emph{Monte Carlo statistical methods}.\hskip 1em plus 0.5em minus 0.4em\relax Springer, 2004, vol.~2.

\bibitem{li2021hidden}
S.~Li, H.~Liu, T.~Dong, B.~Z.~H. Zhao, M.~Xue, H.~Zhu, and J.~Lu, ``Hidden backdoors in human-centric language models,'' in \emph{Proceedings of the 2021 ACM SIGSAC Conference on Computer and Communications Security}, 2021, pp. 3123--3140.

\bibitem{qi2021mind}
F.~Qi, Y.~Chen, X.~Zhang, M.~Li, Z.~Liu, and M.~Sun, ``Mind the style of text! adversarial and backdoor attacks based on text style transfer,'' \emph{arXiv preprint arXiv:2110.07139}, 2021.

\bibitem{jaccard1912distribution}
P.~Jaccard, ``The distribution of the flora in the alpine zone. 1,'' \emph{New phytologist}, vol.~11, no.~2, pp. 37--50, 1912.

\bibitem{pearson1901liii}
K.~Pearson, ``Liii. on lines and planes of closest fit to systems of points in space,'' \emph{The London, Edinburgh, and Dublin philosophical magazine and journal of science}, vol.~2, no.~11, pp. 559--572, 1901.

\bibitem{gu2017badnets}
T.~Gu, B.~Dolan-Gavitt, and S.~Garg, ``Badnets: Identifying vulnerabilities in the machine learning model supply chain,'' \emph{arXiv preprint arXiv:1708.06733}, 2017.

\bibitem{dai2019backdoor}
J.~Dai, C.~Chen, and Y.~Li, ``A backdoor attack against lstm-based text classification systems,'' \emph{IEEE Access}, vol.~7, pp. 138\,872--138\,878, 2019.

\bibitem{du2022ppt}
W.~Du, Y.~Zhao, B.~Li, G.~Liu, and S.~Wang, ``{PPT:} backdoor attacks on pre-trained models via poisoned prompt tuning,'' in \emph{Proceedings of the Thirty-First International Joint Conference on Artificial Intelligence, {IJCAI} 2022, Vienna, Austria, 23-29 July 2022}.\hskip 1em plus 0.5em minus 0.4em\relax ijcai.org, 2022, pp. 680--686.

\bibitem{mei2023notable}
K.~Mei, Z.~Li, Z.~Wang, Y.~Zhang, and S.~Ma, ``Notable: Transferable backdoor attacks against prompt-based nlp models,'' \emph{arXiv preprint arXiv:2305.17826}, 2023.

\bibitem{du2024nws}
W.~Du, T.~Yuan, H.~Zhao, and G.~Liu, ``Nws: Natural textual backdoor attacks via word substitution,'' in \emph{ICASSP 2024-2024 IEEE International Conference on Acoustics, Speech and Signal Processing (ICASSP)}.\hskip 1em plus 0.5em minus 0.4em\relax IEEE, 2024, pp. 4680--4684.

\bibitem{du2024uor}
W.~Du, P.~Li, H.~Zhao, T.~Ju, G.~Ren, and G.~Liu, ``Uor: Universal backdoor attacks on pre-trained language models,'' in \emph{Findings of the Association for Computational Linguistics: ACL 2024}, 2024, pp. 7865--7877.

\bibitem{yang2021rap}
W.~Yang, Y.~Lin, P.~Li, J.~Zhou, and X.~Sun, ``Rap: Robustness-aware perturbations for defending against backdoor attacks on nlp models,'' \emph{arXiv preprint arXiv:2110.07831}, 2021.

\bibitem{cui2022unified}
G.~Cui, L.~Yuan, B.~He, Y.~Chen, Z.~Liu, and M.~Sun, ``A unified evaluation of textual backdoor learning: Frameworks and benchmarks,'' \emph{Advances in Neural Information Processing Systems}, vol.~35, pp. 5009--5023, 2022.

\bibitem{zhang2022fine}
Z.~Zhang, L.~Lyu, X.~Ma, C.~Wang, and X.~Sun, ``Fine-mixing: Mitigating backdoors in fine-tuned language models,'' \emph{arXiv preprint arXiv:2210.09545}, 2022.

\bibitem{zhao2024defense}
X.~Zhao, D.~Xu, and S.~Yuan, ``Defense against backdoor attack on pre-trained language models via head pruning and attention normalization,'' in \emph{Forty-first International Conference on Machine Learning, {ICML} 2024, Vienna, Austria, July 21-27, 2024}.\hskip 1em plus 0.5em minus 0.4em\relax OpenReview.net, 2024.

\bibitem{arora2024here}
A.~Arora, X.~He, M.~Mozes, S.~Swain, M.~Dras, and Q.~Xu, ``Here's a free lunch: Sanitizing backdoored models with model merge,'' \emph{arXiv preprint arXiv:2402.19334}, 2024.

\bibitem{hu2025dup}
M.~Hu, Y.~Ding, Y.~Yang, L.~Chen, Y.~Jia, and S.~Zhao, ``Dup: Detection-guided unlearning for backdoor purification in language models,'' \emph{arXiv preprint arXiv:2508.01647}, 2025.

\bibitem{kong2025neural}
J.~Kong, H.~Fang, S.~Guo, C.~Qing, B.~Chen, B.~Wang, and S.-T. Xia, ``Neural antidote: Class-wise prompt tuning for purifying backdoors in pre-trained vision-language models,'' \emph{arXiv preprint arXiv:2502.19269}, 2025.

\bibitem{wang2025model}
H.~Wang, S.~Guo, J.~He, H.~Liu, T.~Zhang, and T.~Xiang, ``Model supply chain poisoning: Backdooring pre-trained models via embedding indistinguishability,'' in \emph{Proceedings of the ACM on Web Conference 2025}, 2025, pp. 840--851.

\bibitem{devlin2018bert}
J.~Devlin, M.-W. Chang, K.~Lee, and K.~Toutanova, ``Bert: Pre-training of deep bidirectional transformers for language understanding,'' \emph{arXiv:1810.04805}, 2018.

\bibitem{lewis2020bart}
M.~Lewis, Y.~Liu, N.~Goyal, M.~Ghazvininejad, A.~Mohamed, O.~Levy, V.~Stoyanov, and L.~Zettlemoyer, ``Bart: Denoising sequence-to-sequence pre-training for natural language generation, translation, and comprehension,'' in \emph{Proceedings of the 58th annual meeting of the association for computational linguistics}, 2020, pp. 7871--7880.

\bibitem{yang2019xlnet}
Z.~Yang, Z.~Dai, Y.~Yang, J.~G. Carbonell, R.~Salakhutdinov, and Q.~V. Le, ``Xlnet: Generalized autoregressive pretraining for language understanding,'' in \emph{Advances in Neural Information Processing Systems 32: Annual Conference on Neural Information Processing Systems 2019, NeurIPS 2019, December 8-14, 2019, Vancouver, BC, Canada}, 2019, pp. 5754--5764.

\bibitem{lan2019albert}
Z.~Lan, M.~Chen, S.~Goodman, K.~Gimpel, P.~Sharma, and R.~Soricut, ``{ALBERT:} {A} lite {BERT} for self-supervised learning of language representations,'' in \emph{8th International Conference on Learning Representations, {ICLR} 2020, Addis Ababa, Ethiopia, April 26-30, 2020}.\hskip 1em plus 0.5em minus 0.4em\relax OpenReview.net, 2020.

\bibitem{sanh2019distilbert}
V.~Sanh, L.~Debut, J.~Chaumond, and T.~Wolf, ``Distilbert, a distilled version of bert: smaller, faster, cheaper and lighter,'' \emph{arXiv preprint arXiv:1910.01108}, 2019.

\bibitem{zhang2019ernie}
Z.~Zhang, X.~Han, Z.~Liu, X.~Jiang, M.~Sun, and Q.~Liu, ``{ERNIE:} enhanced language representation with informative entities,'' in \emph{Proceedings of the 57th Conference of the Association for Computational Linguistics, {ACL} 2019, Florence, Italy, July 28- August 2, 2019, Volume 1: Long Papers}.\hskip 1em plus 0.5em minus 0.4em\relax Association for Computational Linguistics, 2019, pp. 1441--1451.

\bibitem{maas2011learning}
A.~Maas, R.~E. Daly, P.~T. Pham, D.~Huang, A.~Y. Ng, and C.~Potts, ``Learning word vectors for sentiment analysis,'' in \emph{Proceedings of the 49th annual meeting of the association for computational linguistics: Human language technologies}, 2011, pp. 142--150.

\bibitem{minaee2021deep}
S.~Minaee, N.~Kalchbrenner, and E.~a. Cambria, ``Deep learning--based text classification: a comprehensive review,'' \emph{ACM computing surveys (CSUR)}, vol.~54, no.~3, pp. 1--40, 2021.

\bibitem{zampieri2019semeval}
M.~Zampieri, S.~Malmasi, P.~Nakov, S.~Rosenthal, N.~Farra, and R.~Kumar, ``Semeval-2019 task 6: Identifying and categorizing offensive language in social media (offenseval),'' in \emph{Proceedings of the 13th International Workshop on Semantic Evaluation}, 2019, pp. 75--86.

\bibitem{founta2018large}
A.~Founta, C.~Djouvas, D.~Chatzakou, I.~Leontiadis, J.~Blackburn, G.~Stringhini, A.~Vakali, M.~Sirivianos, and N.~Kourtellis, ``Large scale crowdsourcing and characterization of twitter abusive behavior,'' in \emph{Proceedings of the international AAAI conference on web and social media}, vol.~12, no.~1, 2018.

\bibitem{metsis2006spam}
V.~Metsis, I.~Androutsopoulos, and G.~Paliouras, ``Spam filtering with naive bayes-which naive bayes?'' in \emph{CEAS}, vol.~17.\hskip 1em plus 0.5em minus 0.4em\relax Mountain View, CA, 2006, pp. 28--69.

\bibitem{sakkis2003memory}
G.~Sakkis, I.~Androutsopoulos, G.~Paliouras \emph{et~al.}, ``A memory-based approach to anti-spam filtering for mailing lists,'' \emph{Information retrieval}, vol.~6, no.~1, pp. 49--73, 2003.

\bibitem{zhang2015character}
X.~Zhang, J.~Zhao, and Y.~LeCun, ``Character-level convolutional networks for text classification,'' \emph{NeurIPS}, vol.~28, pp. 649--657, 2015.

\bibitem{mcinnes2018umap}
L.~McInnes, J.~Healy, N.~Saul, and L.~Gro{\ss}berger, ``Umap: Uniform manifold approximation and projection,'' \emph{Open Source Software}, vol.~3, no.~29, p. 861, 2018.

\end{thebibliography}
